\documentclass[10pt,a4paper]{article}
\usepackage[utf8]{inputenc}
\usepackage[english]{babel}
\usepackage{graphpap,amsmath,amsfonts,amssymb,epsfig,makeidx,graphicx,textcomp,cite,citeall,authblk,subfloat,float} 
\usepackage[left=1.5cm, right=1.5cm, top=1.5cm, bottom=1.5cm]{geometry}

\author[1]{{Soham Chandra} \thanks{E-mail addresses: soham.rs@presiuniv.ac.in ; sohamc07@gmail.com}}

\affil[1]{\textit{\normalsize{Department of Physics, Presidency University, 86/1 College Street, Kolkata -700 073, India}}}

\vspace{20pt}

\title{\textbf{Effect of the Uniform Random External Magnetic Field with Spatio-temporal Variation on Compensation in Ising Spin-1/2 Trilayered Square Ferrimagnet}}

\date{} 
\begin{document}
	\maketitle	
	\begin{abstract}
		Trilayered spin-1/2 Ising ferrimagnets are interesting \textit{thin} systems for \textit{compensation phenomenon}. In this work, a Metropolis Monte Carlo study is performed on the magnetic and thermodynamic response of such a system on square Bravais lattice, driven by uniform random external magnetic field with spatio-temporal variations. In two distinct configurations, the surface layers are made up of A and the mid-layer is made up of B atoms in a \textit{ABA} type stacking while in \textit{AAB} type stacking, the top-layer and the mid-layer is made up of A-atoms while the bottom layer is made up of B-atoms. The magnetic coupling between the like atoms (A-A and B-B) is ferromagnetic while between the unlike atoms (A-B), it is antiferromagnetic. For the time-dependent external uniform random field, the mean is always set to zero and the standard deviation is varied until spin-field energy is comparable to the dominant cooperative energy of the system. The findings show that the observed compensation and critical points shift and steady-state magnetic behaviours shift between N-, L-, P- and Q- etc. type of ferrimagnetic behaviours, depending upon the strength of external uniform random field. The compensation phenomenon even vanishes after crossing a finite threshold of standard deviation of the magnetic field for particular choices of the other controlling parameters. Thus islands of ferrimagnetic phase without compensation appear within the phase area with compensation of field-free case, in the 2D Hamiltonian parameter space. For both the configurations, the areas of such islands even grow with increasing standard deviation of the external field, $\sigma$, obeying a scaling relation of the form: $f(\sigma,A(\sigma))=\sigma^{-b}A(\sigma)$ with $b_{ABA}=1.958\pm 0.122$ and $b_{AAB}=1.783\pm 0.118$ . 
	\end{abstract}

\newpage
\twocolumn
\section{Introduction}
\label{sec_intro}
The Random Field Ising Model (RFIM) was introduced by Larkin \cite{Larkin} in 1970. In spite of its simplicity, such systems exhibit many interesting static and dynamic behaviour \cite{Belanger}. From intuitive domain wall arguments, Imry and Ma \cite{Imry} \&  Grinstein and Ma \cite{Grinstein}, suggested that lower critical dimension for RFIM, $d_{l} = 2$ which was consolidated by rigorous mathematics \cite{Fisher-Spencer}, and by Monte Carlo (MC) simulations \cite{Andelman}. In \cite{Houghton}, the dependence of the critical properties of the RFIM on the form of the distribution function of the random field has been emphasized. We have witnessed non-trivial results for different field distributions, e.g. the existence of a tricritical point in the strong disorder regime of the system, present only in the bimodal case \cite{Houghton,Aharony,Andelman2}.
Random field type phenomenology can be found in a variety of experimentally accessible disordered systems, such as: (a) structural phase transitions in random alloys \cite{Childress}; (b) frustration introduced by the disorder in interacting many body systems and several aspects of electronic transport in disordered insulators \cite{Efros}; (c) commensurate charge-density-wave systems with impurity pinning \cite{Fisher}; (d) systems near the metal-insulator transition \cite{Kirkpatrick,Pastor}; (e) melting of intercalates in layered compounds such as $TiS_{2}$ \cite{Suter} and (f) binary fluid mixtures in random porous media \cite{Maher}. These examples obviously lend credibility to such a simple model and attract the attention of the experimentalists. An interesting article by Sethna et al. \cite{Sethna} and references therein throw light on RFIM in the context of hysteresis. The simulational results and subsequent analyses explain how this model may describe some real phenomena e.g. different kinds of noises in magnets.


Mermin-Wagner theorem \cite{Mermin} had conceived the absence of intrinsic long range ferromagnetic (FM) and antiferromagnetic (AFM) order in 2D magnetic materials. Contradicting, recent experiments have established the presence of intrinsic 2D magnetism: (a) In atomically thin $Cr_{2}Ge_{2}Te_{6}$ \cite{Gong} and $CrI_{3}$ \cite{Huang}, we find the existence of finite low temperature long-range FM order; (b) In monolayer $VSe_{2}$, strong FM order is observed at room temperatures \cite{Bonilla}; (c) Strong magnetic anisotropy is responsible for the observed long-range AFM order in atomically thin $FePS_{3}$ \cite{Tian,Wang}. Apart from intrinsic magnetism, the focus also is on \textit{tunable} magnetism. Successful techniques in inducing FM order in intrinsically non-magnetic low-dimensional materials are charge or carrier doping \cite{Zhang,Cao,Seixas,Miao}. Application of electric field \cite{Jiang} in experiments and strain engineering \cite{Vatansever} by numerical calculations are also established to be successful in AFM to FM transition in monolayers. Another successful method in switchable magnetism is $Li$-intercalation. In a recent work by DFT based calculations \cite{Kabiraj}, induced ferrimagnetism by intercalation with $Li$, $Mg$ and $Li-Mg$ mixture in pristine, naturally AFM, $FeO_{2}$ monolayer (predicted by computational exfoliation from its bulk \cite{Mounet,Mounet-data}) is studied. Interested readers may find a few more interesting theoretical and computational works on 2D FM and AFM materials in References \cite{Hu,Zhu,Haastrup,Miao2}. That is why, recent scientific and technological interests in the behavior of \textit{thin} systems, where one dimension is significantly reduced than the other two, are growing. Many interesting results of magnetism in thin systems (e.g., ribbons and films) under RFIM have come up because of numerous experiments \cite{Puppin,Yang,Ryu,Merazzo,Lee,Lima,Bohn} in this direction. Sophisticated experimental techniques, e.g. atomic layer deposition (ALD) \cite{George}, pulsed laser deposition (PLD) \cite{Singh}, molecular beam epitaxy (MBE) \cite{Herman} and metalorganic chemical vapor deposition (MOCVD) \cite{Stringfellow} have made growth of bilayered \cite{Stier}, trilayered \cite{Leiner} and multilayered \cite{Sankowski,Maitra} systems a reality. 


Equilibrium (field-free and in presence of static fields) studies by numerical methods on the layered Ising ferrimagnetic systems on different lattice geometries has been performed in the recent past \cite{Oitmaa,Lv,Fadil,Diaz,Chandra}. In some of such systems, for certain combinations of the coupling strengths, we find two temperatures with zero bulk magnetization. One is the Critical temperature with zero sublattice magnetizations, consequently zero bulk magnetization. The other temperature, the \textit{Compensation temperature}, lower than the critical temperature, has vanishing bulk magnetization but the sublattice magnetizations has non-zero values. In \cite{Diaz}(a), by MFA and EFA and in \cite{Diaz}(b), by MC simulations with Wolff single cluster Algorithm, the authors have shown that under certain range of interaction strengths, different temperature dependencies of sublattice magnetisations cause the compensation point to appear in both ABA and AAB configurations. In \cite{Chandra}(a), it was hinted that there may exist underlying mathematical relations between Inverse absolute of reduced residual magnetisation (IARRM for brevity, which may be considered as an interesting physical quantity for such kind of systems) and parameters of the trilayered, $s=1/2$, Ising system and in \cite{Chandra}(b), certain functional forms describing the systematics of compensation, are proposed for both the AAB and ABA configurations, which agree fairly well with accepted numerical results of \cite{Diaz}(b). In \cite{Chandra}(c), for a triangular trilayered spin-1/2 ferrimagnet, the magnetic description by traditional Monte Carlo Simulation is shown to be in very good agreement with the description provided by IARRM and Temperature interval between Critical and Compensation temperatures (TICCT). The results in \cite{Chandra}(c) actually strengthens the conjecture proposed in \cite{Chandra}(b).\\
\indent Disorder is crucial in the description of spin models as departures from ideal systems is only natural. In the system of our study, disorder may appear in different aspects, such as: (a) in the number of spins interacting with a particular one may be different (dilution or creation of bond(s); interactions with other than nearest neighbours etc.) (b) the interaction strength between pairs of spins may be different (distances between them may become different at some of the sites) (c) change in the nature of spins in the ordered structure (A atoms getting doped by B or vice versa or by atoms with spin values other than $1/2$). Such various types of disorders may have effects that \textit{vary with time and be uncorrelated among the lattice sites}. A uniform random external magnetic field, with spatio-temporal variations, may effectively model such kinds of disordered trilayered ferrimagnetic systems with a dynamic Hamiltonian. In the literature, such studies are yet to be numerically performed on the system of this study. The scope of the present work is to shed light on the effects, that a site-dependent and time-varying uniform random external field may have, on the compensation phenomenon of the ABA and AAB type trilayered spin-$1/2$ Ising ferrimagnets. It will be interesting to observe how the phase separation curve reacts to the different strengths of the external field. This eventually leads to interesting kind of phase diagrams in the Hamiltonian parameter space, which for the present case, has the standard deviation (interchangably, strength) of the external uniform random field as another controlling variable.\\
\indent The rest of the paper is organized as follows. The model and the technical details of the simulation scheme is described in Section \ref{sec_model}. The analysis of simulational data and results are presented in Section \ref{sec_results}. Section \ref{sec_summary} contains the summary of the work.\\
\section{Model and Simulation Protocol}
\label{sec_model}
\begin{figure*}[!htb]
	\begin{center}
		\begin{tabular}{c}
			\textbf{(a)} \resizebox{7cm}{!}{\includegraphics[angle=0]{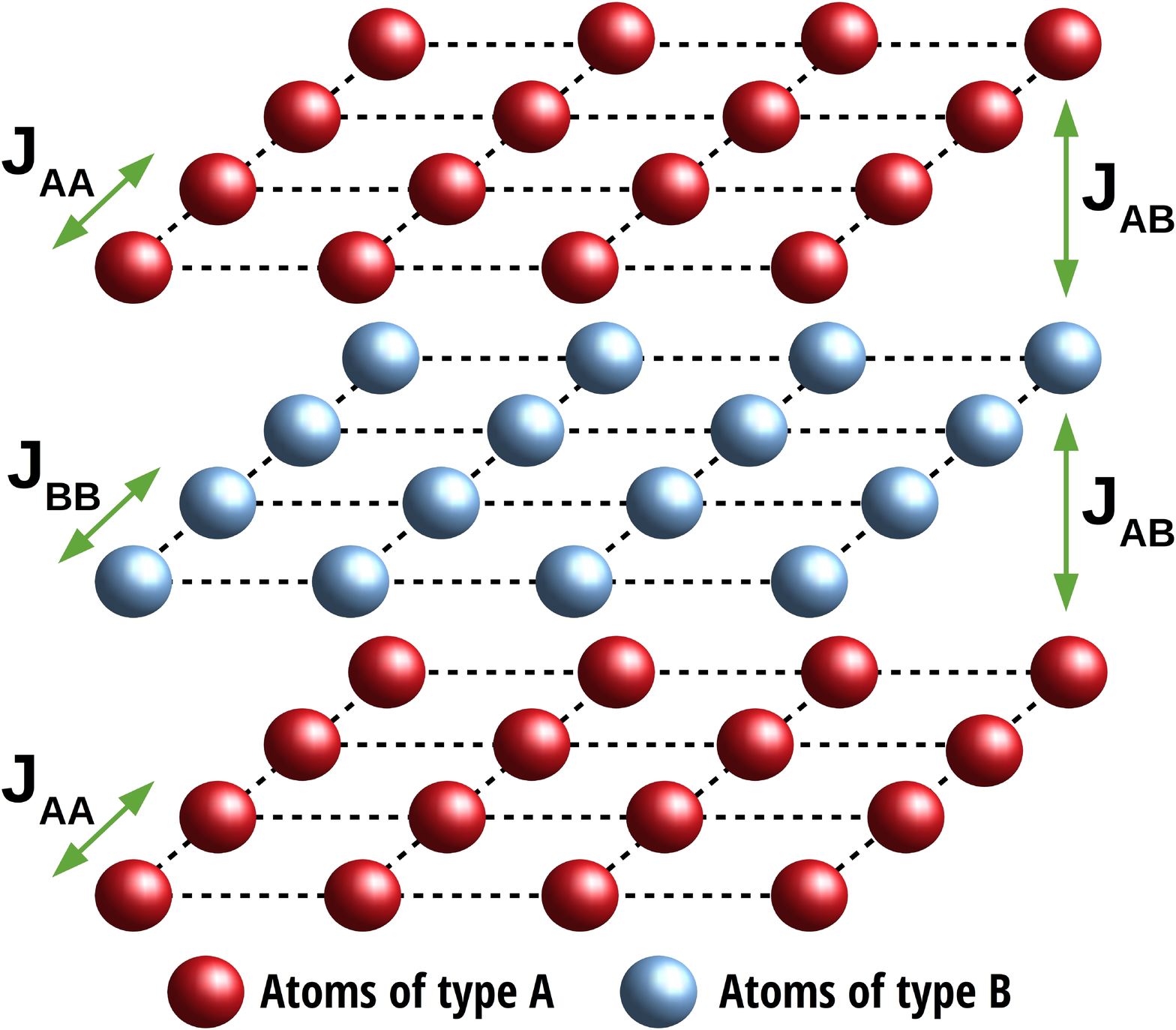}}
			\textbf{\hskip 1.5cm (b)}
			\resizebox{7cm}{!}{\includegraphics[angle=0]{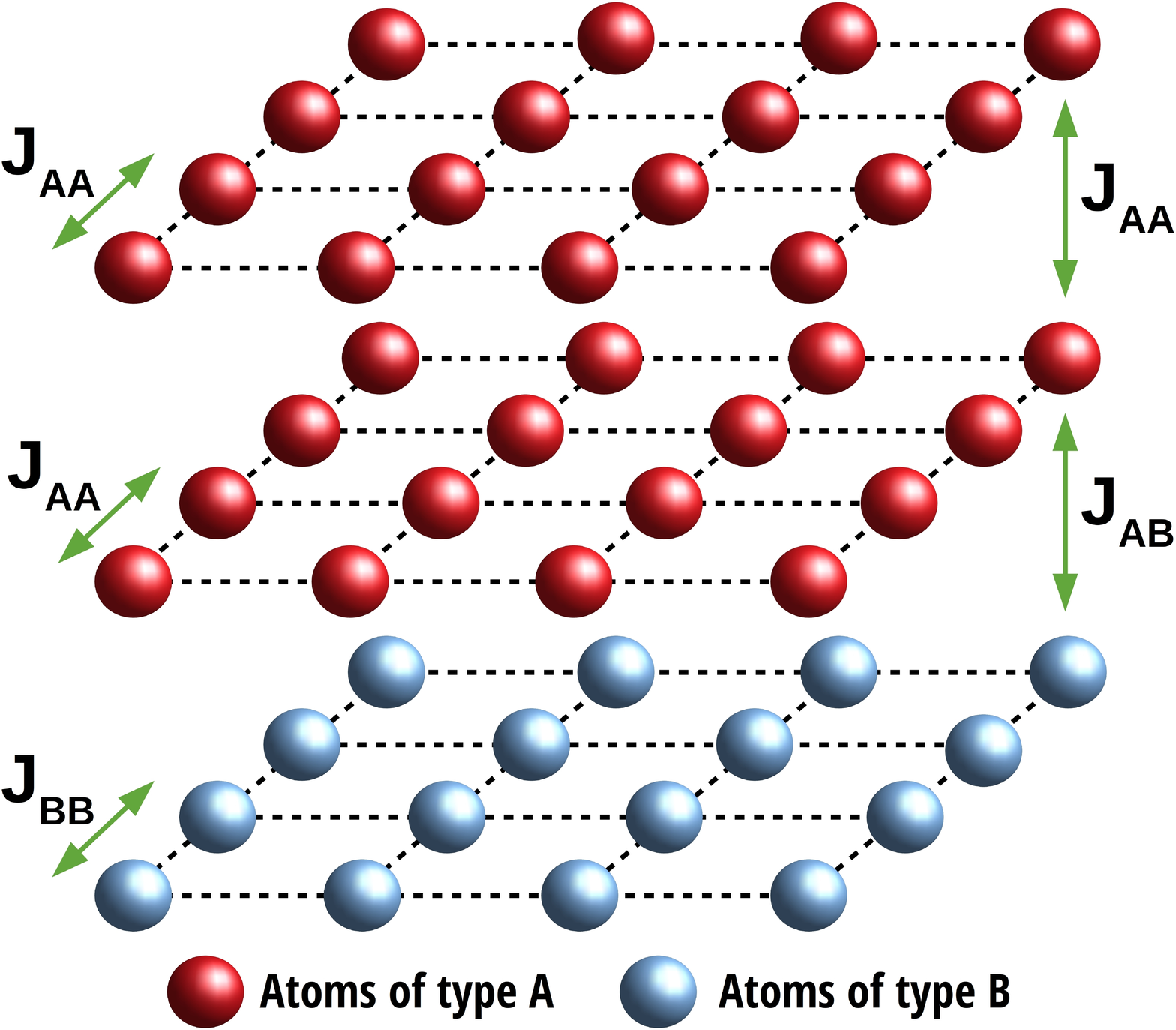}}
		\end{tabular}
		\caption{ (Colour Online) Miniaturised versions ($3\times4\times4$) of (a) ABA and (b) AAB square trilayered ferrimagnet with two types of theoretical atoms, $A$ and $B$. Each of the sublattices of the ferrimagnetic systems are formed on square lattice. The actual simulation is carried out on a system with $N_{sites}=3\times100\times100$ .}
		\label{fig_lattice_structure}
	\end{center}
\end{figure*}

\indent The ferrimagnetic Ising superlattice in this study (with each site having spin value, $s=1/2$), contains three magnetic sub-layers on square lattice with the following details:
\begin{itemize}
	\item[(a)] Each alternate layer is exhaustively composed of by either A or B type of atoms with \textit{no coupling between spins on top and bottom layers} [Fig.-\ref{fig_lattice_structure}].  
	
	\item[(b)] The system has mixed interactions between atoms throughout the bulk:\\
	A-A $\to$ Ferromagnetic\\
	B-B $\to$ Ferromagnetic\\
	A-B $\to$ Anti-ferromagnetic

	\item[(c)] At each of the sites, $i$, on every layer, the $z$-component of spins, $S_{i}^{z}$ couples with a uniform random external magnetic field, $h_{i}(t)$. This external field varies in time at a particular lattice site and at a frozen time instant, the values of this field are different from one site to another.
\end{itemize}

As we have considered the spins to interact Ising-like, in-plane as well as inter-plane, the time dependent Hamiltonian for the trilayered ferrimagnetic system is:
\begin{eqnarray}
\nonumber
& &H(t) = - 
J_{11}\sum_{<t,{t}^{\prime }>}S_{t}^{z}S_{{t}^{\prime }}^{z} - J_{22}\sum_{<m,{m}^{\prime }>}S_{m}^{z}S_{{m}^{\prime }}^{z} \\\nonumber
& - &
J_{33} \sum_{<b,{b}^{\prime }>}S_{b}^{z}S_{{b}^{\prime }}^{z} - 
J_{12} \sum_{<t,m>}S_{t}^{z}S_{m}^{z} - 
J_{23} \sum_{<m,b>}S_{m}^{z}S_{b}^{z} \\
\label{eq_Hamiltonian}
&-& \sum_{i}h_{i}(t)S_{i}^{z}
\end{eqnarray}
where $\langle t,{t}^{\prime }\rangle$, $\langle m,{m}^{\prime }\rangle$, $\langle b,{b}^{\prime }\rangle$ denote nearest-neighbor pairs in the top, mid and bottom layers respectively and $\langle t,m\rangle$, $\langle m,b\rangle$ are, respectively, pairs of nearest-neighbor sites in adjacent layers, top \& mid and mid \& bottom layers. At the right of Equation [\ref{eq_Hamiltonian}], the first three terms are for the intra-planar ferromagnetic contributions. The fourth and fifth terms arise out of the nearest neighbour inter-planar interactions, between top and mid layers and mid and bottom layers, respectively. The sixth term is the \textit{spin-field interaction term} of all the spins to the external random magnetic field, at time instant $t$. The summation index, $i$, runs over all the spins in the system. To satisfy the type of interactions, we need: $J_{AA}>0$ , $J_{BB}>0$, and $J_{AB}<0$. For an ABA type system, $J_{11}=J_{33}=J_{AA}$; $J_{22}=J_{BB}$ and $J_{12}=J_{23}=J_{AB}$. For an AAB type system, $J_{11}=J_{22}=J_{12}=J_{AA}$; $J_{33}=J_{BB}$ and $J_{23}=J_{AB}$. We've considered periodic boundary conditions in-plane and open boundary conditions along the vertical.


The Metropolis single spin-flip algorithm \cite{Landau, Binder} was employed to simulate the model. Each of the three planes has $L^{2}$ sites where the linear size, $L$ is 100. Each site is labelled by an integer index, say $i$, and the $z$-components of spin projections, $S_{i}^{z}$ $(S_{i}^{z}=\pm1)$ contribute to the interactions. At each site $i$, a \textit{local, time-varying random field} $h_{i}$ couples with the spin. In \cite{Diaz}(b), for $L\geqslant60$, the authors found the value of the compensation temperature to be practically constant for the system of this study. So the size of the lattice, considered here, is sufficient in obtaining statistically reliable results. The system was initiated at a high temperature paramagnetic phase, with randomly selected half of the total spin projections, $S_{i}^{z}=+1$ and the rest with $S_{i}^{z}=-1$ (Using $1$ instead of $1/2$ fixes up the energy scale). At a fixed temperature $T$, the Metropolis rate \cite{Metropolis, Newman} , of Equation [\ref{eq_metropolis}], governs the spin flipping from $S_{i}^{z}$ to $-S_{i}^{z}$:
\begin{equation}
\label{eq_metropolis}
P(S_{i}^{z} \to -S_{i}^{z}) = \text{min} \{1, \exp (-\Delta E/k_{B}T)\}
\end{equation}
where the associated change in internal energy in flipping the $i$-th spin projection from $S_{i}^{z}$ to $-S_{i}^{z}$, is $\Delta E$ . Similar $3L^{2}$ individual, random single-spin updates constitute One Monte Carlo sweep (MCS) of the entire system and this \textit{one MCS} is the unit of time in this study.\\
\indent At every temperature step, the system goes through $10^{5}$ MCS. The last configuration of the system at the just previous temperature acts as the starting configuration. For the first $5\times10^{4}$ MCS, the system is allowed to reach \textit{equilibrium} (which is sufficient for equilibration [Refer to Figure \ref{fig_field_mag_time} and discussions therein]) in a field-free environment. After that the external field is switched on and kept switched on for the next $5\times10^{4}$ MCS. So for the system, the \textit{exposure time interval} in the field, $\delta$ is $5\times10^{4}$. The temperatures of the systems are measured in units of $J_{BB}/k_{B}$. The tactics for observation, at first, include fixing the standard deviation (sd) or \textit{randomness} of the field. For each of the fixed values of the sd of the field, the system was observed for seven equidistant values of $J_{AA}/J_{BB}$, from $0.04$ to $1.0$ with an interval of $0.16$. For each fixed value of $J_{AA}/J_{BB}$, $J_{AB}/J_{BB}$ was varied from $-0.04$ to $-1.0$ with a decrement of $-0.16$ at each step.\\
\indent For each combination of $J_{AA}/J_{BB}$ and $J_{AB}/J_{BB}$, the time averages of the following quantities are calculated at each of the temperature steps $(T)$ and fields, in the following manner:\\
\textbf{(1) Sublattice magnetisations} for top, mid and bottom layers calculated, identically, at time instant say, $t$, after equilibration, denoted by $M_{q}(T,t)$, by:
\begin{equation}
M_{q}(T,t)=\frac{1}{L^{2}}\sum_{x,y=1}^{L} \left( S_{q}^{z}(T,t)\right)_{xy}
\end{equation}
Then we get the time averaged sublattice magnetizations at temperature, $T$, as:
\begin{equation}
\langle M_{q}(T)\rangle =\frac{1}{\delta}\int_{t_{0}}^{t_{0}+\delta} M_{q}(T,t)dt
\end{equation}
where $q$ is to be replaced by $t,m\text{ or }b$ for top, mid and bottom layers. The \textbf{order parameter}, $O_{T}$, for the trilayer at temperature, $T$ is defined as:
\begin{equation}
O_{T}=\frac{1}{3}(\langle M_{t}(T)\rangle+\langle M_{m}(T)\rangle+\langle M_{b}(T)\rangle)
\end{equation}
\\
\textbf{(2)} After attaining equilibrium, we calculate \textbf{fluctuation of the order parameter,} $\Delta O(T)$ at temperature, $T$ as follows \cite{Robb}:
\begin{equation}
{\Delta O}(T)=\sqrt{\dfrac{1}{\delta} \int_{t_{0}}^{t_{0}+\delta} \left[M(T,t)-O_{T}\right]^{2}dt }
\end{equation}
where $M(T,t)$ is the total magnetisation of the whole system, at temperature, $T$, calculated at the (\textit{t-th time instant}). On the similar lines,  $\langle E \rangle_{T}$, the time averaged value of cooperative energy, per site at temperature, $T$, is determined for the two configurations by:
\begin{eqnarray}
\nonumber 
& \langle E \rangle_{T}^{ABA} & = \dfrac{-1}{3L^{2}\delta}  \int_{t_{0}}^{t_{0}+\delta} dt [ J_{AA}(\sum_{<t,{t}^{\prime }>}S_{t}^{z}S_{{t}^{\prime }}^{z} 
\\\nonumber 
&+& \sum_{<b,{b}^{\prime }>}S_{b}^{z}S_{{b}^{\prime }}^{z}) + J_{BB} \sum_{<m,{m}^{\prime }>}S_{m}^{z}S_{{m}^{\prime }}^{z} 
\\
&+& J_{AB} (\sum_{<t,m>}S_{t}^{z}S_{m}^{z}+\sum_{<m,b>}S_{m}^{z}S_{b}^{z})] 
\end{eqnarray}
and
\begin{eqnarray}
\nonumber 
& \langle E \rangle_{T}^{AAB} & = \dfrac{-1}{3L^{2}\delta}  \int_{t_{0}}^{t_{0}+\delta} dt [ J_{AA}(\sum_{<t,{t}^{\prime }>}S_{t}^{z}S_{{t}^{\prime }}^{z}
\\\nonumber 
&+& \sum_{<m,{m}^{\prime }>}S_{m}^{z}S_{{m}^{\prime }}^{z} \sum_{<t,m>}S_{t}^{z}S_{m}^{z} ) \\
&+& J_{BB} \sum_{<b,{b}^{\prime }>}S_{b}^{z}S_{{b}^{\prime }}^{z} + J_{AB} \sum_{<m,b>}S_{m}^{z}S_{b}^{z} ]
\end{eqnarray}
These relations enable us to calculate the fluctuation of the cooperative energy per site at temperature, $T$, by the following formula:
\begin{equation}
{\Delta E}(T)=\sqrt{\dfrac{1}{\delta} \int_{t_{0}}^{t_{0}+\delta} \left[E(T,t)-\langle E \rangle_{T}\right]^{2}dt }
\end{equation}
with $E(T,t)$ being the instantaneous cooperative energy, per site, for the system at time instant $t$ and at temperature, $T$, residing within the exposure interval of $\delta$. The sharp peaks in the fluctuations, allow us to detect the pseudo-critical temperatures. Around this temperature close range simulations were performed with temperature interval of $0.02$ to land up on the \textit{reported} critical temperatures with an accuracy of, $\Delta T_{crit}=0.04$ . The compensation temperature ($<T_{crit}$), where the average magnetisation again becomes zero, is determined by linear interpolation from the two neighbouring points across the zero of magnetization in the plots of order parameter vs. temperature [e.g. Figure \ref{fig_mag_fr_afr}(a)]. The errors associated with the magnetizations and fluctuations are estimated by the Jackknife method \cite{Newman}. 
\section{Results and discussions}
\label{sec_results}
\subsection{Characteristics of the External field}
\label{subsec_char_field}
\indent The local, uniform random external magnetic field values $h_{i}(t)$ at any site, $i$ at time instant $t$, are drawn from the following symmetric probability distribution:
 \begin{equation}
	\label{eq_prob_dist_uniform}
	P_{uniform}(h_{i}(t)) = \begin{cases}
	\frac{1}{\sqrt{12}\sigma} \,\, &|h_{i}(t)| \le \sqrt{3}\sigma \\
	0 \,\, &|h_{i}(t)| > \sqrt{3}\sigma
	\end{cases}
\end{equation}

Here, $\sigma$ is the \textit{standard deviation} (sd) of the uniform random distribtion.

\indent The external field is also considered to have the following characteristics:
\begin{itemize}
	\item[(a)] The values of the external field are	uncorrelated for different sites at a particular time instant. Also at a particular lattice site, the values of the external field are	uncorrelated for different time instants. So these conditions can be conveniently written as :  $h_{p}(t)h_{q}(t^{\prime})=a(t)\text{ }\delta_{pq}\text{ }\delta (t-t^{\prime})$, where $p,q$ are two different lattice sites and $t,t^{\prime}$ are two different time instants. 
	
	\item[(b)] The following conditions are also trivially met:
	\begin{itemize}
		\item[(i)] After $t=t_{0}$ (when the field is switched ON), the spatial mean (equivalently, bulk average) of the above symmetric distribution of the random field, at any fixed time instant $t$, is zero as: $$\sum_{p} h_{p}(t) =0$$.\\
		Consequently, $$ \sum_{p,q} h_{p}(t)h_{q}(t)\delta_{pq} =3L^{2}\sigma^{2}$$.
		
		\item[(ii)] At the $p$-th site, the temporal mean of the local field $h_{p}(t)$ over the exposure interval, $\delta$, is also \textit{zero}: $\langle h_{p}(t)\rangle=\dfrac{1}{\delta}\int_{t_{0}}^{t_{0}+\delta}  h_{p}(t)dt=0$ .
		It has been checked the duration of the exposure interval satisfies the above condition.
	\end{itemize}	
	
\end{itemize}

\indent The reliability of implementation of such a probability distribution at a few randomly chosen time instants within the exposure interval, is then checked by the Cumulative Distribution Function (CDF), the Kernel Density Estimate (KDE) and the Histogram. Interested readers are referred to \cite{Rosenblatt-Parzen} for derivations of KDE and to \cite{Deisenroth} for discussions on CDF. The Histogram is the most intuitive  and traditional one in this regard. While the properly normalised CDF's show half of the total events (drawing of the values of fields) happen just before reaching $0$, for all the sampled time instants. Kernel density estimates indicate the number density in field intervals at all the selected time instants.

\subsection{Thermodynamic Response}
\label{subsec_response}
\begin{figure*}[!htb]
	\begin{center}
		\begin{tabular}{c}
			
			\resizebox{9.5cm}{!}{\includegraphics[angle=0]{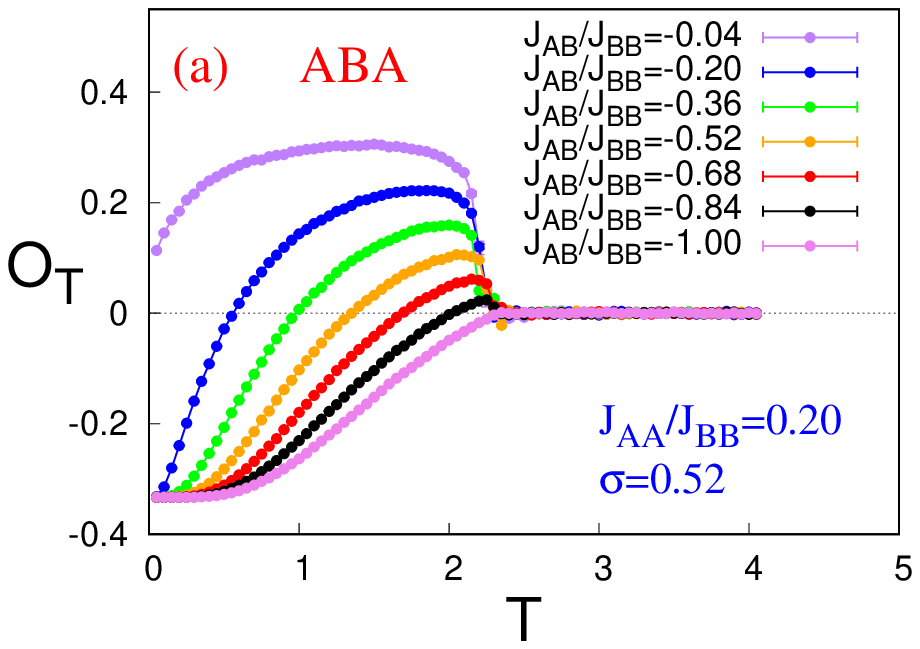}}
			\resizebox{9.5cm}{!}{\includegraphics[angle=0]{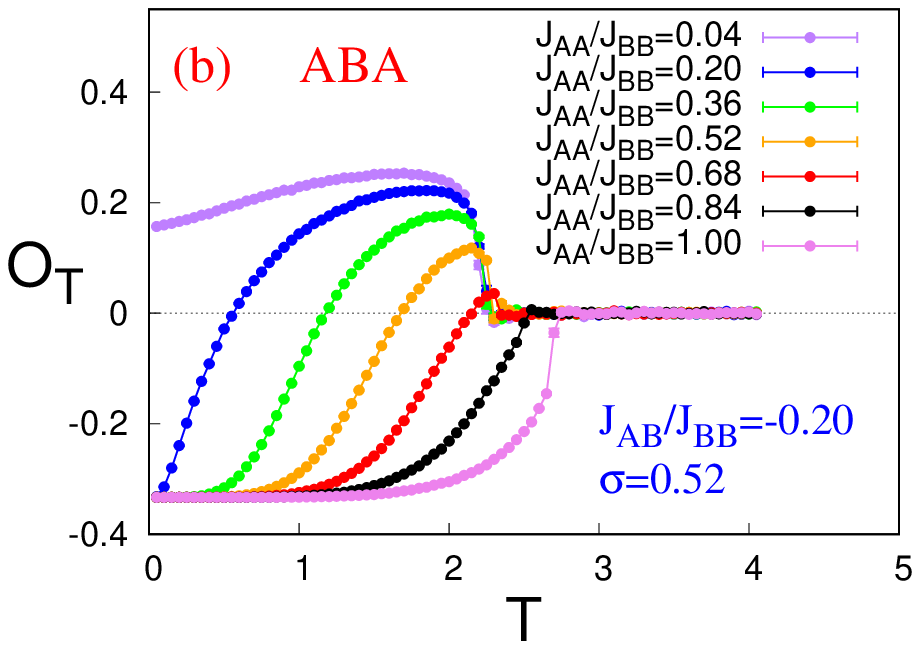}}\\
			
			\resizebox{9.5cm}{!}{\includegraphics[angle=0]{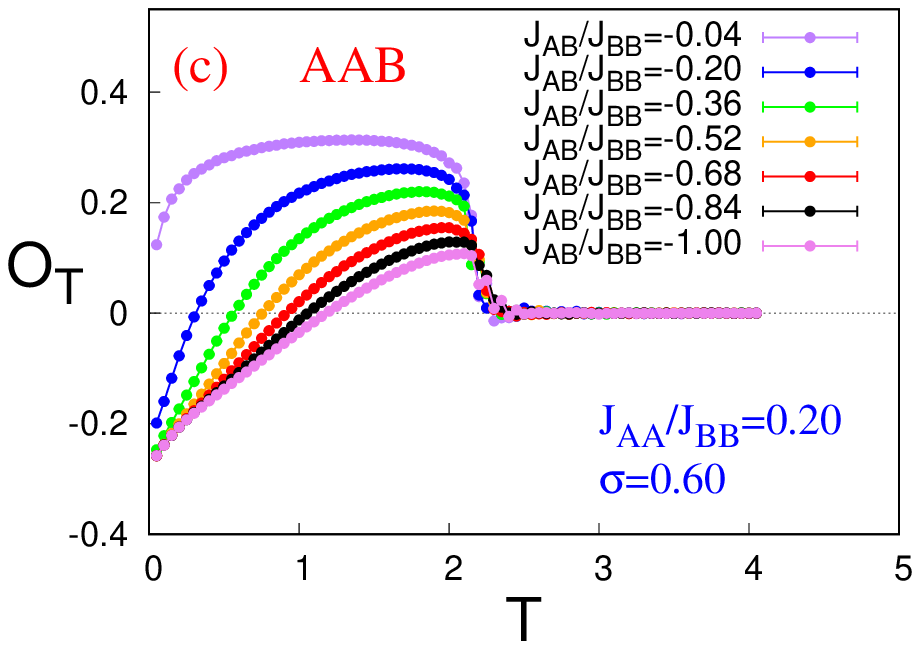}}
			\resizebox{9.5cm}{!}{\includegraphics[angle=0]{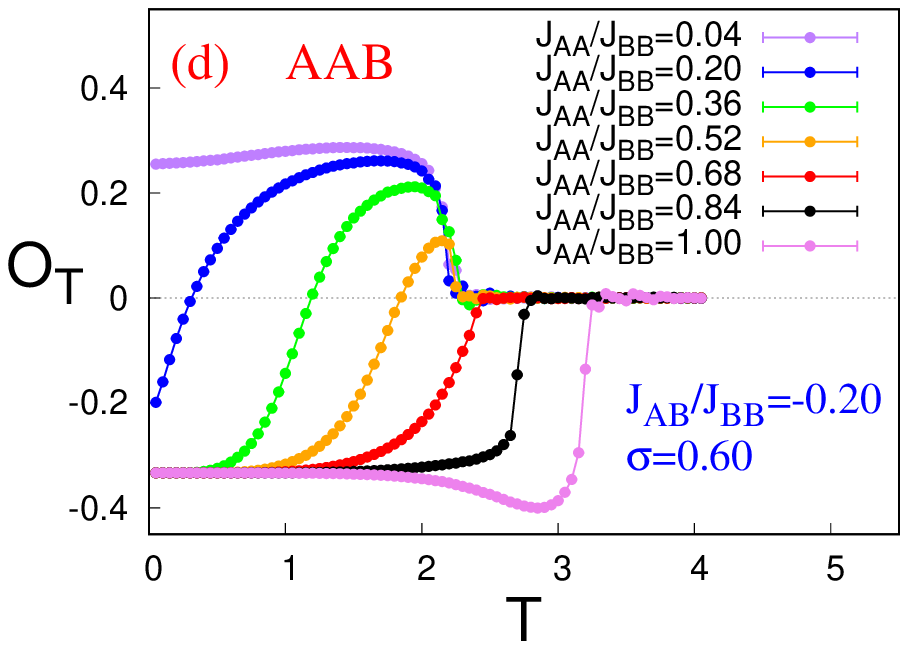}}
			
		\end{tabular}
		\caption{ (Colour Online) Order parameter (i.e. time averaged total magnetisation) versus reduced temperature for: (a) ABA: $J_{AA}/J_{BB}=0.20$ and variable $J_{AB}/J_{BB}$ for $\sigma=0.52$; (b) ABA: $J_{AB}/J_{BB}=-0.20$ and variable $J_{AA}/J_{BB}$ for $\sigma=0.52$; (c) AAB: $J_{AA}/J_{BB}=0.20$ and variable $J_{AB}/J_{BB}$ for $\sigma=0.60$; (d) AAB: $J_{AB}/J_{BB}=-0.20$ and variable $J_{AA}/J_{BB}$ for $\sigma=0.60$ . In all the cases Compensation and Critical temperatures shift towards higher temperature end with increase in any of the coupling ratios. The field-driven vanishing of compensation is also present for the weakest combination of coupling strengths.}
		\label{fig_mag_fr_afr}
	\end{center}
\end{figure*}

\begin{figure*}[!htb]
	\begin{center}
		\begin{tabular}{c}
			
			\resizebox{8.5cm}{!}{\includegraphics[angle=0]{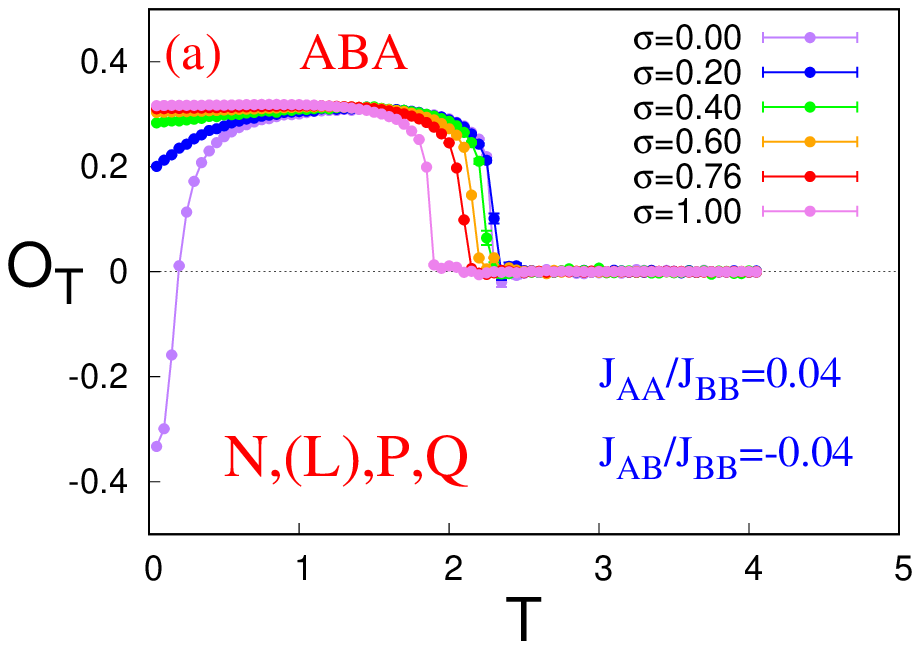}}
			\resizebox{8.5cm}{!}{\includegraphics[angle=0]{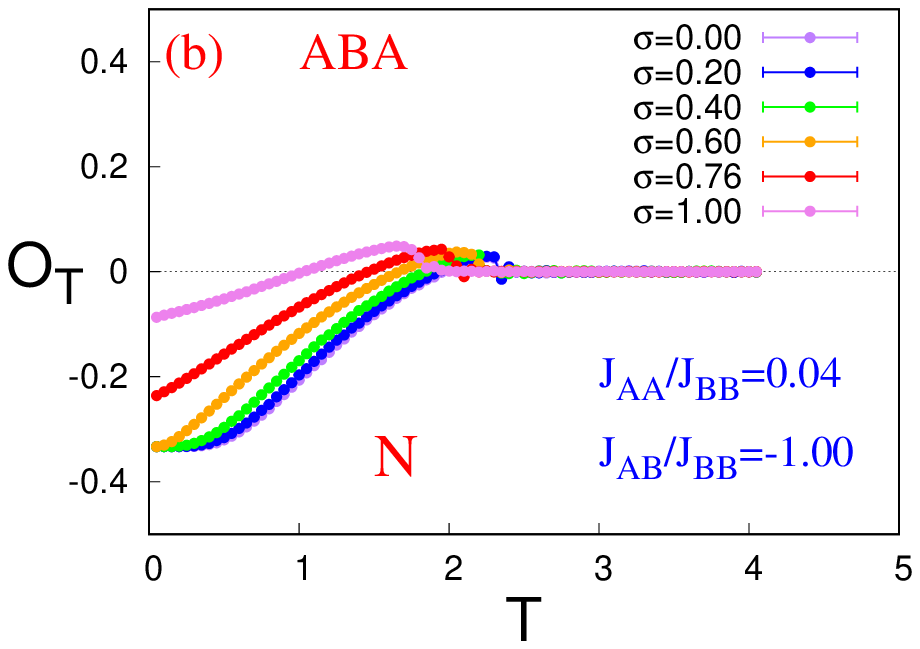}}\\
			
			\resizebox{8.5cm}{!}{\includegraphics[angle=0]{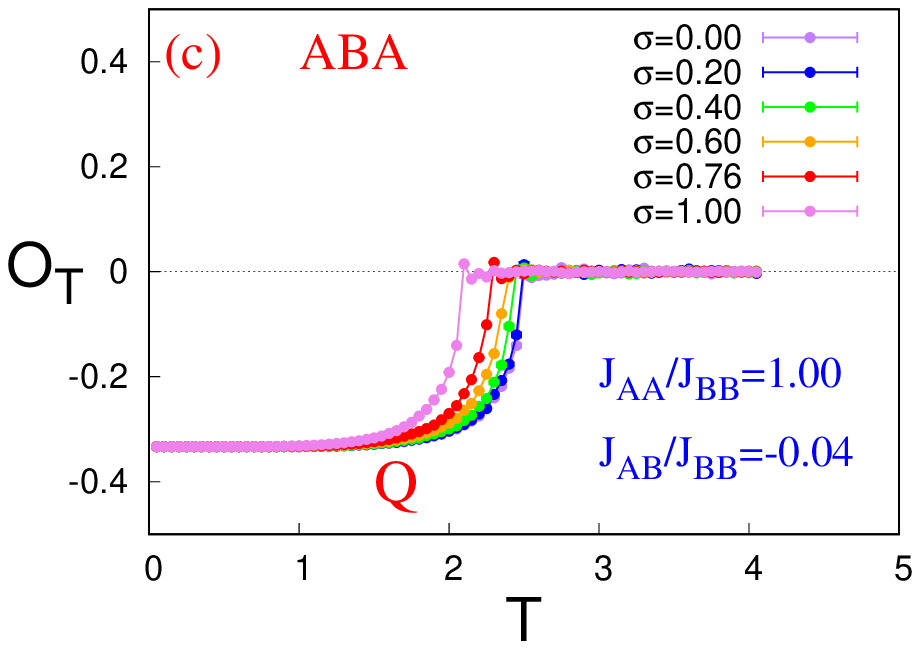}}
			\resizebox{8.5cm}{!}{\includegraphics[angle=0]{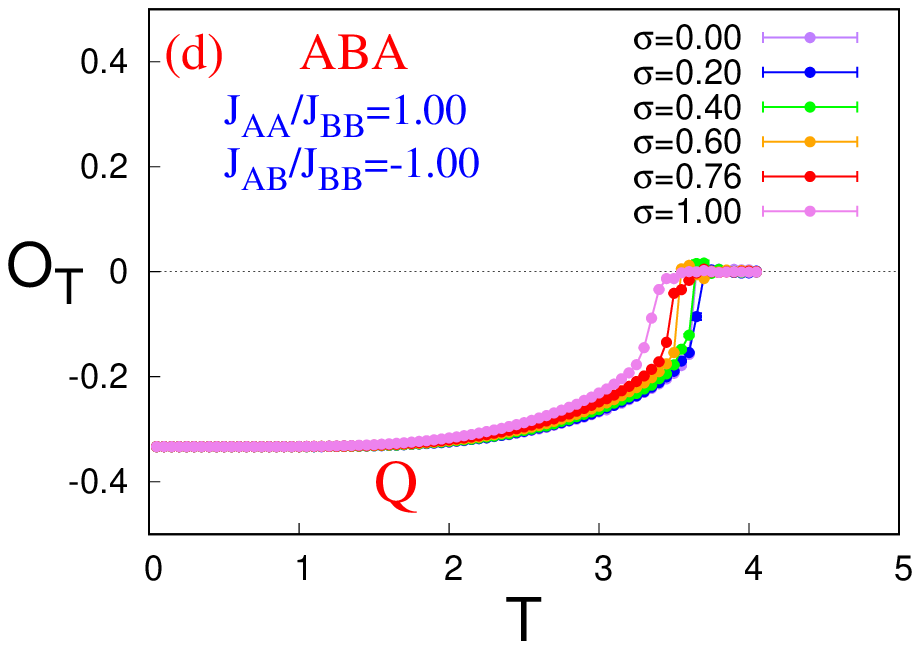}}\\
			
			\resizebox{8.5cm}{!}{\includegraphics[angle=0]{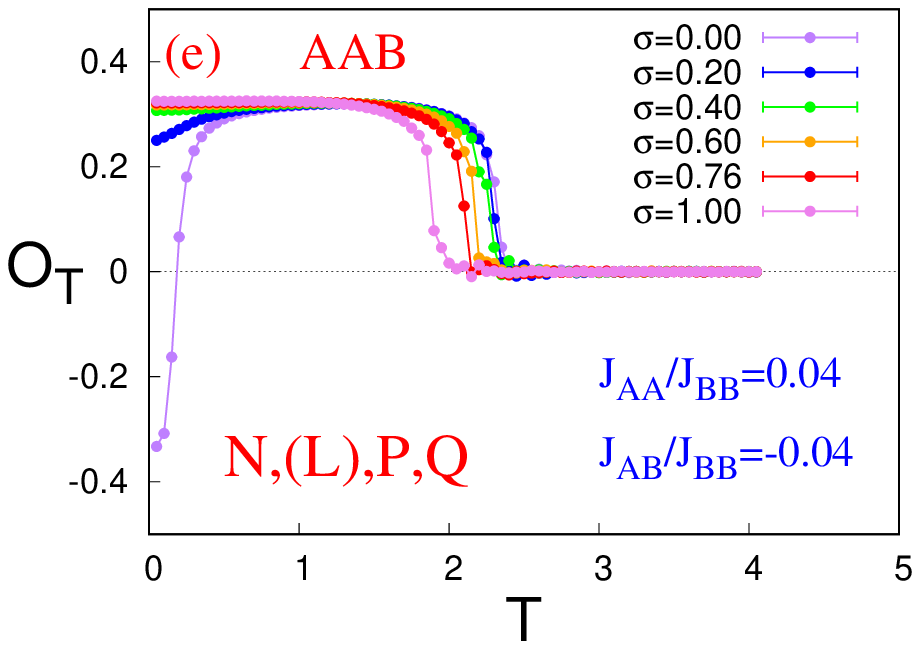}}
			\resizebox{8.5cm}{!}{\includegraphics[angle=0]{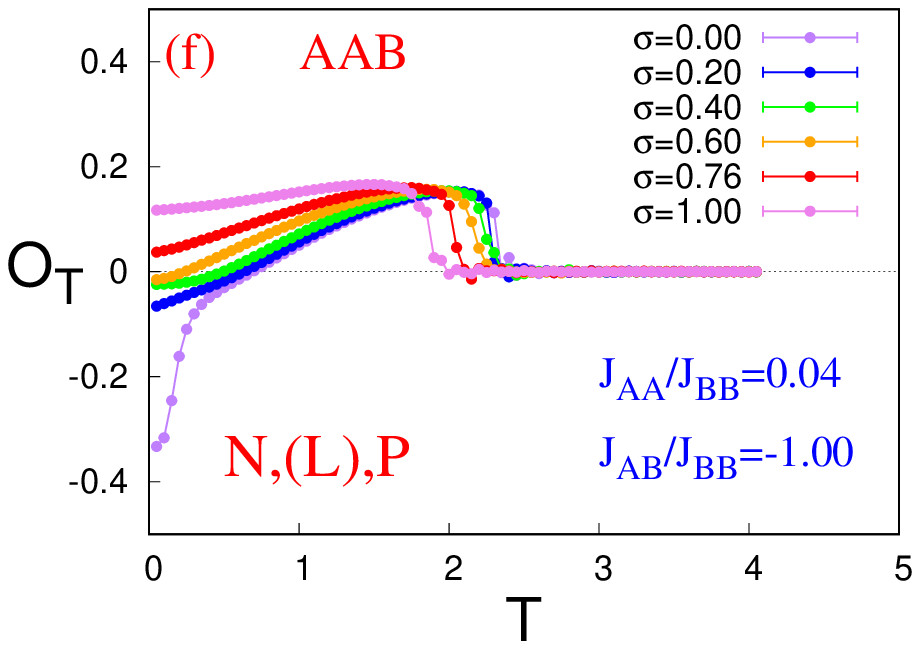}}\\
			
			\resizebox{8.5cm}{!}{\includegraphics[angle=0]{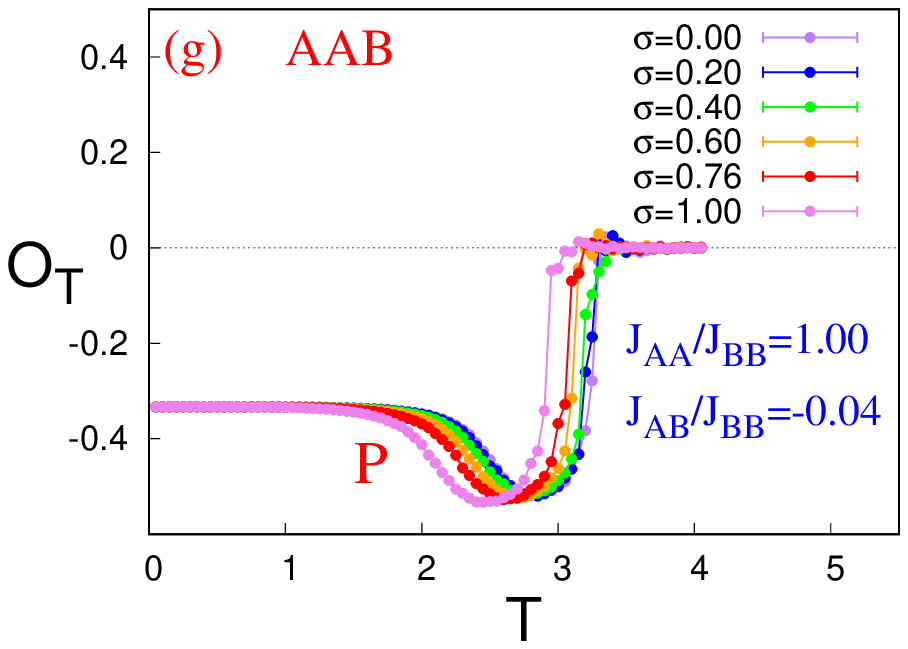}}
			\resizebox{8.5cm}{!}{\includegraphics[angle=0]{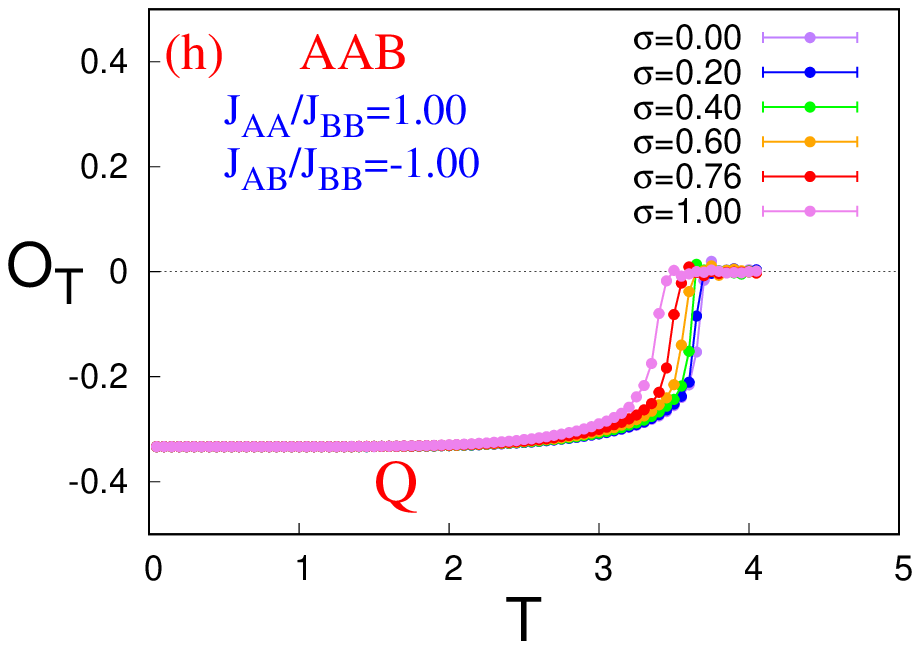}}
		\end{tabular}
		\caption{ (Colour Online) Magnetic response of the trilayered system for a few selected cases with: (a)-(d) ABA and (e)-(h) AAB. The shift of both, the compensation (where it is present) and critical temperatures towards the low temperature ends and shift of the magnetic behaviours between N,L,P,Q etc. type of ferrimagnetism, with increase in standard deviation of the uniform random external magnetic field, are clearly visible in all these plots. The type L within brackets is explicitly not seen in the plots but encountered in-transition. In (a) and (e): we witness the field-driven vanishing of compensation from $\sigma=0.20$ and upwards. Where, the errorbars are not visible, they are smaller than the area of the point-markers. All these plots are obtained for a system of $3\times100\times100$ sites. }
		\label{fig_mag_response}
	\end{center}
\end{figure*}

The zero-field magnetic response \cite{Diaz,Chandra} for the ABA and AAB type systems showed us, if $J_{AA}/J_{BB}$ is kept fixed and the \textit{magnitude} of $J_{AB}/J_{BB}$ is increased or vice-versa, both, the critical and the compensation temperatures for such systems, increases. This feature of shift of compensation and critical temepratures is retained when we apply an external uniform random field with desired characteristics of Section \ref{subsec_char_field} [Refer to Figure \ref{fig_mag_fr_afr}]. For any combination of coupling strengths, with increase in the value of standard deviation (or randomness) of the external field distribution, the compensation and critical, both the temperatures decrease [Refer to Figure \ref{fig_mag_response}]. With increment in the randomness of the external field, the magnitude of decrement for the compensation temperatures is much more pronounced than the corresponding magnitude of decrement of critical temperature, with or without compensation. The nature ferrimagnetic magnetisation versus temperature dependences change with a change in the standard deviation of the field and even a \textit{field driven absence of compensation phenomenon} can also be observed in Figures \ref{fig_mag_response}(a)\&(b) for ABA and AAB configrations respectively. This salient feature is very robust, as it is always present irrespective of which coupling ratio is kept fixed among $J_{AA}/J_{BB}$ and $J_{AB}/J_{BB}$. 

Now a few comments regarding the \textit{field-driven changes} in the nature of the magnetization curves of Figure \ref{fig_mag_response}, are in order after references \cite{Neel,Chikazumi,Strecka}. In Figure \ref{fig_mag_response}(a) for the ABA configuration, the magnetic response changes from type-$N$ $(\sigma =0)$ to type-$P$ $(\sigma ={0.20,0.40,0.60,0.76})$ to type-$Q$ $(\sigma =1.00)$. While going from from type-$N$ to type-$P$, the curve passes through type-$L$ where \textit{the magnetization at the lowest temperature is zero}. Exactly similar conclusions are drawn from the Figure \ref{fig_mag_response}(e) for the AAB configuration. For the ABA configurations, in Figure \ref{fig_mag_response}(b), the magnetic response is confined to type-$N$  and in Figures \ref{fig_mag_response}(c)\&(d), the magnetic responses are all type-$Q$ for all the fields of observation. Now for the AAB configuration, in Figure \ref{fig_mag_response}(f), the transition happens from type-$N$ $(\sigma={0.00,0.20,0.40,0.60})$ to type-$P$ $(\sigma={0.76,1.00})$ via type-$L$. In Figure \ref{fig_mag_response}(g), all the magnetic responses are of type-$P$ and in Figure \ref{fig_mag_response}(h), all the magnetic responses are of type-$Q$, for the AAB configuration.

Now it is worthy of looking at the fluctuation of the \textit{order parameter} and fluctuation of the \textit{cooperative energy per site} at this stage, for how they react according to the randomness of the external field as a function of temperature.
\begin{figure*}[!htb]
	\begin{center}
		\begin{tabular}{c}
			
			\resizebox{8.2cm}{!}{\includegraphics[angle=0]{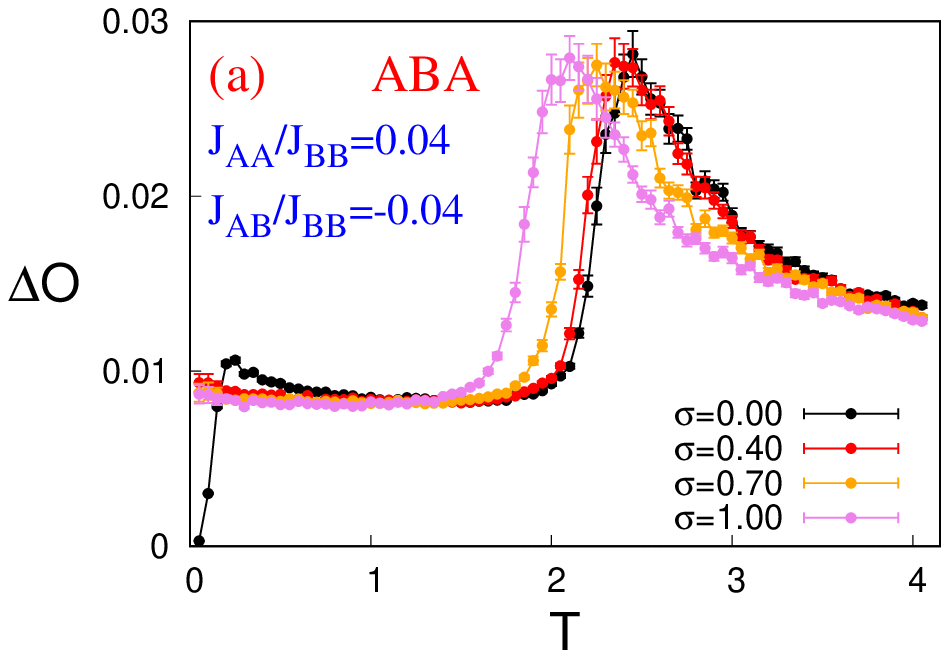}}
			
			\resizebox{8.2cm}{!}{\includegraphics[angle=0]{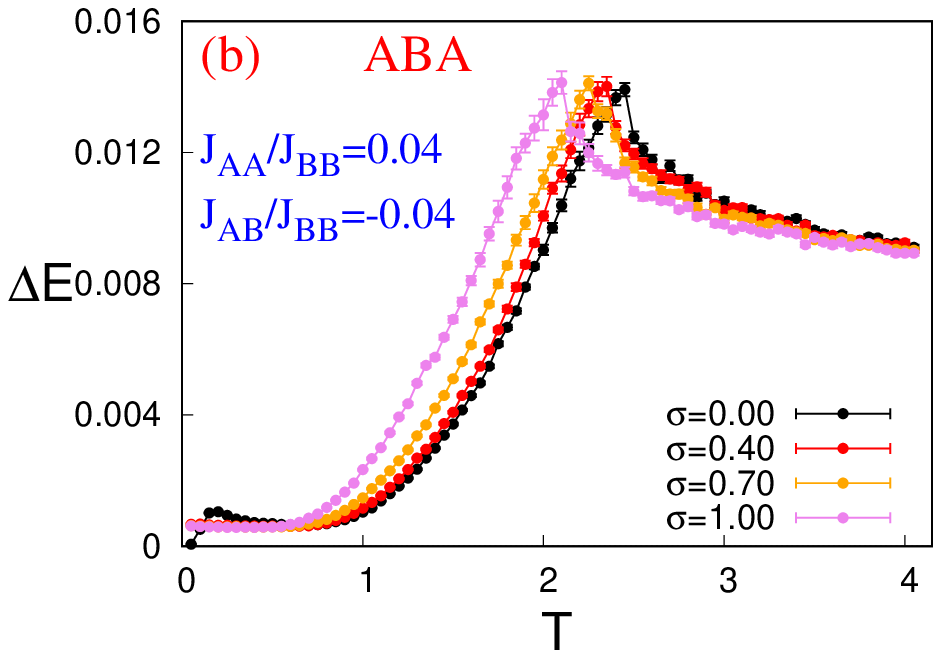}}\\
			
			\resizebox{8.2cm}{!}{\includegraphics[angle=0]{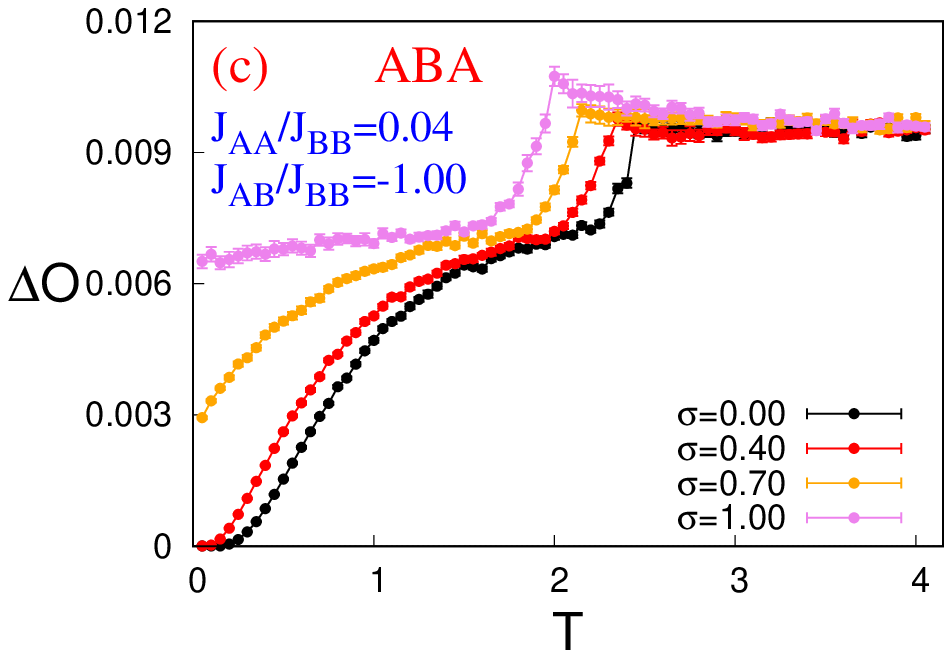}}
			
			\resizebox{8.2cm}{!}{\includegraphics[angle=0]{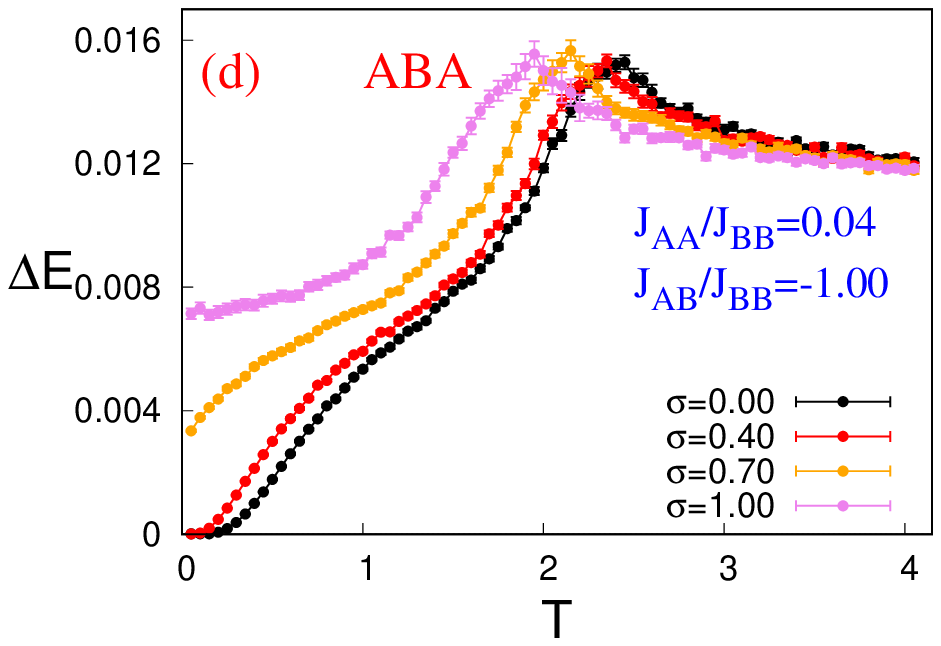}}\\
			
			\resizebox{8.2cm}{!}{\includegraphics[angle=0]{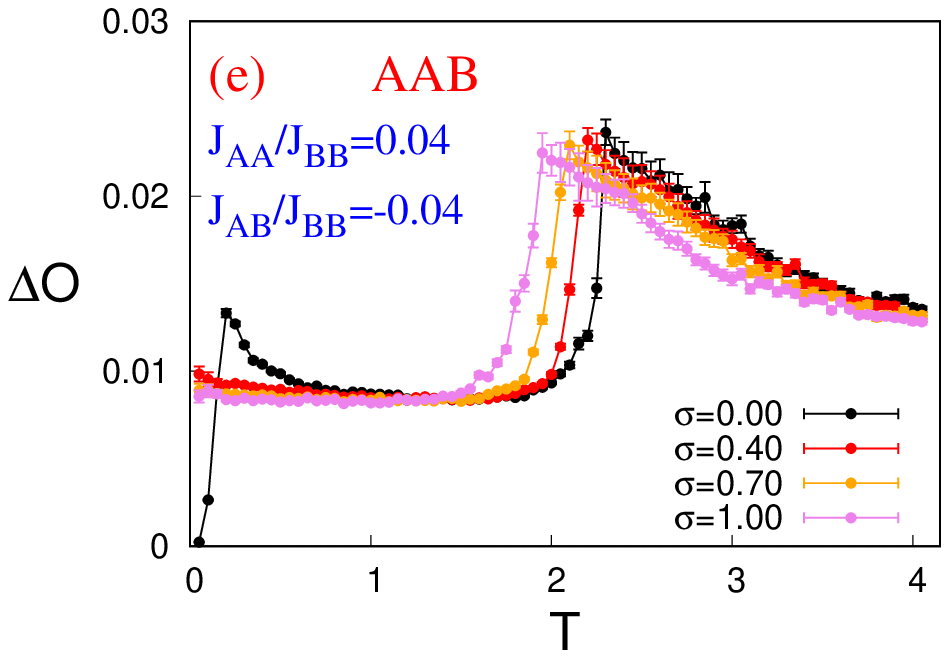}}
			
			\resizebox{8.2cm}{!}{\includegraphics[angle=0]{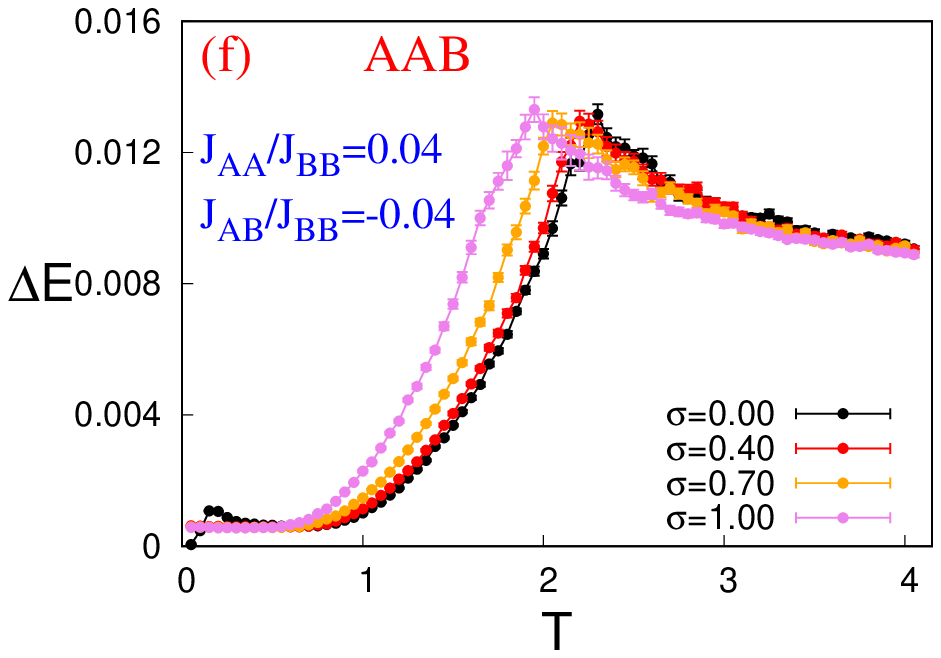}}\\
			
			\resizebox{8.2cm}{!}{\includegraphics[angle=0]{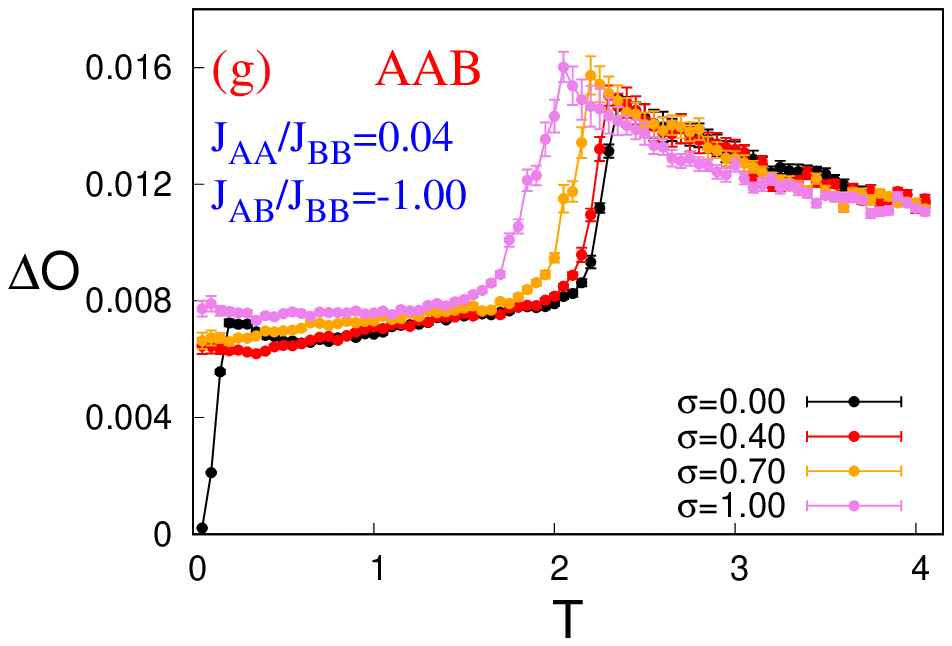}}
			
			\resizebox{8.2cm}{!}{\includegraphics[angle=0]{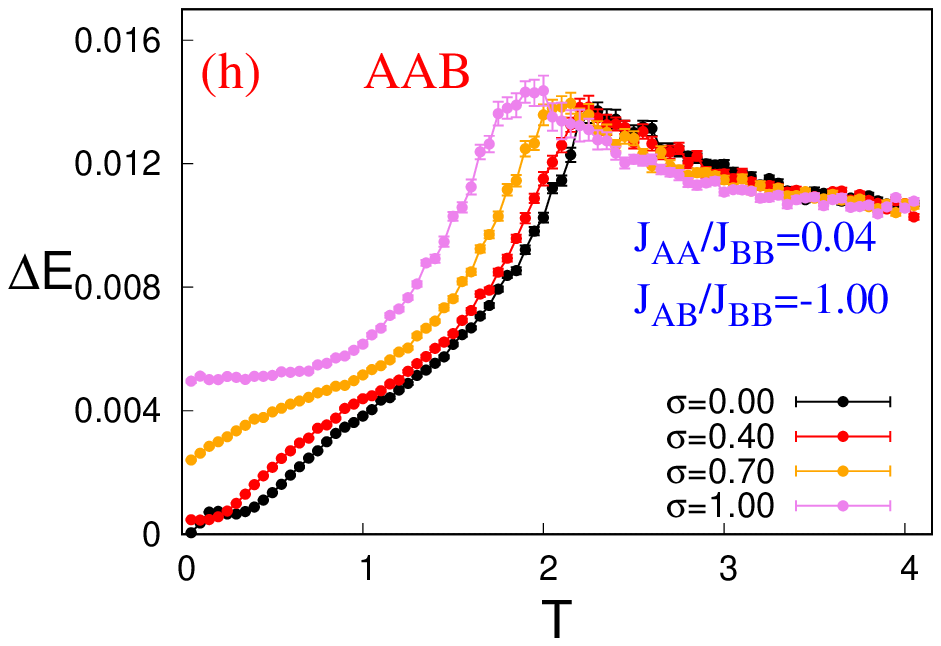}}
			
		\end{tabular}
		\caption{ (Colour Online) Temperature dependence of Fluctuation of order parameter, $\Delta O$ and Fluctuation of cooperative energy per site, $\Delta E$, for: ABA in (a)-(d) and AAB in (e)-(h) with ${J_{AA}}/{J_{BB}}=0.04$ and ${J_{AB}}/{J_{BB}}=-0.04$ and with ${J_{AA}}/{J_{BB}}=0.04$ and ${J_{AB}}/{J_{BB}}=-1.00$. Where, the errorbars are not visible, they are smaller than the area of the point-markers. All these plots are obtained for a system of $3\times100\times100$ sites. The nature of the curves prominently shows the shift of critical temperatures and even reason for absence of compensation can be understood from the low temperature segment of the curves.}
		\label{fig_fluc_mageng}
	\end{center}
\end{figure*}
In Figure \ref{fig_fluc_mageng}, the zero-field cases are also shown to emphasise on the departures in presence of the field. In the zero field cases, both the fluctuations of order parameter and energy, had a plateau with a smeared peak at the position of compensation. Here the field strength is taken for $7$ different values along with the zero-field one, which bears necessary informations. We can see the compensation and critical temperatures move towards lower temperature values, with increase in the field strength. As we make the strength of the external field larger in comparison to the cooperative part of the Hamiltonian, the shifts in both the temepratures become readily detectable. Even one can clearly see the compensation gradually vanishing as the smeared peaks at the low temperature segments flatten out. This is a signature of field-driven vanishing of compensation. At the lowest temperature, the increase of both the fluctuations readily suggests considerable loss of magnetic ordering with the increase of the strength or randomness of the external field.  

\begin{figure*}[!htb]
	\begin{center}
		\begin{tabular}{c}
			
			\resizebox{5.75cm}{!}{\includegraphics[angle=0]{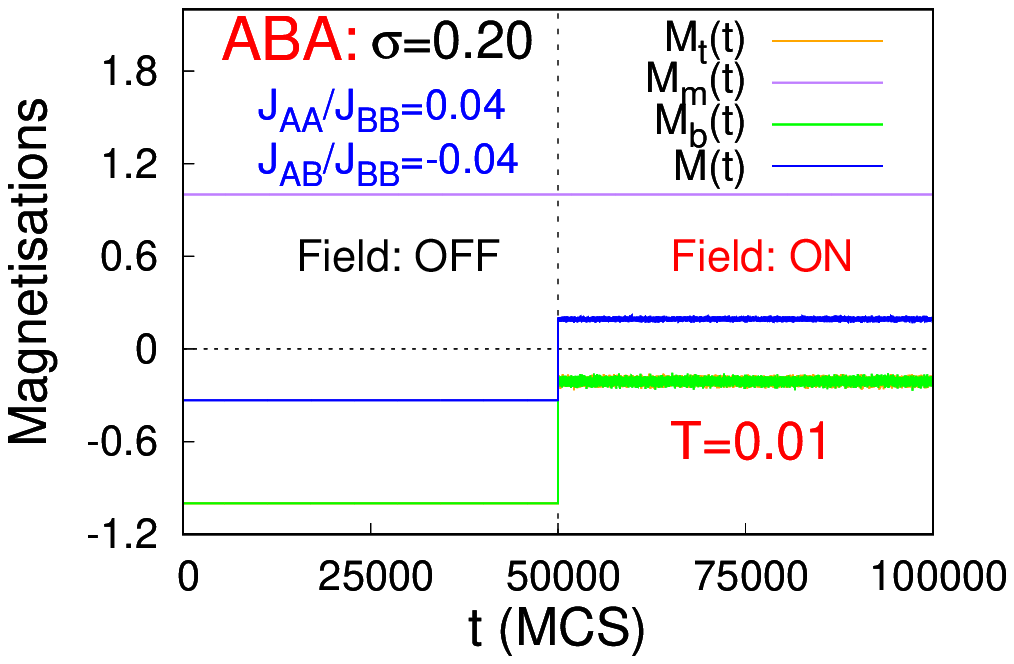}}
			\resizebox{5.75cm}{!}{\includegraphics[angle=0]{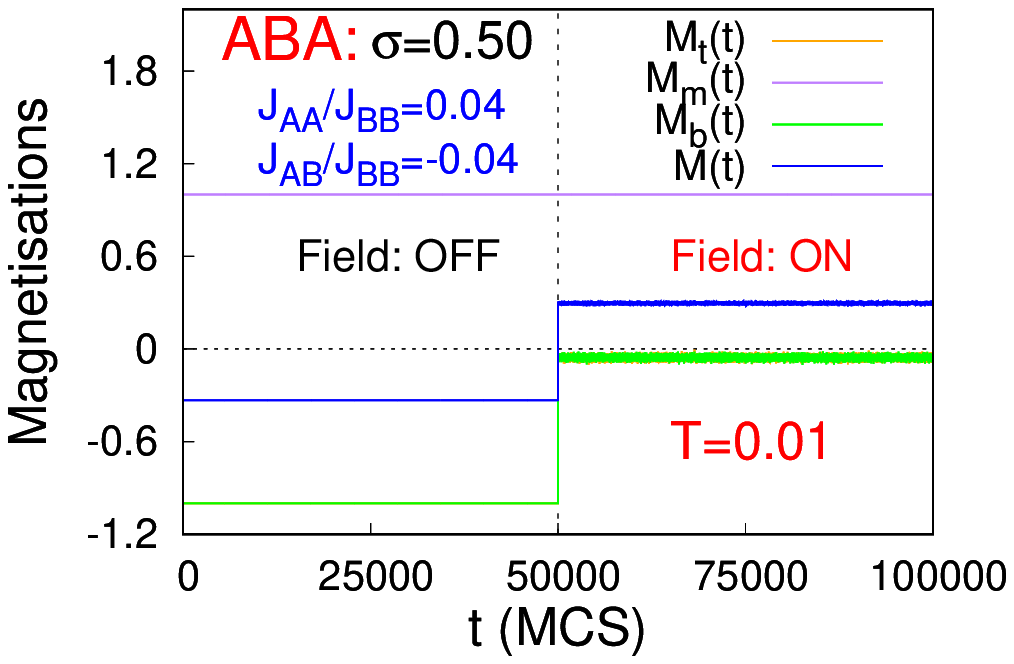}}
			\resizebox{5.75cm}{!}{\includegraphics[angle=0]{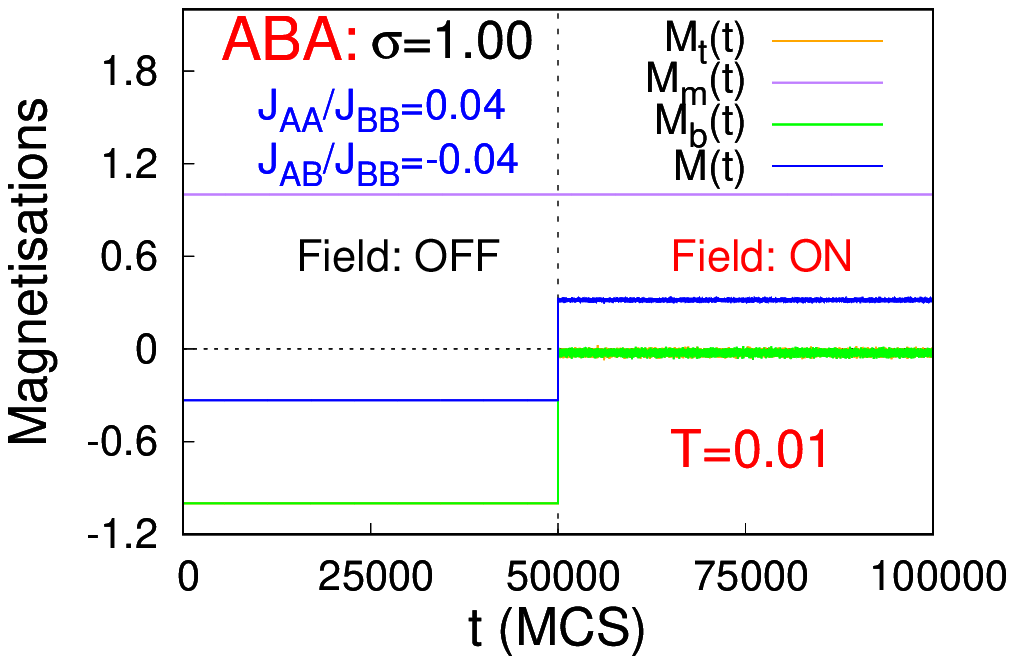}}\\
			
			\resizebox{5.75cm}{!}{\includegraphics[angle=0]{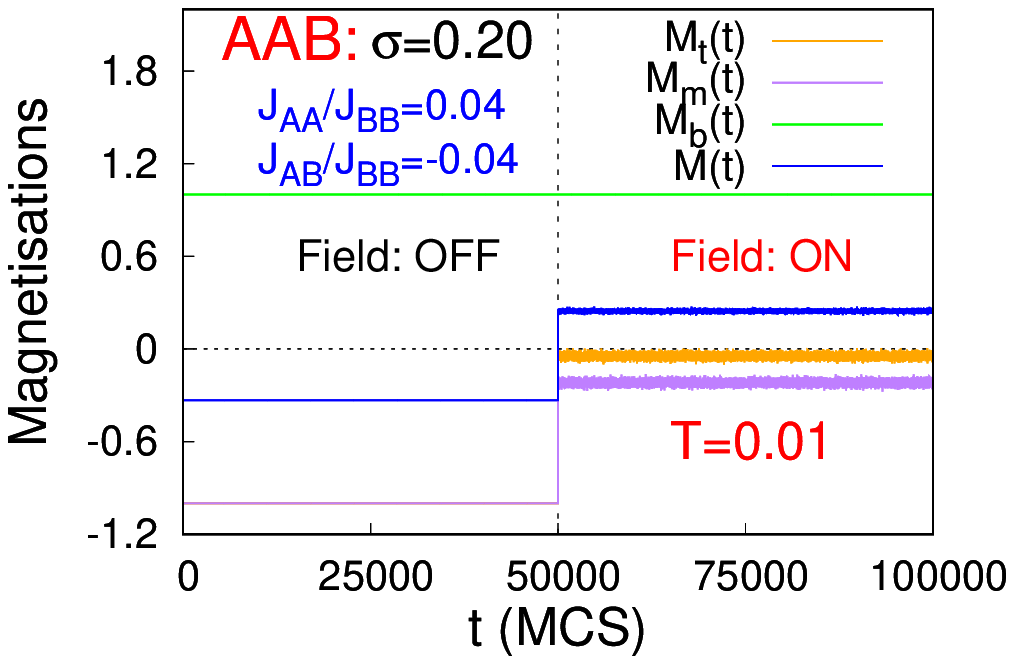}}
			\resizebox{5.75cm}{!}{\includegraphics[angle=0]{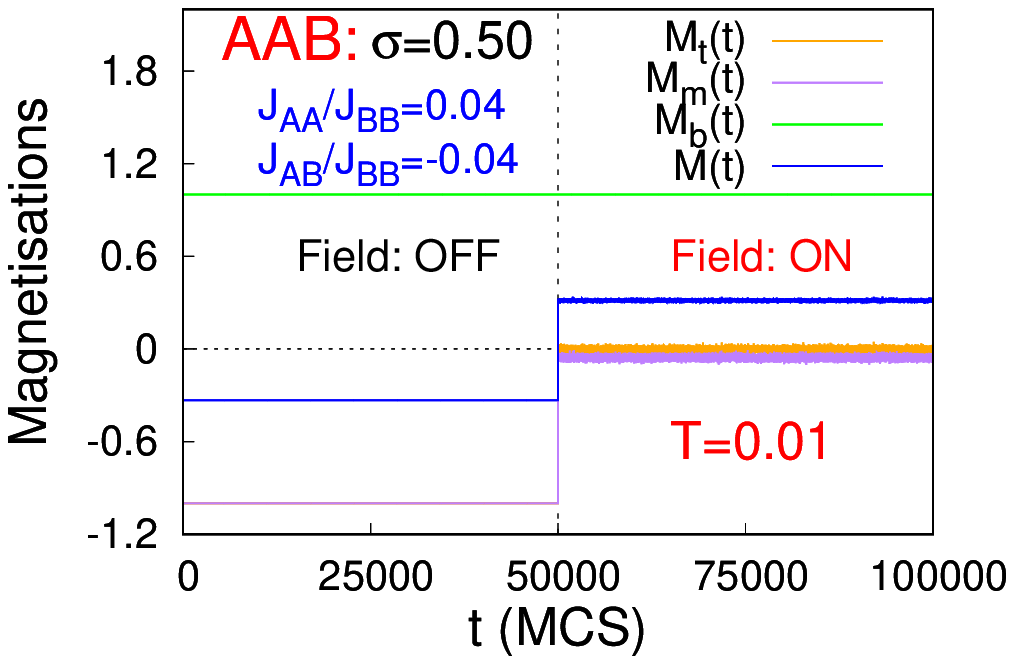}}
			\resizebox{5.75cm}{!}{\includegraphics[angle=0]{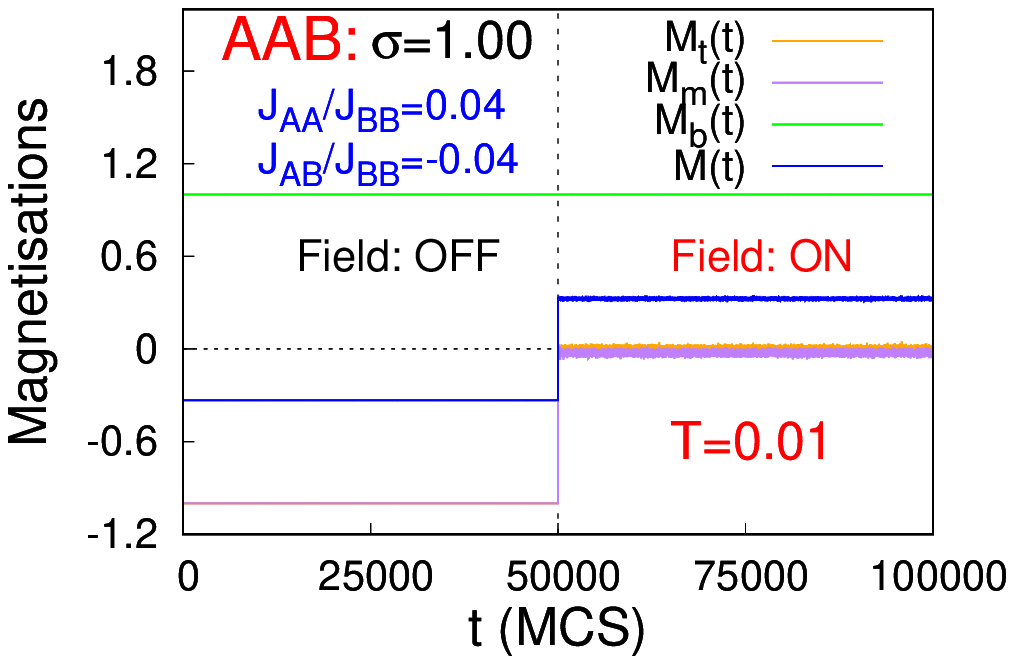}}\\
			
			\large {\textbf{(A) Top panel: $J_{AA}/J_{BB}=0.04$ and $J_{AB}/J_{BB}=-0.04$}}\\
			
			\resizebox{5.75cm}{!}{\includegraphics[angle=0]{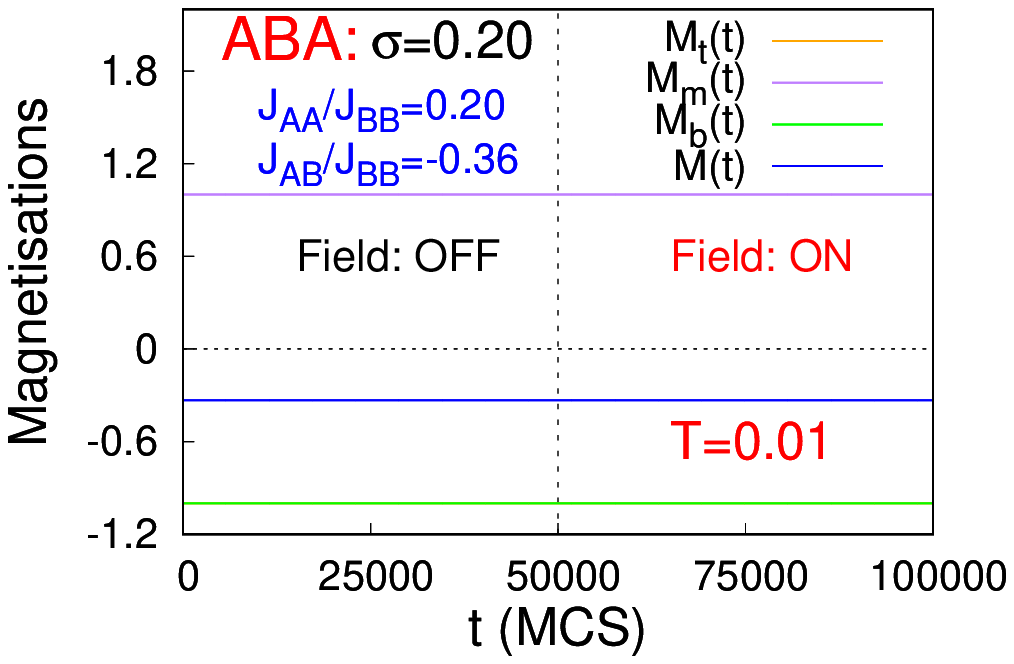}}
			\resizebox{5.75cm}{!}{\includegraphics[angle=0]{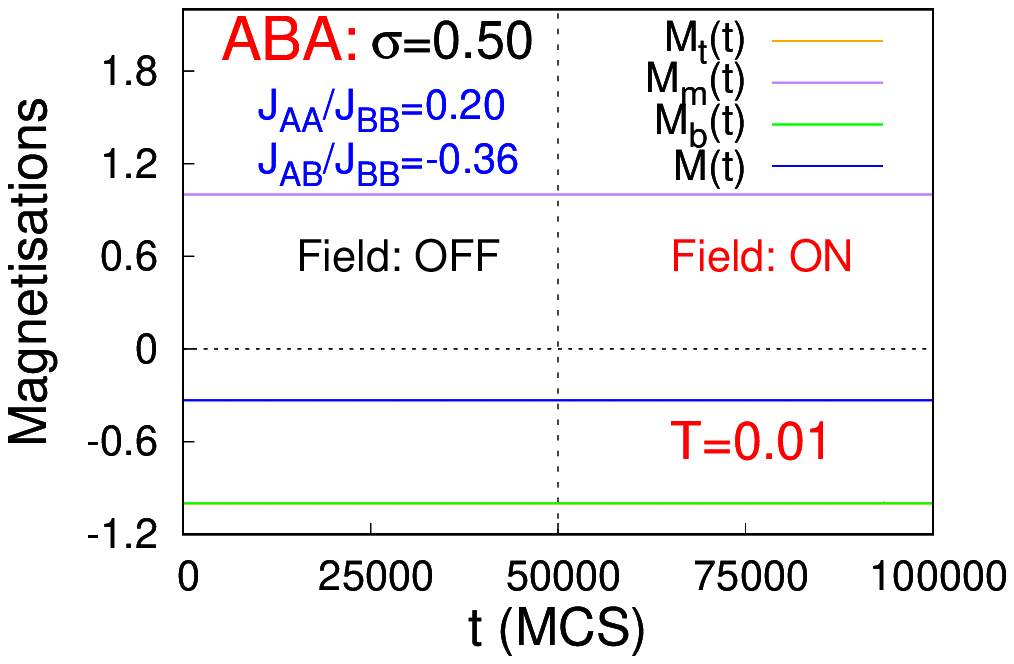}}
			\resizebox{5.75cm}{!}{\includegraphics[angle=0]{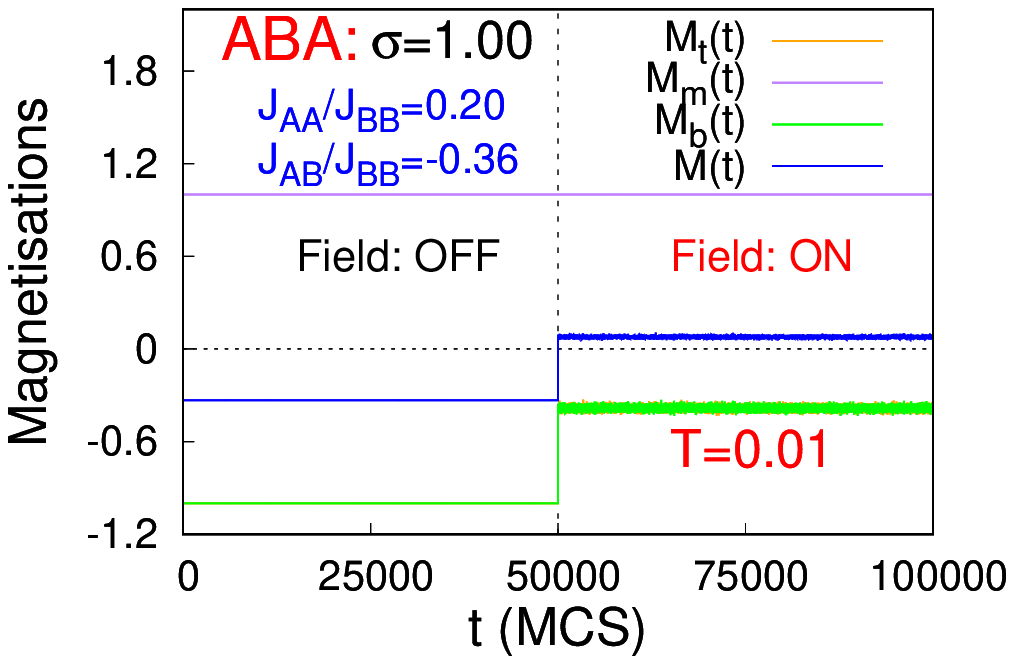}}\\

			\resizebox{5.75cm}{!}{\includegraphics[angle=0]{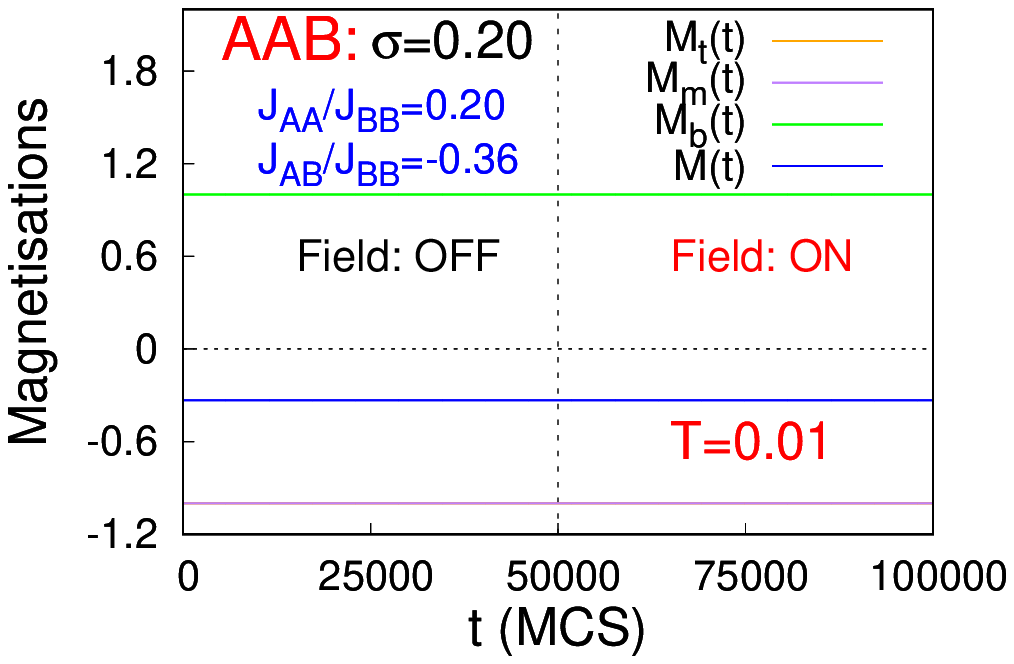}}
			\resizebox{5.75cm}{!}{\includegraphics[angle=0]{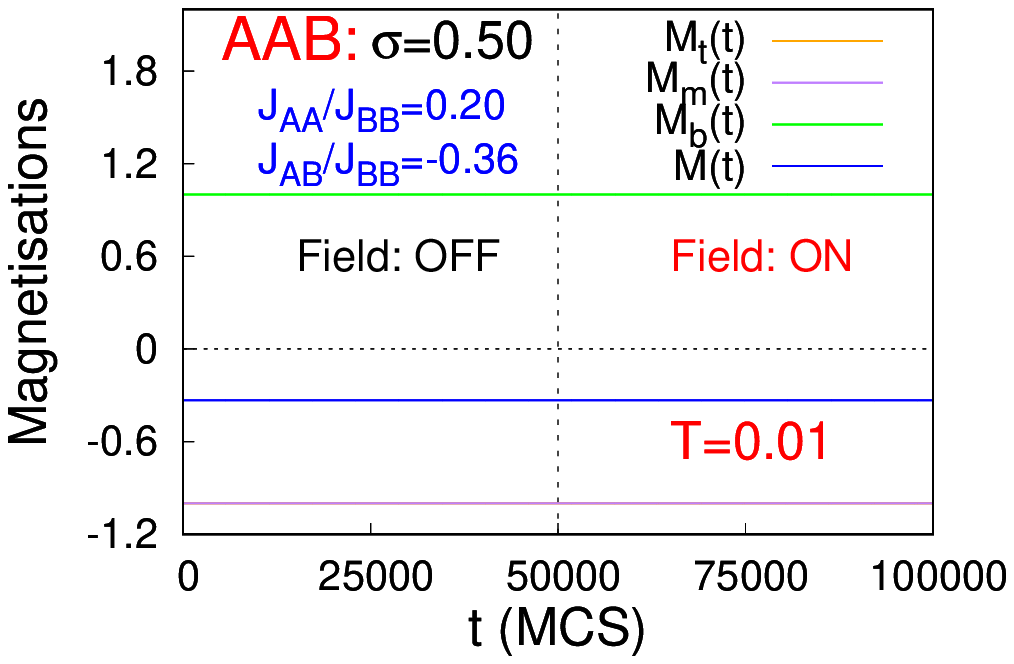}}
			\resizebox{5.75cm}{!}{\includegraphics[angle=0]{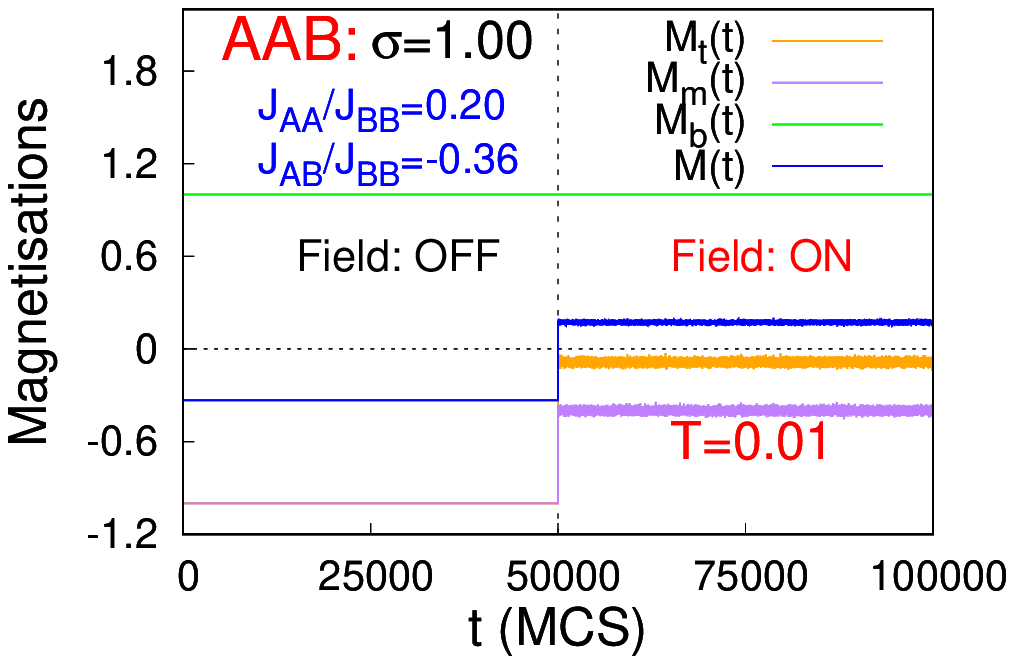}}\\

			\large {\textbf{(B) Middle panel: $J_{AA}/J_{BB}=0.20$ and $J_{AB}/J_{BB}=-0.36$}}\\
			
			\resizebox{5.75cm}{!}{\includegraphics[angle=0]{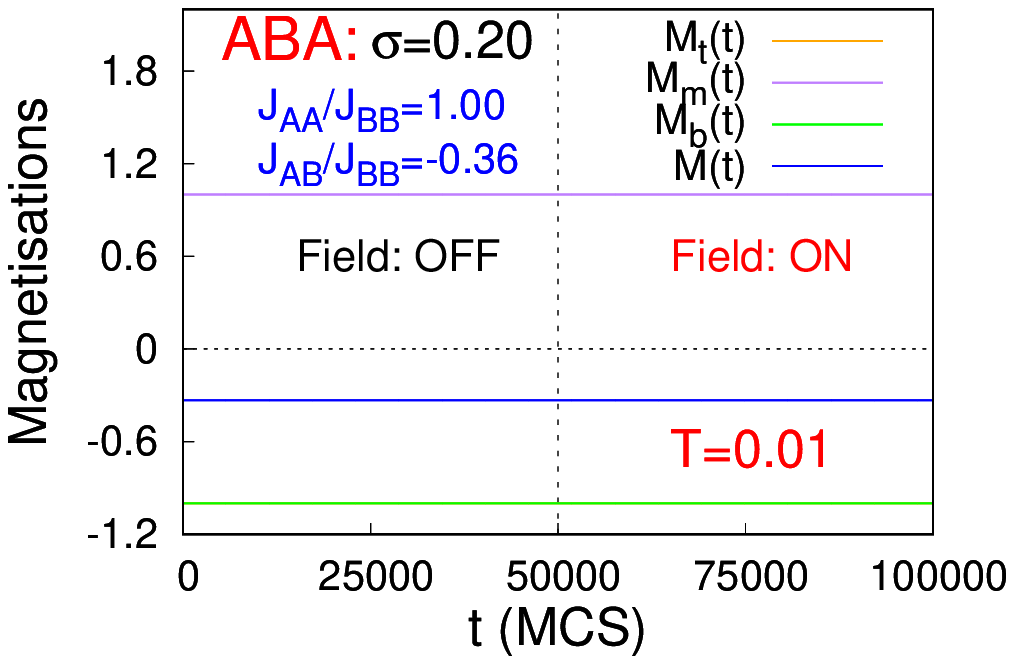}}
			\resizebox{5.75cm}{!}{\includegraphics[angle=0]{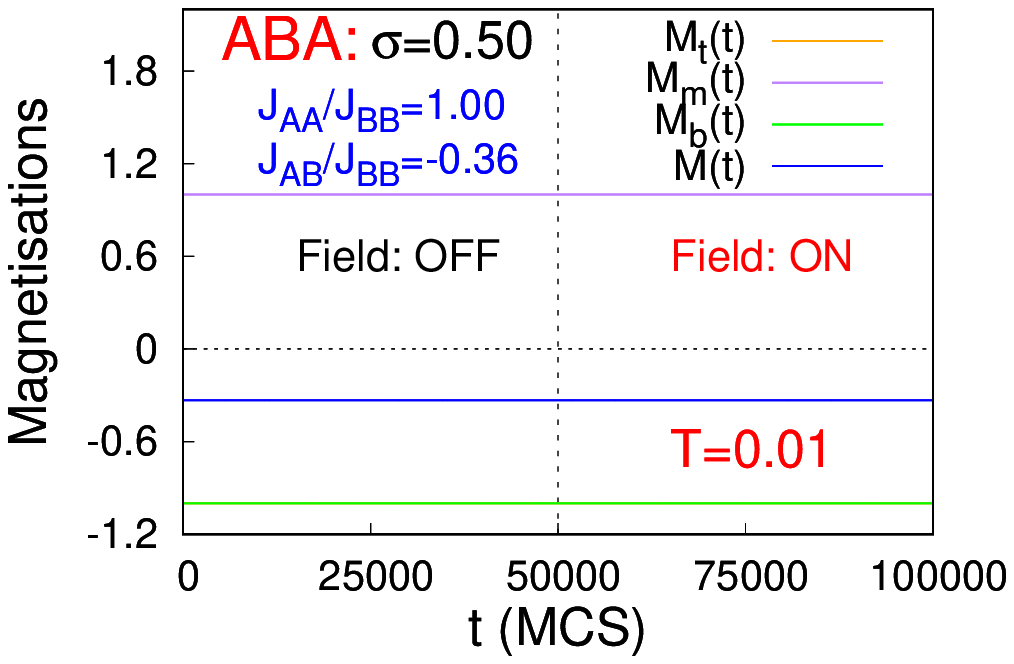}}
			\resizebox{5.75cm}{!}{\includegraphics[angle=0]{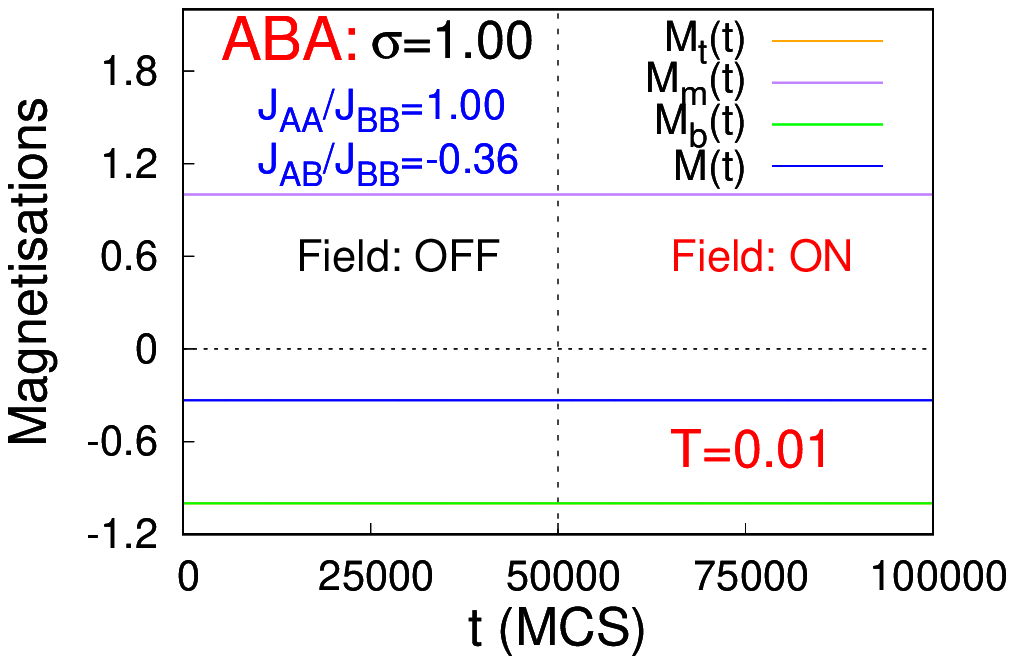}}\\

			\resizebox{5.75cm}{!}{\includegraphics[angle=0]{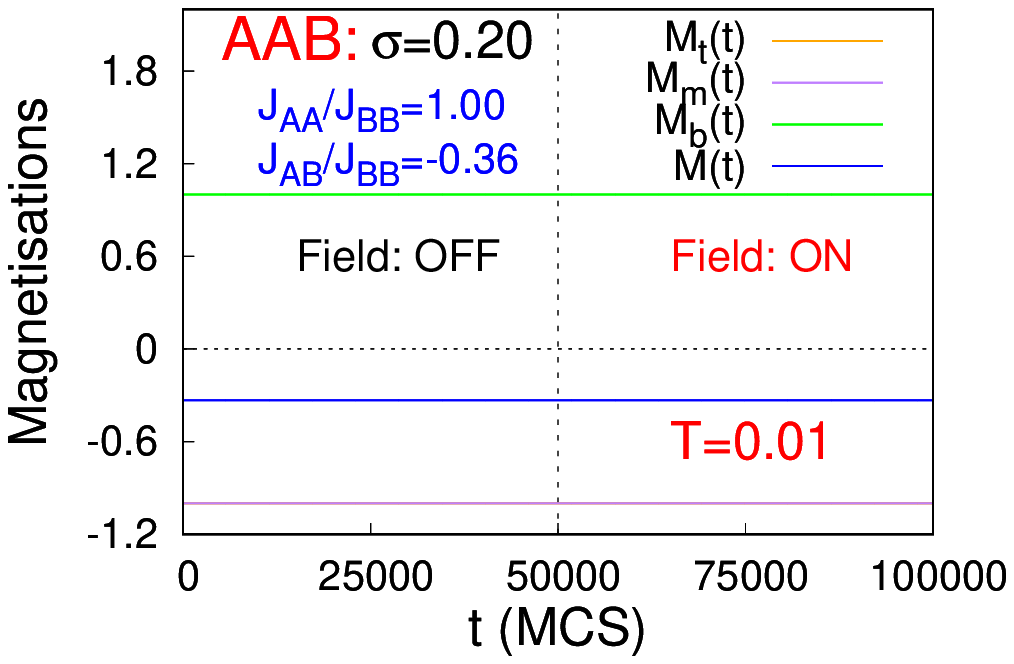}}
			\resizebox{5.75cm}{!}{\includegraphics[angle=0]{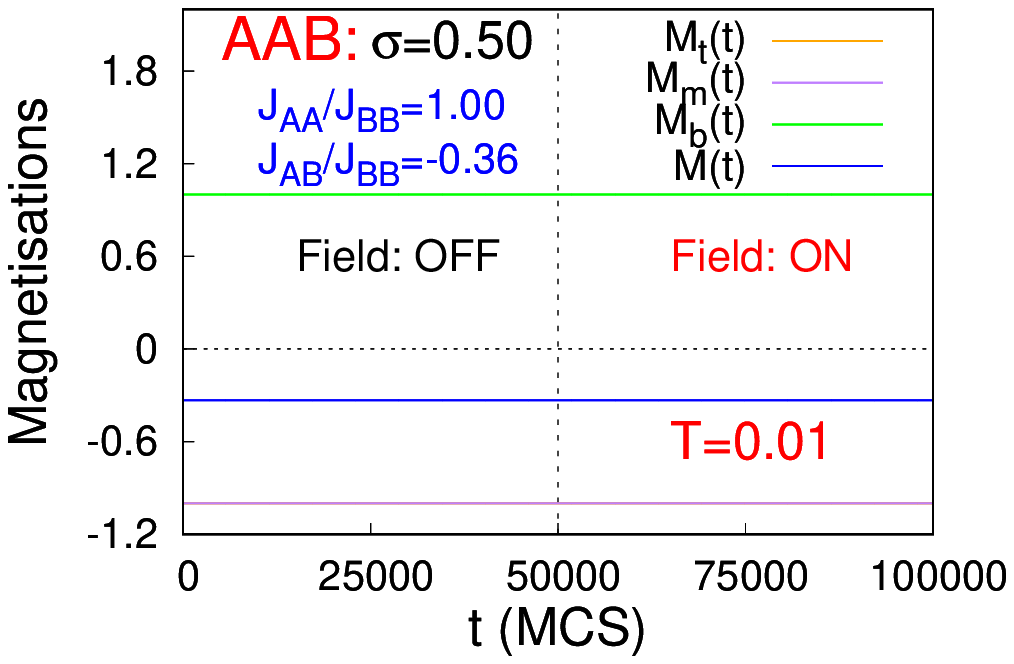}}
			\resizebox{5.75cm}{!}{\includegraphics[angle=0]{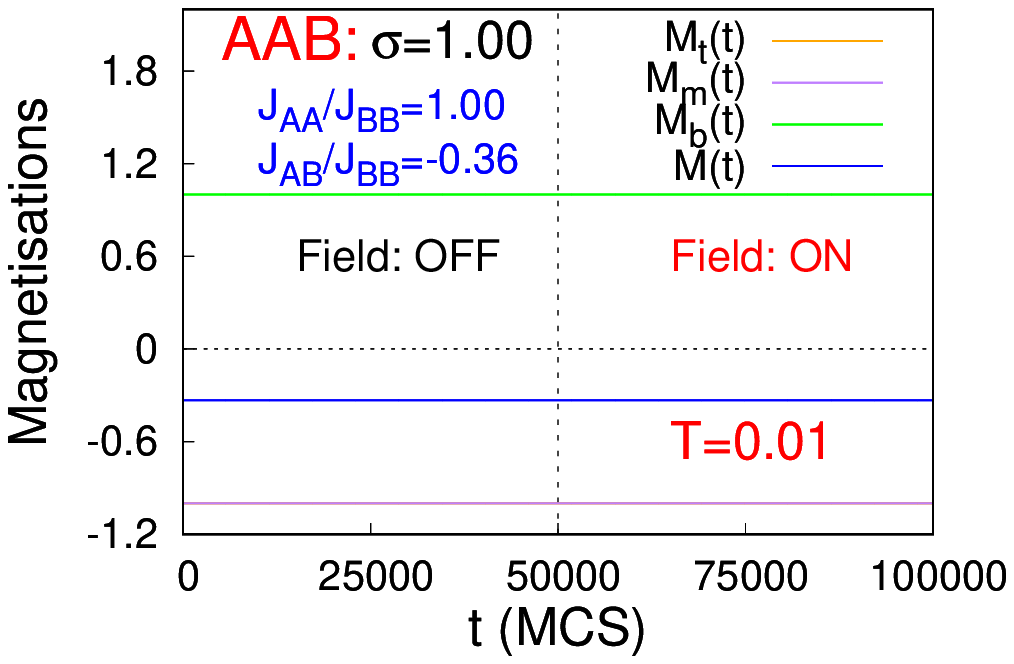}}\\			
			
			\large {\textbf{(C) Bottom panel: $J_{AA}/J_{BB}=1.00$ and $J_{AB}/J_{BB}=-0.36$}}\\
		\end{tabular}
		\caption{ (Colour Online) Plots of Magnetisations for individual layers and average magnetisation of the bulk versus time in MCS where $M_{t}(t)$: Magnetization of the top layer; $M_{m}(t)$: Magnetization of the mid layer; $M_{b}(t)$: Magnetization of the bottom layer are all functions of time,$t$, in units of MCS. For the ABA stacking, the magnetisation curves for the surface layers (orange and green) overlap for the most of the times.}
		\label{fig_field_mag_time}
	\end{center}
\end{figure*}

\indent The question arises now: What exactly does lead to the \textit{field driven absence of compensation}? According to the definition, the total magnetisation can change signature even in the close vicinity of $0$K so the compensation temperature can even be equal to the absolute zero. Let us observe the Magnetisation of individual layers and the total magnetisation of the bulk, as a function of time at a very low temperature $T=0.01$. Here, three particular combinations of coupling ratios: $J_{AA}/J_{BB}$ and $J_{AB}/J_{BB}$ for three values of sd of the external uniform random magnetic field are investigated. The results for both the ABA and AAB configurations are in Figure \ref{fig_field_mag_time}.
One can see in all these cases that about and till $t= 5\times 10^{4}$ MCS, the sublattice as well as total magnetisation of the system has attained equilibrium (as in this time interval, there is no time dependent part in the Hamiltonian). Starting from a completely randomised high temperature state, this equilibrium is achieved by \textit{slow} cooling (in steps of $\Delta T= 0.05$). As the external field at $t=t_{0}=5 \times 10^{4}$ MCS, is switched ON all the sublattice magnetisations and correspondingly the total magnetisation, starts to react in response to the external field. The steady state, is achieved rather quickly, and the system behaves similarly in the rest of the exposure interval. So the choice of exposure interval is potent in revealing the characteristics of the system. We observe whenever the per site in-plane interaction energies are comparable to the magnitude of per site spin-field interaction energies [Refer to the top panel and $\sigma=1.00$ in the middle panel of Figure \ref{fig_field_mag_time}] of the A- layers, the A-layers significantly lose magnetic ordering. In top panel of Figure \ref{fig_field_mag_time} for ABA, when the spin-field energies per site are greater than the cooperative energy per site, it causes the surface layers to become nearly completely random at even such a low temperature. Consequently the total magnetisation doesn't change sign below the Critical temperature which leads to the absence of compensation phenomenon. 

For the ABA system, the surface layers behave almost identically as the mid B-layer affects their behaviour in identical way through nearest neighbour Ising interaction. But in AAB system, the bottom B-layer provides magnetic stability to the mid A-layer. Thus the magnetic behaviour of the mid A-layer is a little less affected than the top A-layer. In top panel of Figure \ref{fig_field_mag_time} for AAB, when the spin-field energies per site are greater than the cooperative energy per site of the A-layers, like in the previous case, the A layers lose much of the magnetic ordering at such a low temperature and the top A-layer is more affected. Consequently the combined magnetisation of A-layers is unable to cancel the magnetisation of the bottom B-layer. It leads to the absence of compensation phenomenon for this configuration.

Such type of an event can be termed as \textit{field-driven vanishing of compensation} in the Ising spin-1/2 trilayers. It is a dynamic phenomenon as the Hamiltonian's time dependent part is responsible for it.  

\subsection{Lattice Morphology}
\label{subsec_morphology}

\indent As it is said before, the increase in the fluctuations in the low temperature segments in Figure \ref{fig_fluc_mageng} signifies the loss of magnetic ordering in the system as a whole with increase in the randomness of the external field. But, how do \textit{the magnetisations of the individual layers} react to the external field? The magnetization versus time plots [Figure \ref{fig_field_mag_time}] for the sublayers as well as the bulk has shown the relevant energy scales where the compensation becomes absent. The spin-density plots or the lattice morphology at teperature $T=0.01$ are now used to establish the reason. 

\begin{figure*}[!htb]
	\begin{center}
		\begin{tabular}{c}
			\phantom{}
			\hspace{0.0cm} \textbf{Top layer} \hspace{3.5cm} \textbf{Mid layer} \hspace{3.0cm} \textbf{Bottom layer} \\

			$\mathbf{(a)\text{ }ABA:\text{ } J_{AA}/J_{BB}=0.04;\text{ } J_{AB}/J_{BB}=-0.04 \text{ and } \sigma=0.50}$\\

			\resizebox{5.0cm}{!}{\includegraphics[angle=0]{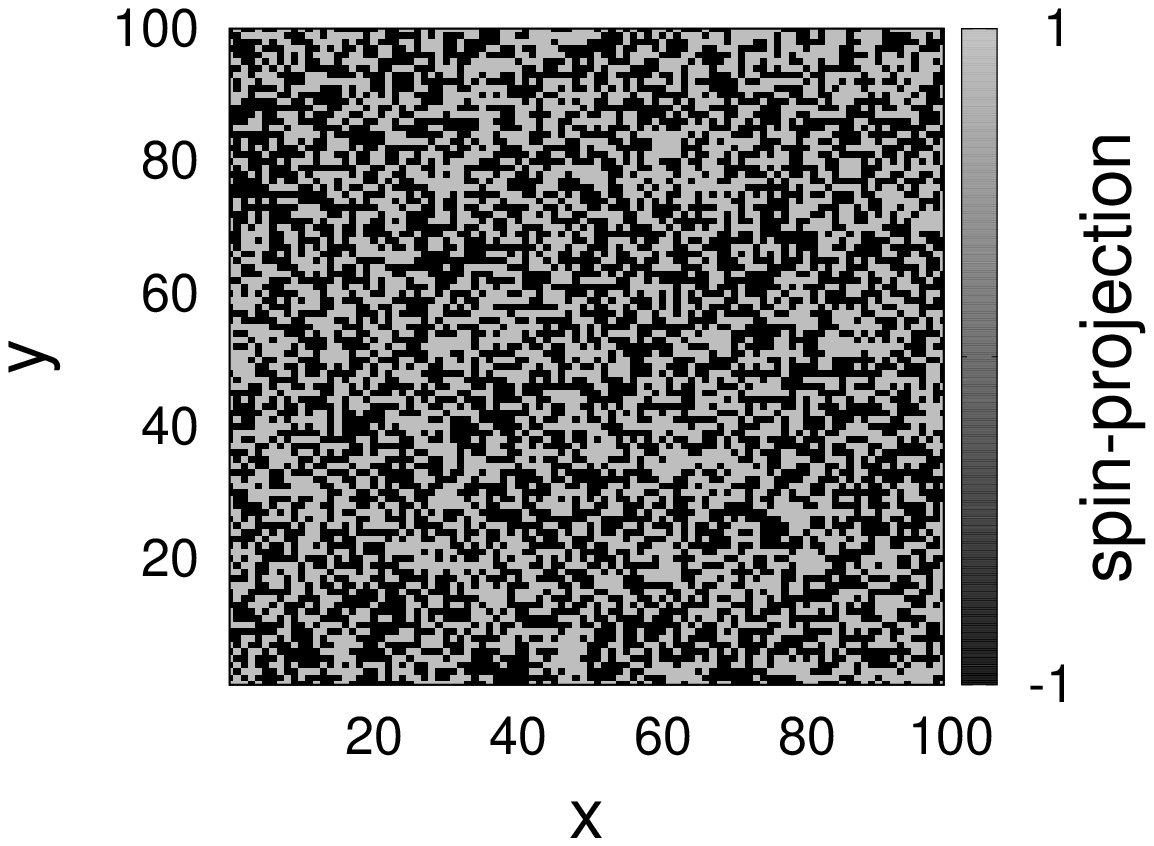}}

			\resizebox{5.0cm}{!}{\includegraphics[angle=0]{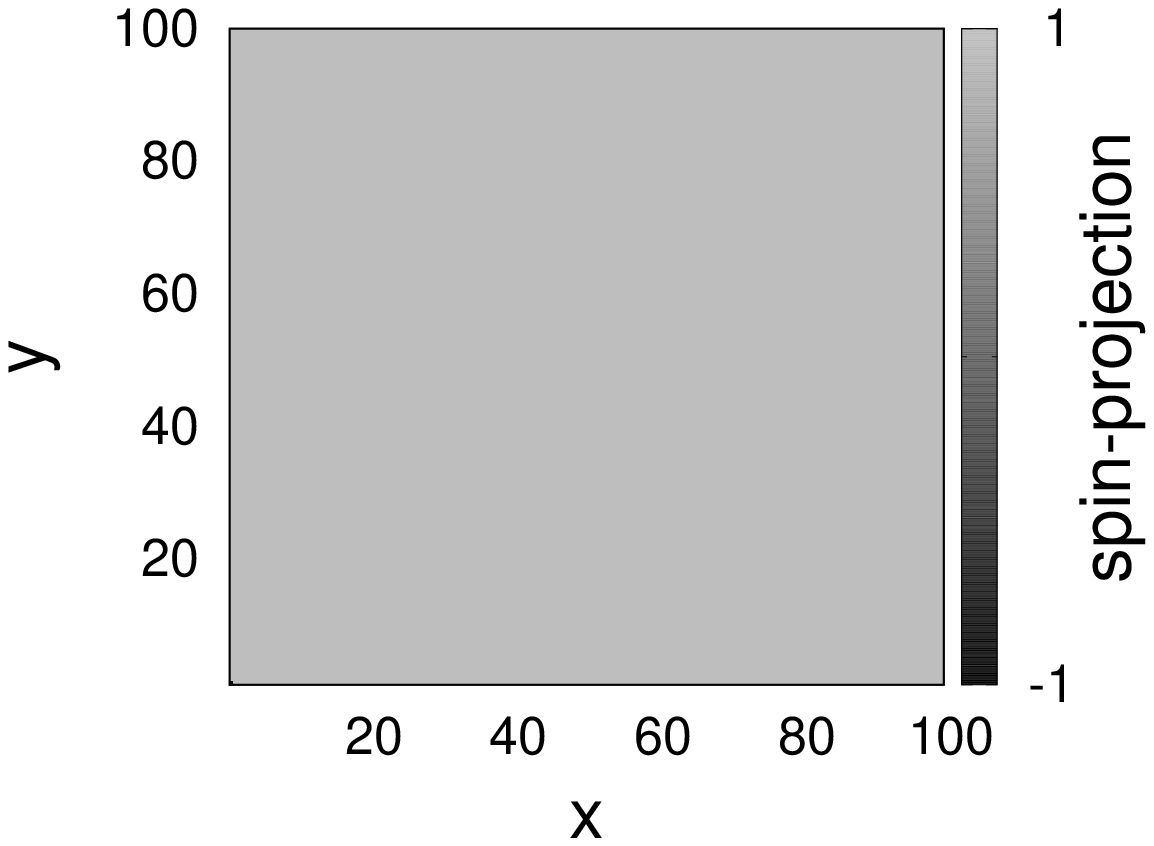}}

			\resizebox{5.0cm}{!}{\includegraphics[angle=0]{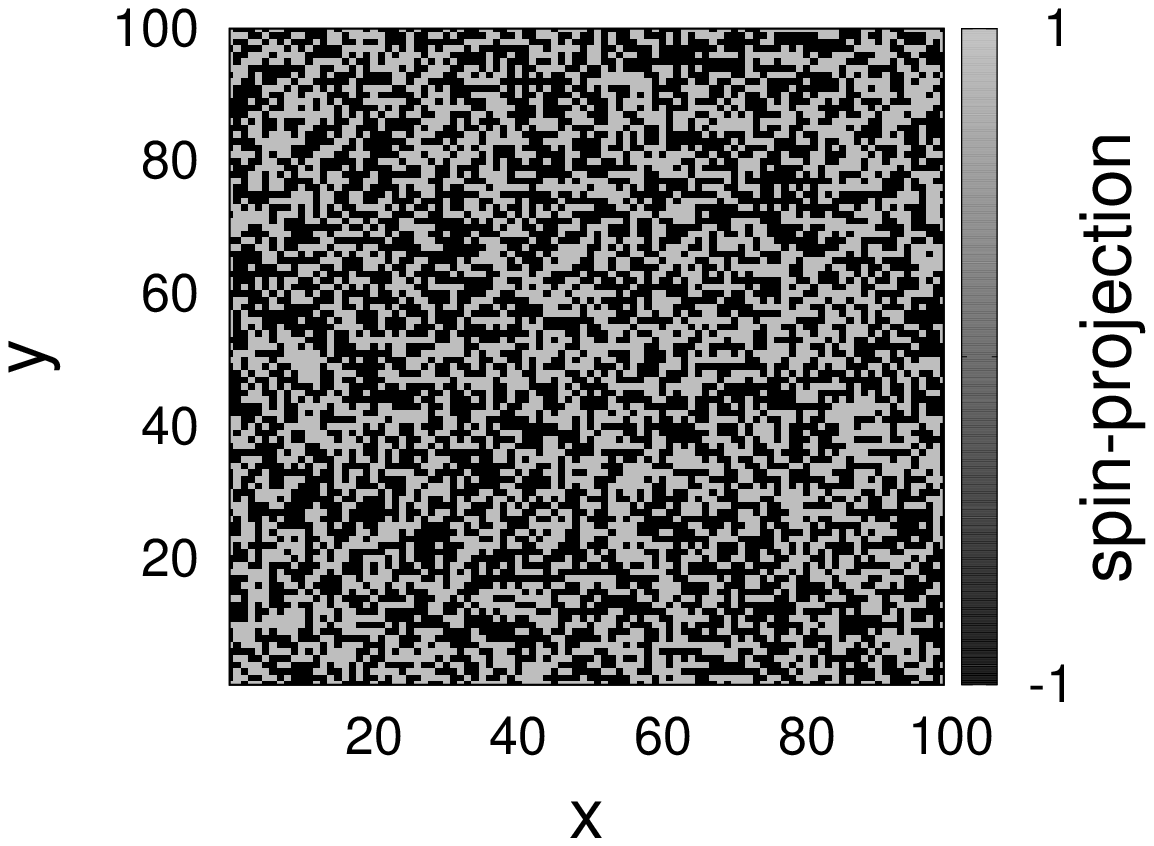}}\\

			\phantom{}
			\hspace{0.0cm} $M_{t}(t_{morph})=-0.039$ \hspace{1.5cm} $M_{m}(t_{morph})=+1.00$ \hspace{1.75cm} $M_{b}(t_{morph})=-0.056$\\

			\\

			$\mathbf{(b)\text{ }ABA:\text{ } J_{AA}/J_{BB}=0.20;\text{ } J_{AB}/J_{BB}=-0.36 \text{ and } \sigma=0.50}$\\
			
			\resizebox{5.0cm}{!}{\includegraphics[angle=0]{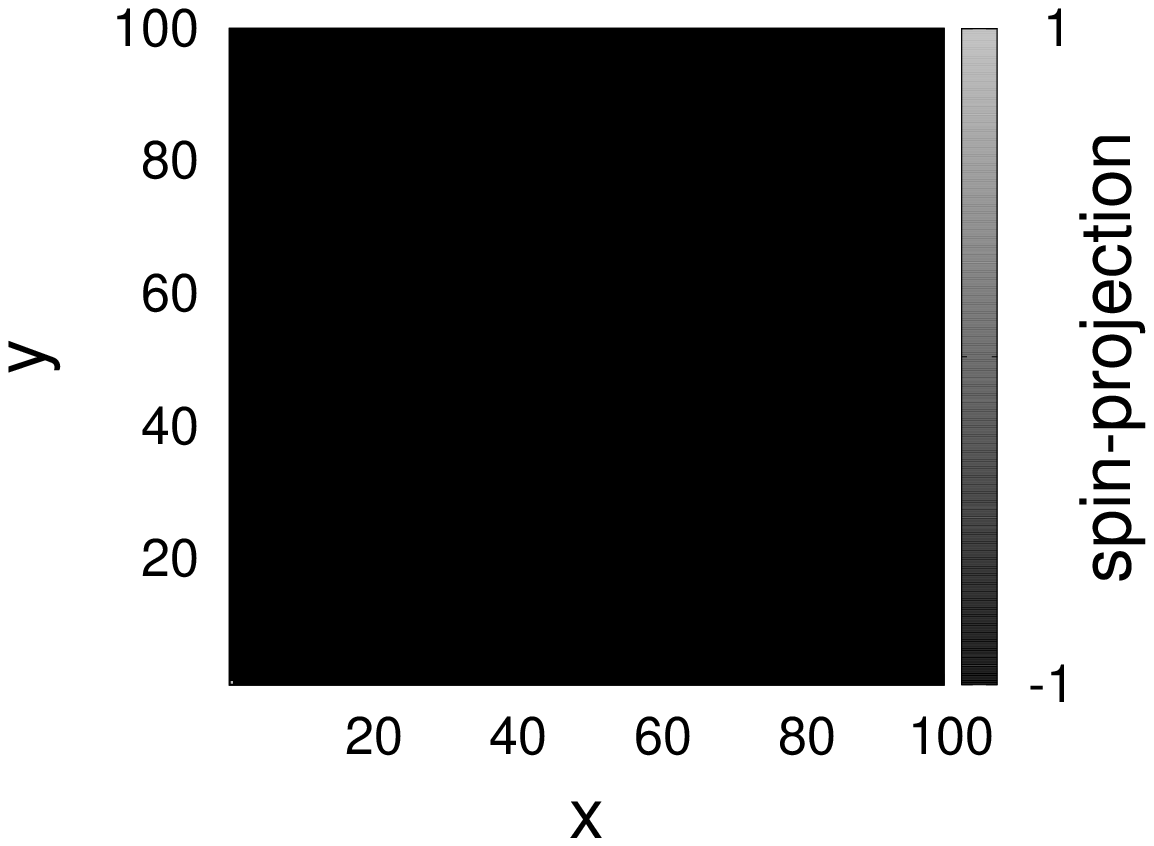}}
			
			\resizebox{5.0cm}{!}{\includegraphics[angle=0]{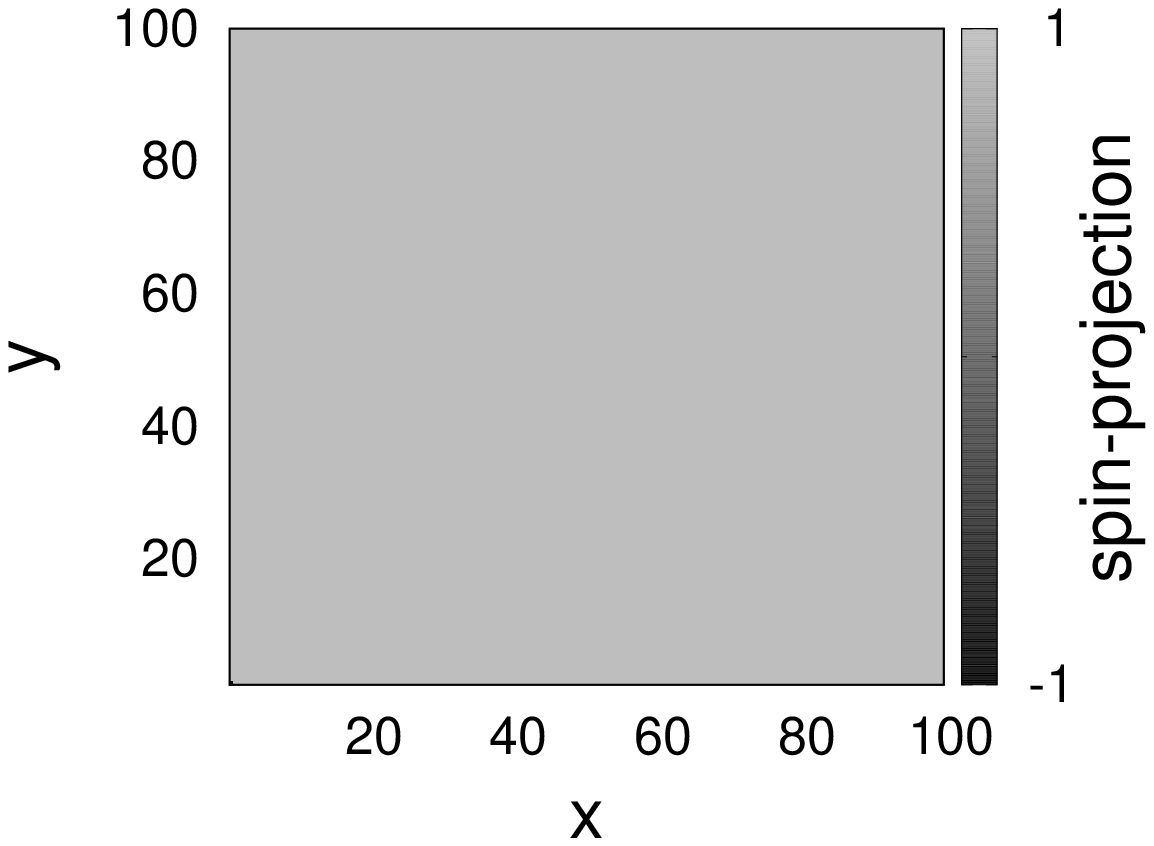}}
			
			\resizebox{5.0cm}{!}{\includegraphics[angle=0]{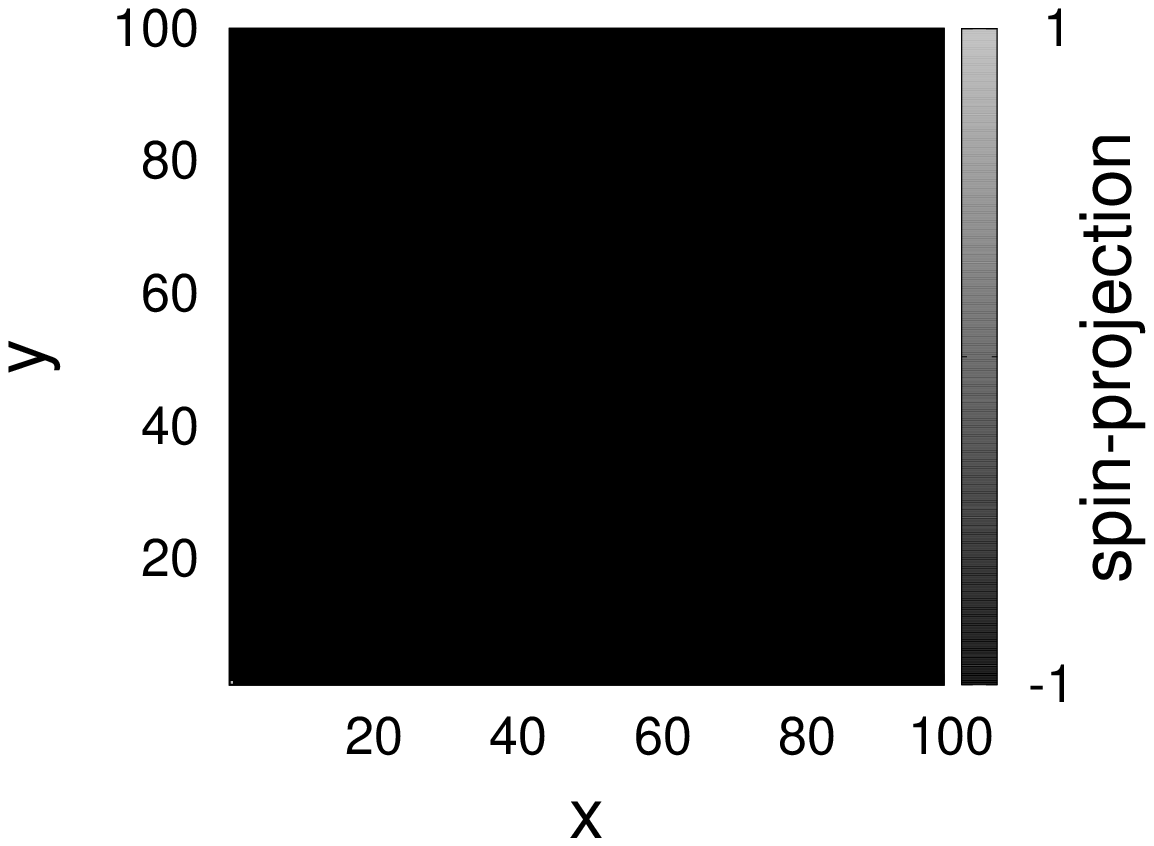}}\\
			
			\phantom{}
			\hspace{0.0cm} $M_{t}(t_{morph})=-1.00$ \hspace{1.5cm} $M_{m}(t_{morph})=+1.00$ \hspace{1.75cm} $M_{b}(t_{morph})=-1.00$\\
			
			\\
			
			$\mathbf{(c)\text{ }ABA:\text{ } J_{AA}/J_{BB}=0.20;\text{ } J_{AB}/J_{BB}=-0.36 \text{ and } \sigma=1.00}$\\
			
			\resizebox{5.0cm}{!}{\includegraphics[angle=0]{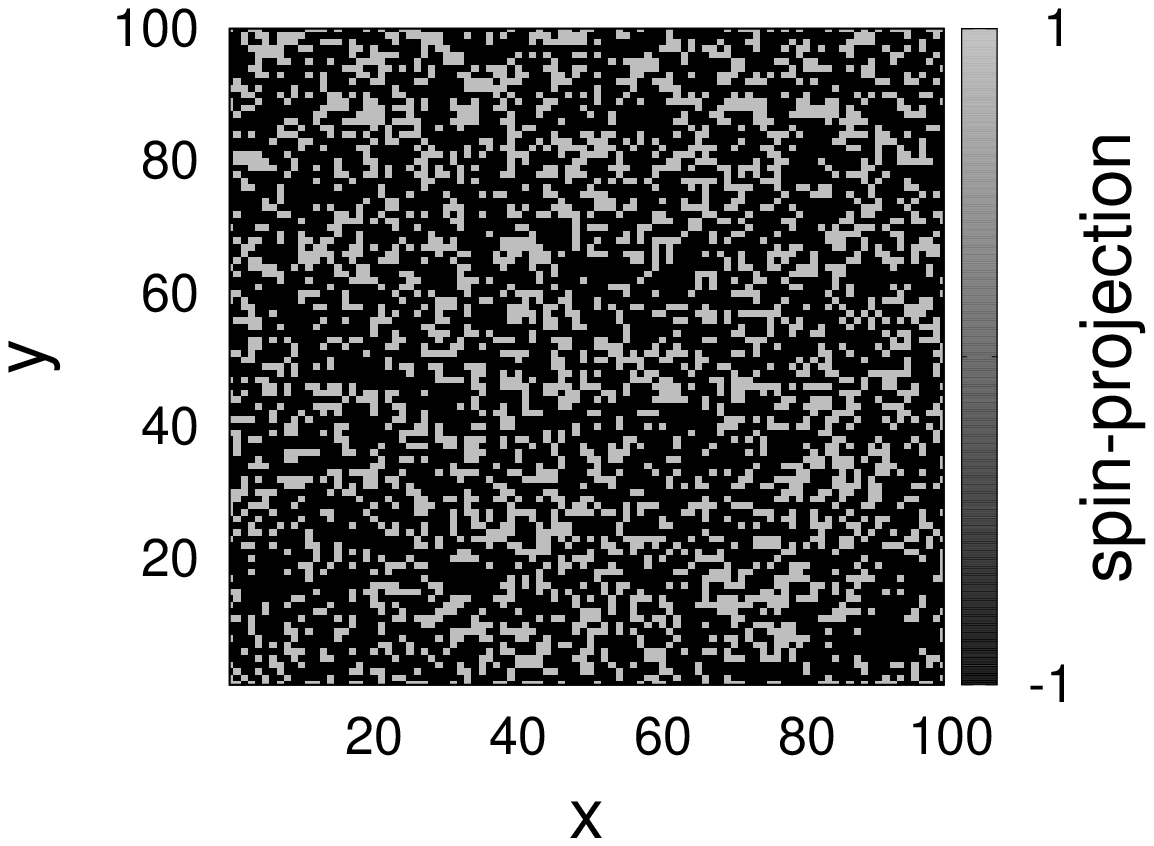}}
			
			\resizebox{5.0cm}{!}{\includegraphics[angle=0]{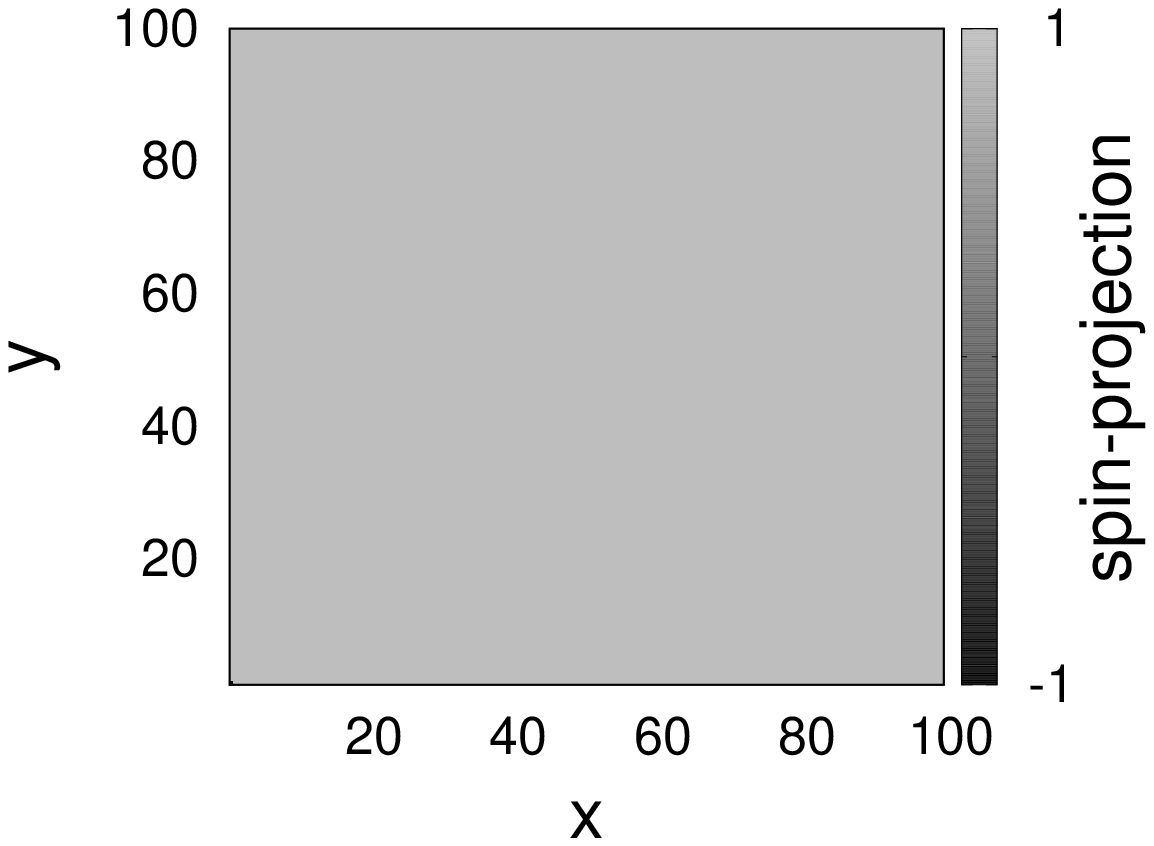}}
			
			\resizebox{5.0cm}{!}{\includegraphics[angle=0]{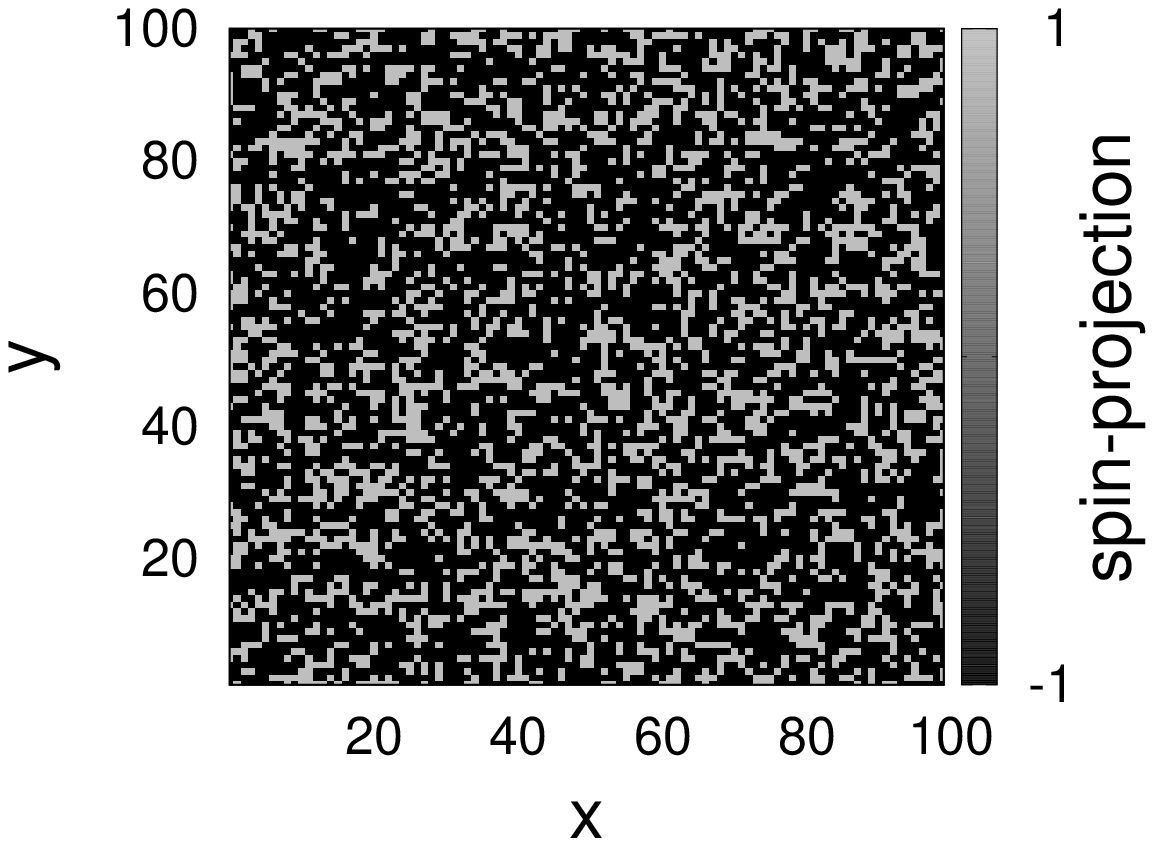}}\\
			
			\phantom{}
			\hspace{0.0cm} $M_{t}(t_{morph})=-0.402$ \hspace{1.5cm} $M_{m}(t_{morph})=+1.00$ \hspace{1.75cm} $M_{b}(t_{morph})=-0.375$\\
			
			\\
						
			$\mathbf{(d)\text{ }ABA:\text{ } J_{AA}/J_{BB}=0.04;\text{ } J_{AB}/J_{BB}=-1.00 \text{ and } \sigma=1.00}$\\
			
			\resizebox{5.0cm}{!}{\includegraphics[angle=0]{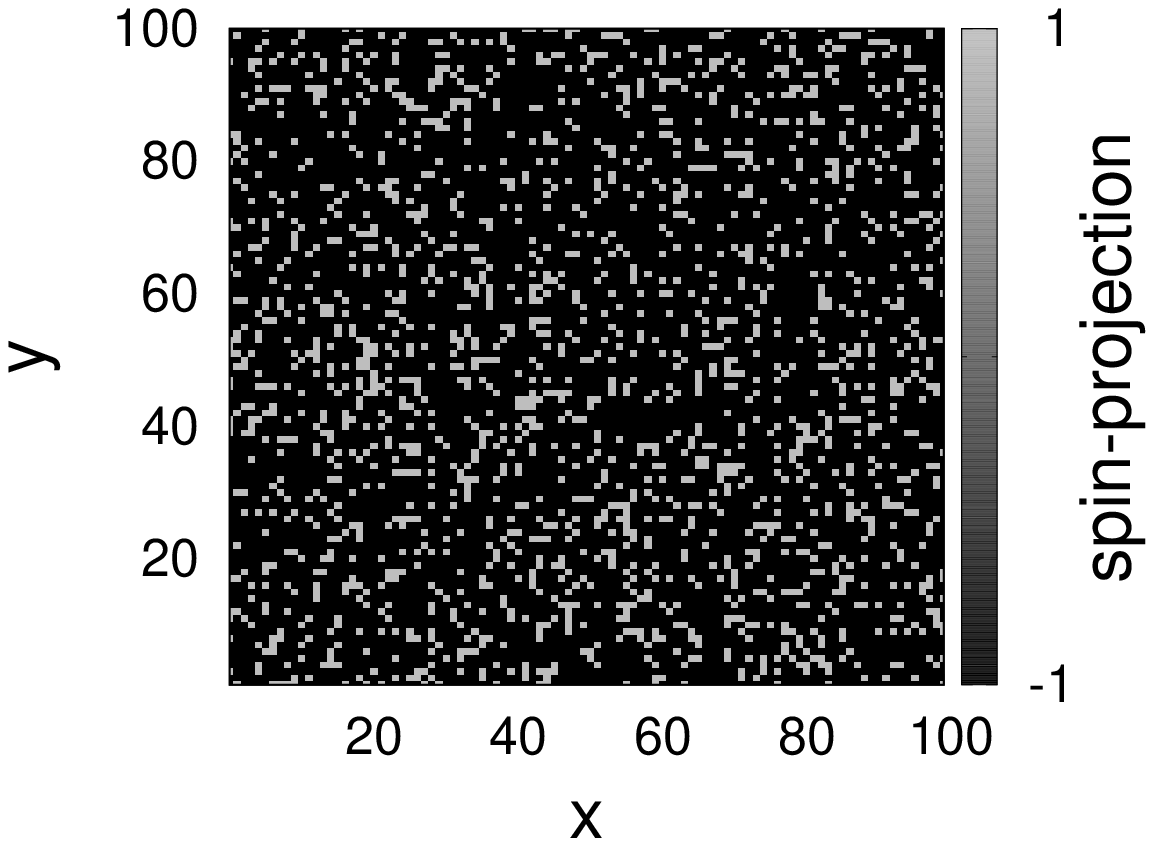}}
			
			\resizebox{5.0cm}{!}{\includegraphics[angle=0]{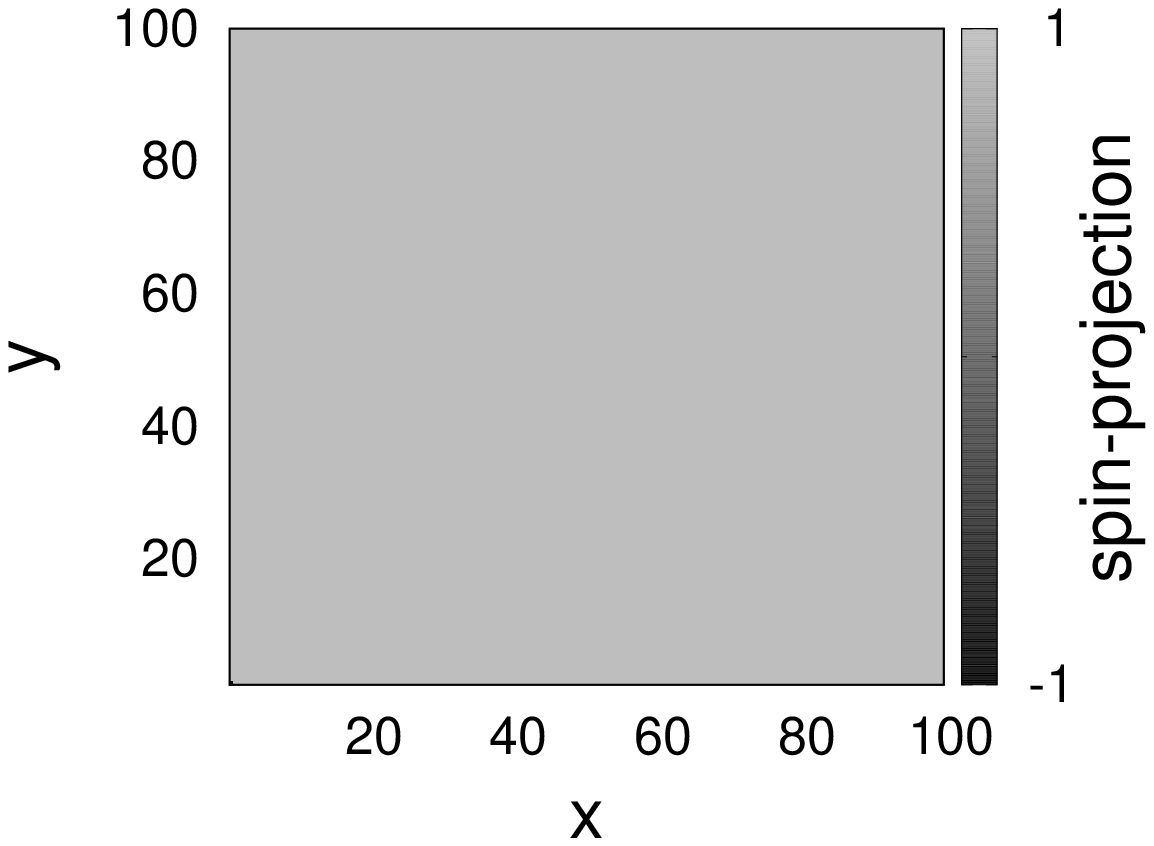}}
			
			\resizebox{5.0cm}{!}{\includegraphics[angle=0]{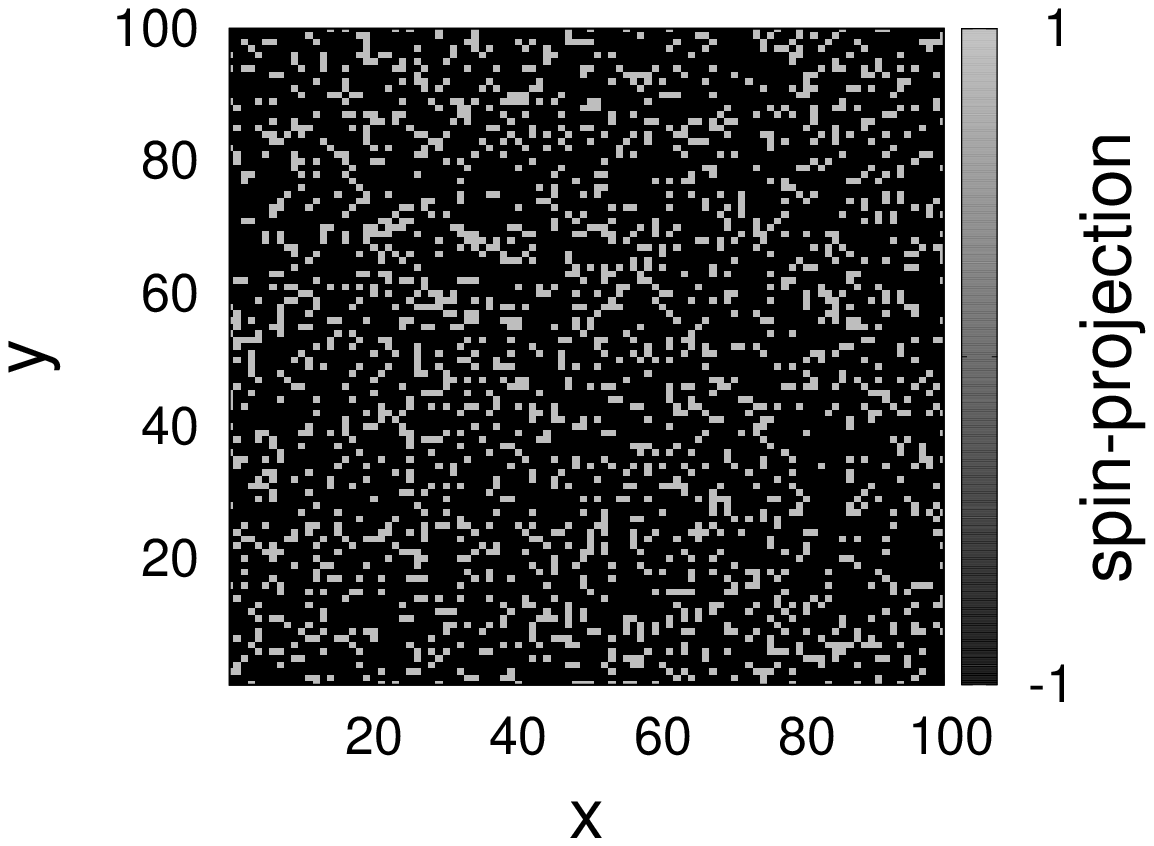}}\\
			
			\phantom{}
			\hspace{0.0cm} $M_{t}(t_{morph})=-0.649$ \hspace{1.5cm} $M_{m}(t_{morph})=+1.00$ \hspace{1.75cm} $M_{b}(t_{morph})=-0.638$\\
			
			\\
		\end{tabular}
		\caption{ \textbf{For ABA configuration}: Lattice morphologies of \textbf{top layer (at Left)}; \textbf{mid layer (at Middle)} and \textbf{bottom layer (at Right)} at $t=t_{morph}=10^{5}$ $MCS$ for a few selected coupling strengths and external field. The destruction of compensation in the cases (a) and (c) is due to the significant reduction of magnetic ordering in the top and bottom layers i.e. \textit{surface layers}. In the case (d), the increase in the value of total magnetisation from $-0.333$ (in field-free cases) to $-0.096$ (with $\sigma=1.00$) at the lowest temperature is due to the partial loss of magnetic order and it is responsible for the change in the nature of evolution of total magnetisation of the system.}
		\label{fig_morphology_ABA}
	\end{center}
\end{figure*}

\textbf{For ABA type composition:} in Figure \ref{fig_morphology_ABA}(a): with $J_{AA}/J_{BB}=0.04$ and $J_{AB}/J_{BB}=-0.04$, and with the introduction of the external field with a very low value ($\sigma=0.20$), the compensation vanishes. In the steady state, as the spin-field terms take effect the local spin configurations of the top and bottom layers lose most of the magnetic ordering [Please refer to the instantaneous values below the spin-density plots]. It is similar for the higher values of standard deviations. 
In Figure \ref{fig_morphology_ABA}(b): with $J_{AA}/J_{BB}=0.20$ and $J_{AB}/J_{BB}=-0.36$, within the exposure interval to the field, as the spin-field terms take effect the local spin configurations of the top and bottom layers lose most of the magnetic ordering for $\sigma=1.0$. Here, the per site cooperative energy of the surface A-layers are comparable with the spin-field energy per site. The changes in magnetization of the surface layers are not-detectable when $\sigma$ is $0.2$ or $0.5$, i.e. when per site spin-field energies are much lower compared to the cooperative energy. The same argument can be extended for the mid-layer where no detectable changes are seen from equilibrium towards steady state values of magnetisation for all the three standard deviations as the energies of spin-field terms are much lesser than the exchange energies ($J_{BB}$ is taken as the dominant coupling strength, equal to $1$, for all cases).

\begin{figure*}[!htb]
	\begin{center}
		\begin{tabular}{c}
			\phantom{}
			\hspace{0.0cm} \textbf{Top layer} \hspace{3.5cm} \textbf{Mid layer} \hspace{3.0cm} \textbf{Bottom layer} \\

			$\mathbf{(a)\text{ }AAB:\text{ } J_{AA}/J_{BB}=0.04;\text{ } J_{AB}/J_{BB}=-0.04 \text{ and } \sigma=0.50}$\\
			
			\resizebox{5.0cm}{!}{\includegraphics[angle=0]{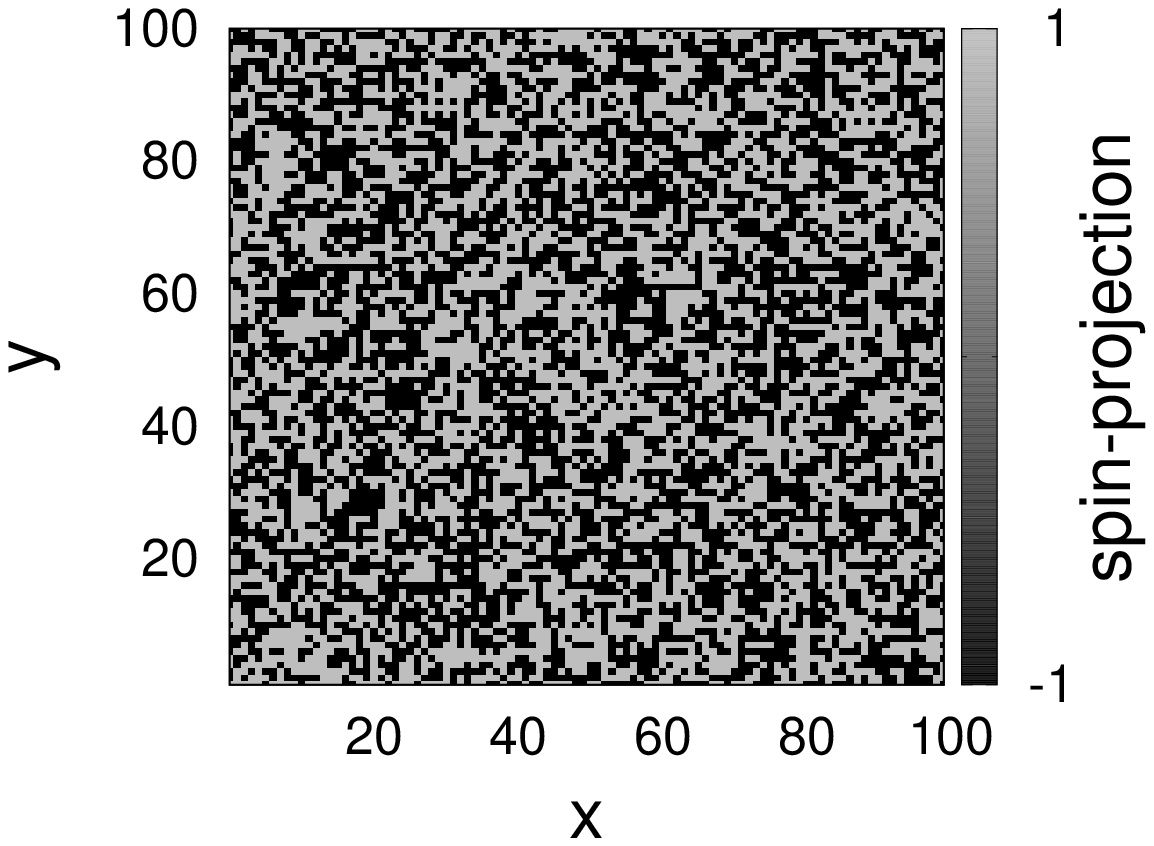}}
			
			\resizebox{5.0cm}{!}{\includegraphics[angle=0]{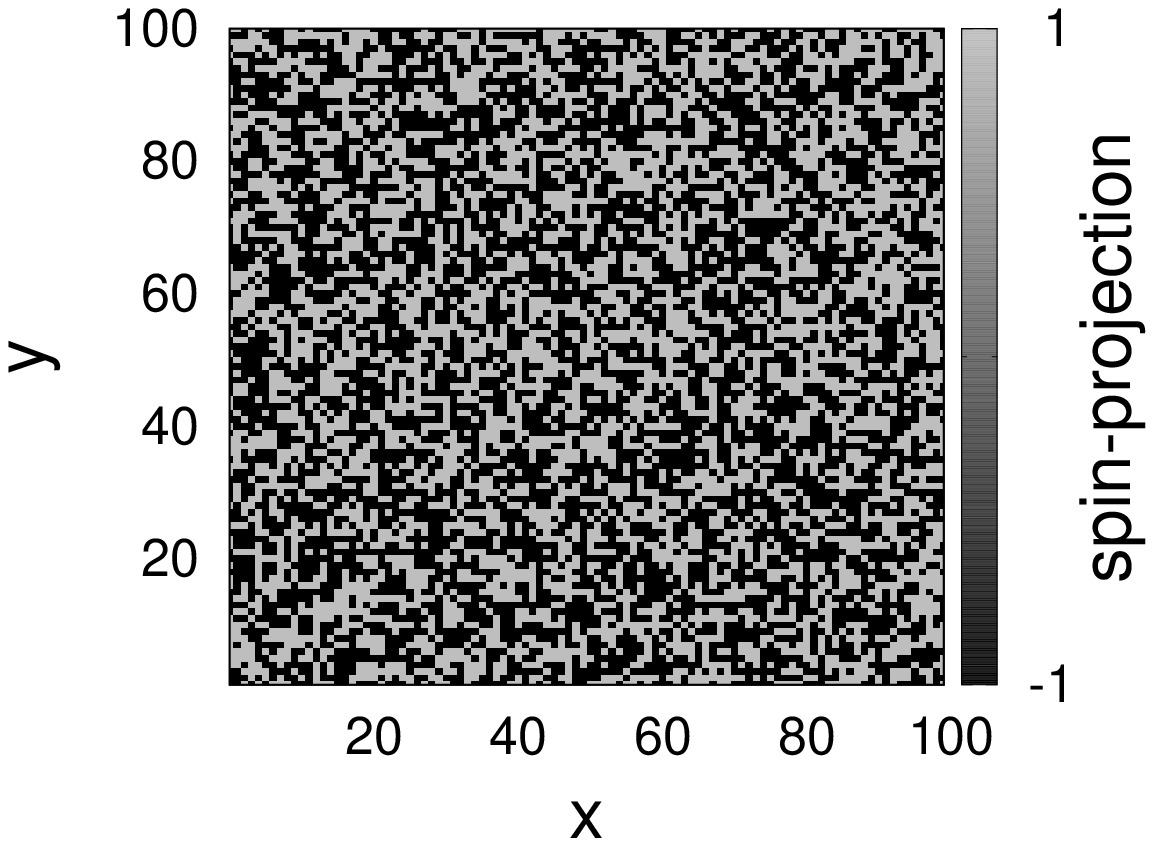}}
			
			\resizebox{5.0cm}{!}{\includegraphics[angle=0]{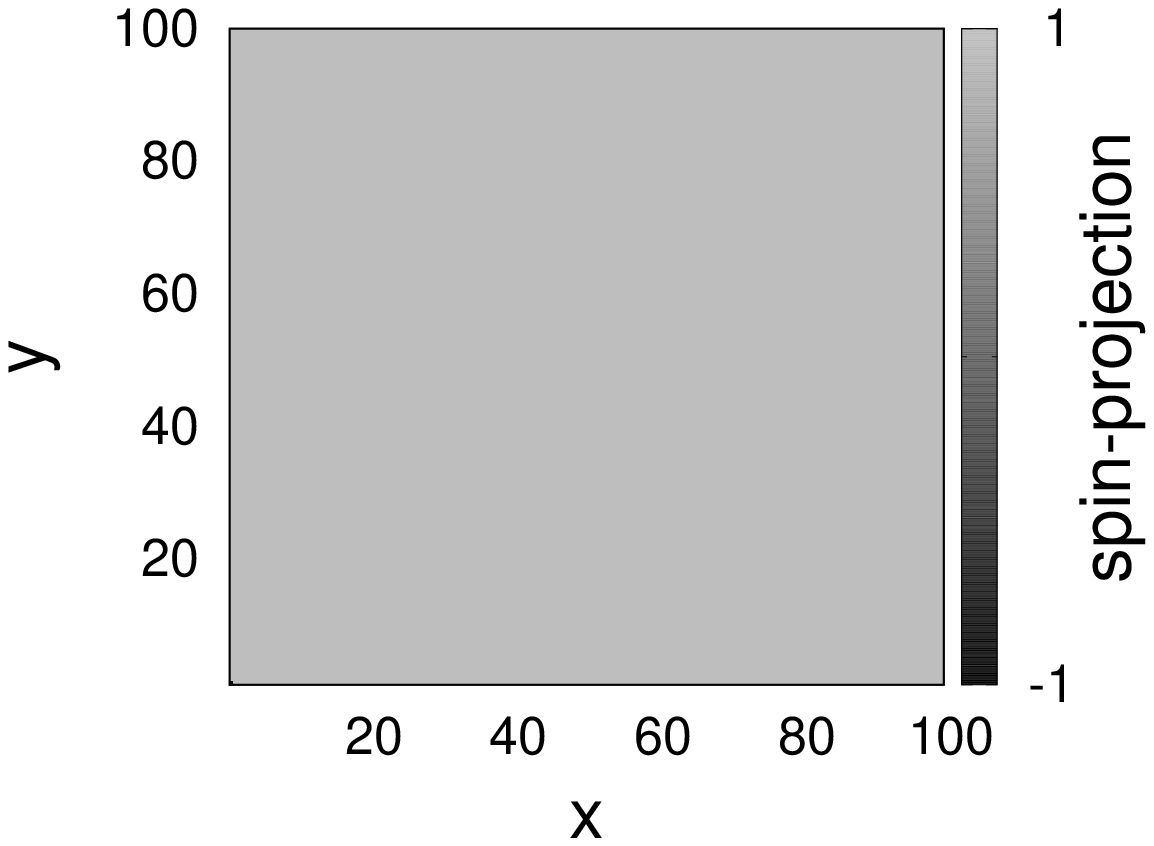}}\\
			
			\phantom{}
			\hspace{0.0cm} $M_{t}(t_{morph})=+0.003$ \hspace{1.5cm} $M_{m}(t_{morph})=-0.053$ \hspace{1.75cm} $M_{b}(t_{morph})=+1.00$\\
			
			\\

			$\mathbf{(b)\text{ }AAB:\text{ } J_{AA}/J_{BB}=0.20;\text{ } J_{AB}/J_{BB}=-0.36 \text{ and } \sigma=0.50}$\\
			
			\resizebox{5.0cm}{!}{\includegraphics[angle=0]{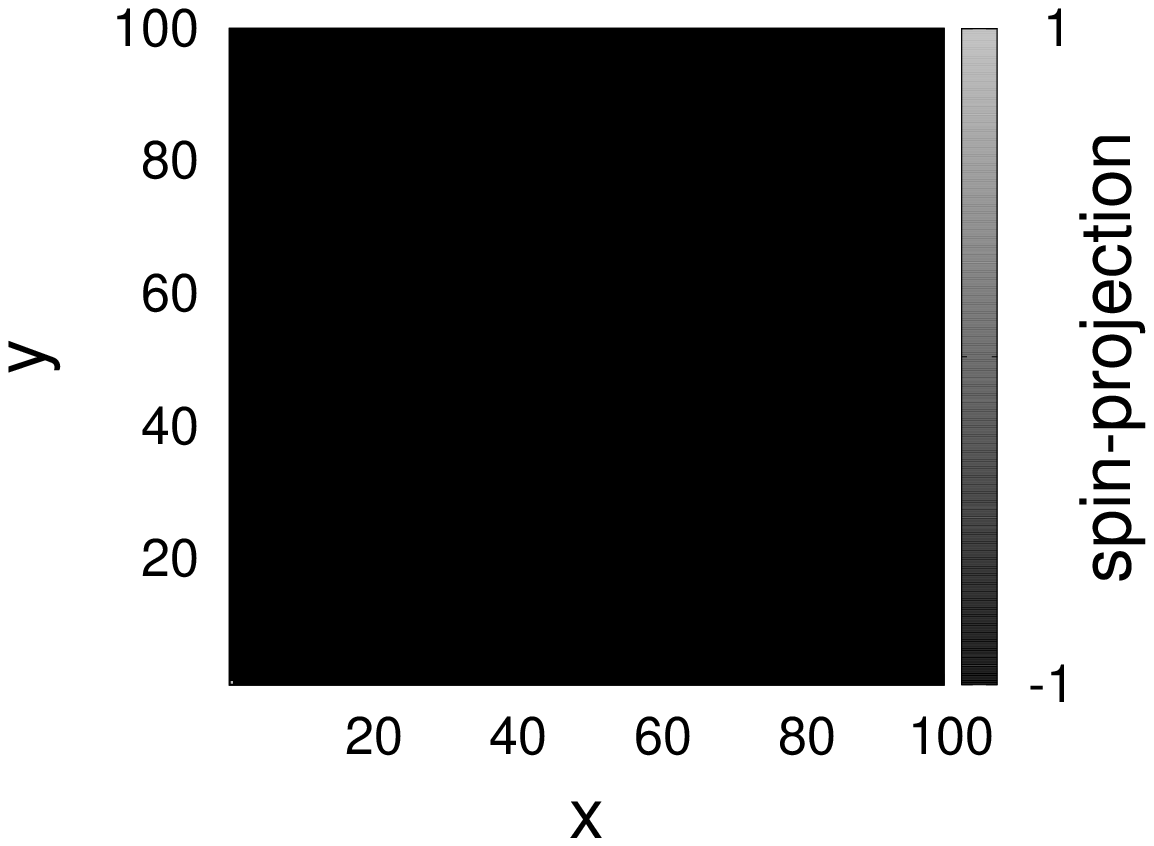}}
			
			\resizebox{5.0cm}{!}{\includegraphics[angle=0]{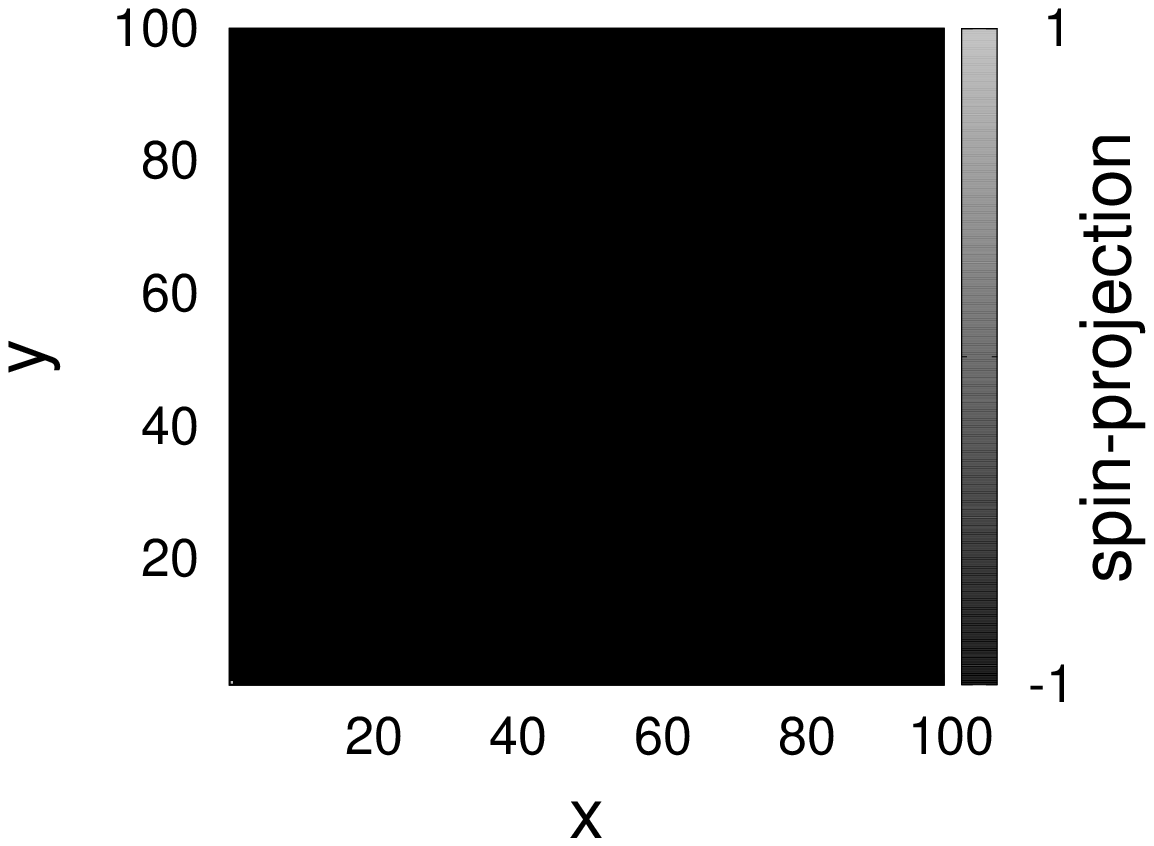}}
			
			\resizebox{5.0cm}{!}{\includegraphics[angle=0]{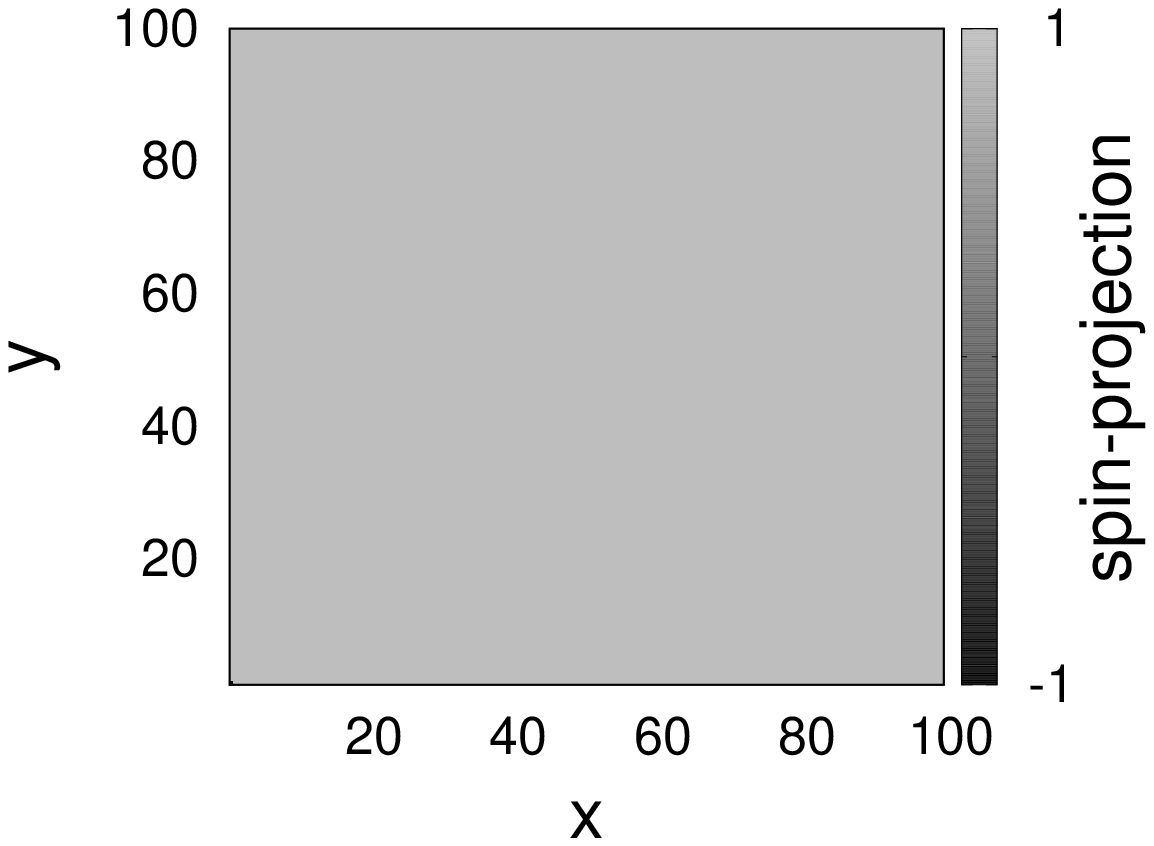}}\\
			
			\phantom{}
			\hspace{0.0cm} $M_{t}(t_{morph})=-1.00$ \hspace{1.5cm} $M_{m}(t_{morph})=-1.00$ \hspace{1.75cm} $M_{b}(t_{morph})=+1.00$\\
			
			\\
			
			$\mathbf{(c)\text{ }AAB:\text{ } J_{AA}/J_{BB}=0.20;\text{ } J_{AB}/J_{BB}=-0.36 \text{ and } \sigma=1.00}$\\
			
			\resizebox{5.0cm}{!}{\includegraphics[angle=0]{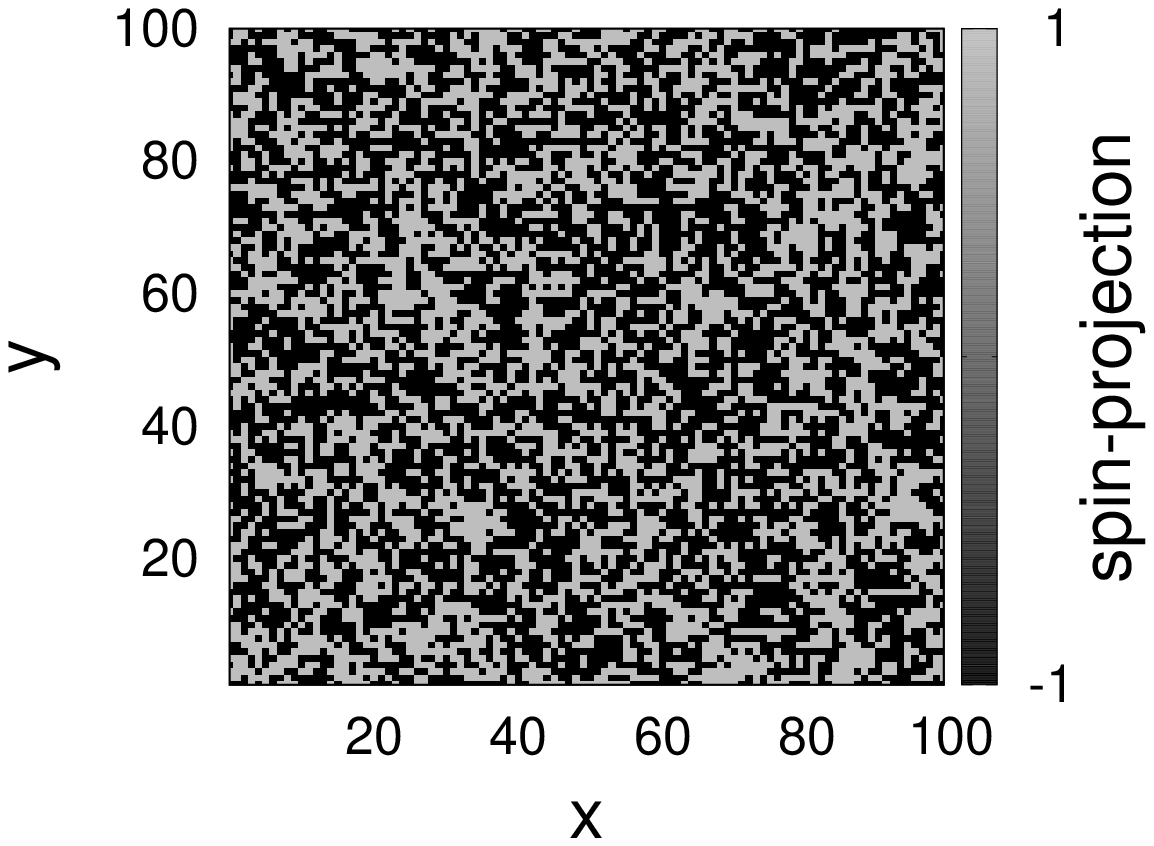}}
			
			\resizebox{5.0cm}{!}{\includegraphics[angle=0]{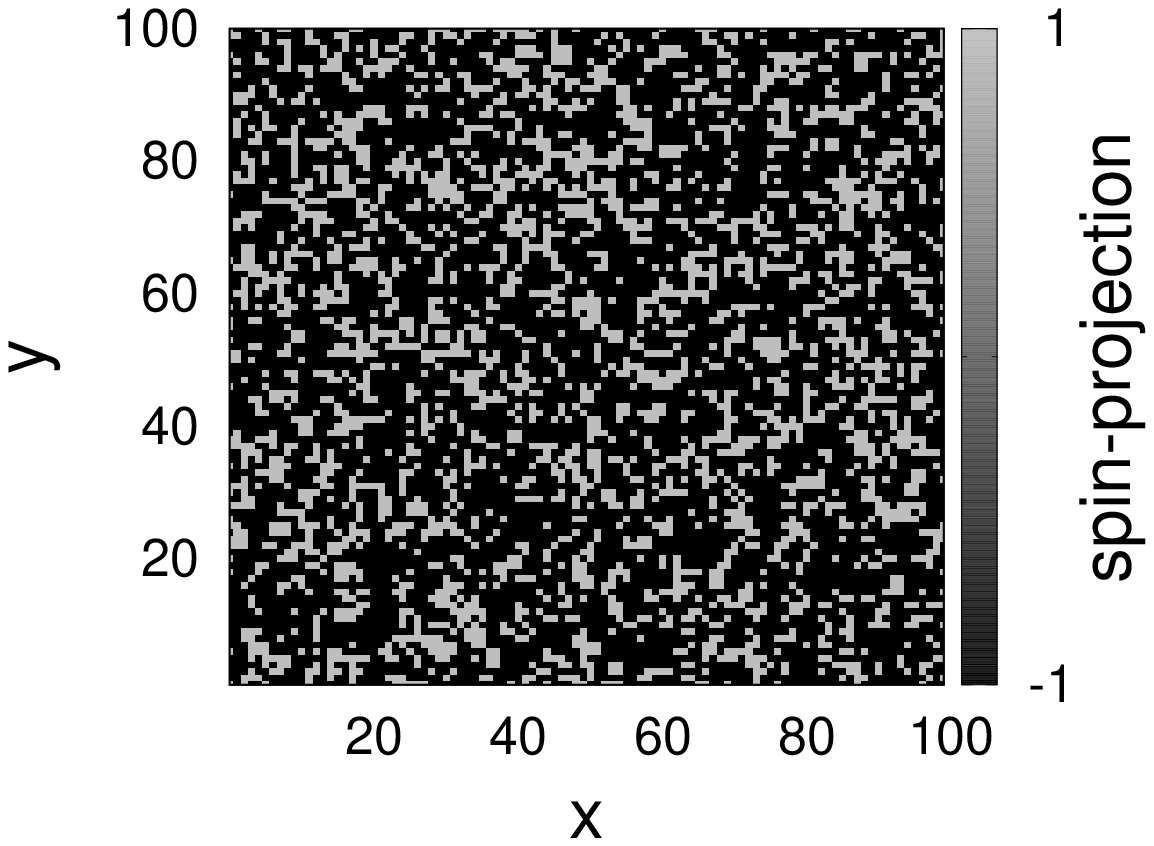}}
			
			\resizebox{5.0cm}{!}{\includegraphics[angle=0]{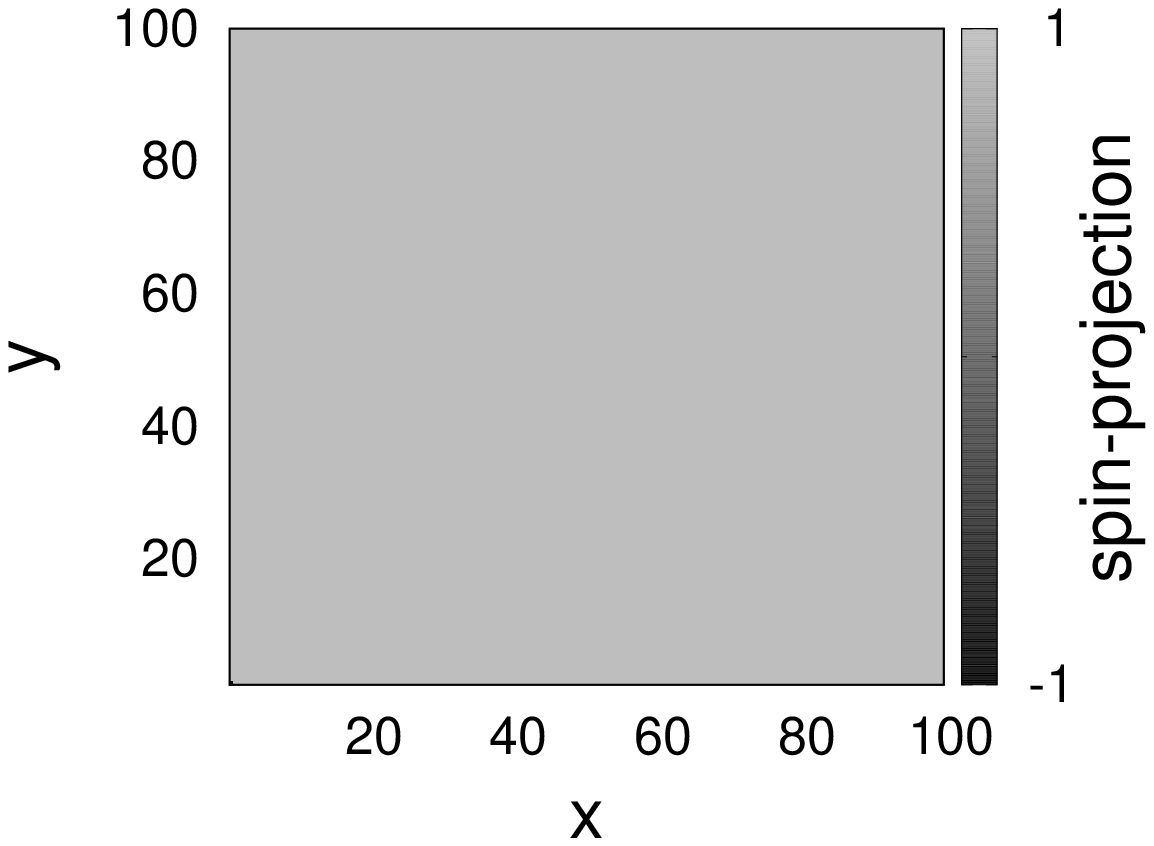}}\\
			
			\phantom{}
			\hspace{0.0cm} $M_{t}(t_{morph})=-0.101$ \hspace{1.5cm} $M_{m}(t_{morph})=-0.407$ \hspace{1.75cm} $M_{b}(t_{morph})=+1.00$\\
			
			\\
			
			$\mathbf{(d)\text{ }AAB:\text{ } J_{AA}/J_{BB}=0.20;\text{ } J_{AB}/J_{BB}=-1.00 \text{ and } \sigma=1.00}$\\
			
			\resizebox{5.0cm}{!}{\includegraphics[angle=0]{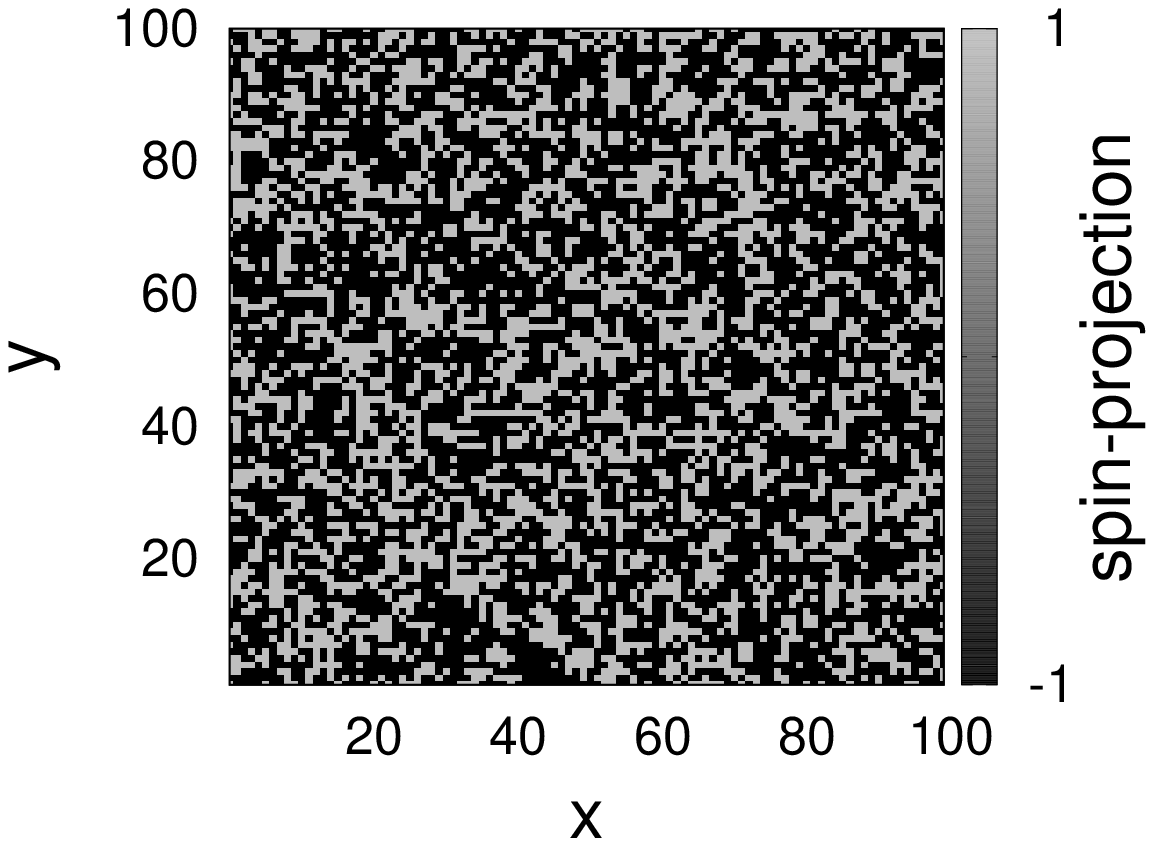}}
			
			\resizebox{5.0cm}{!}{\includegraphics[angle=0]{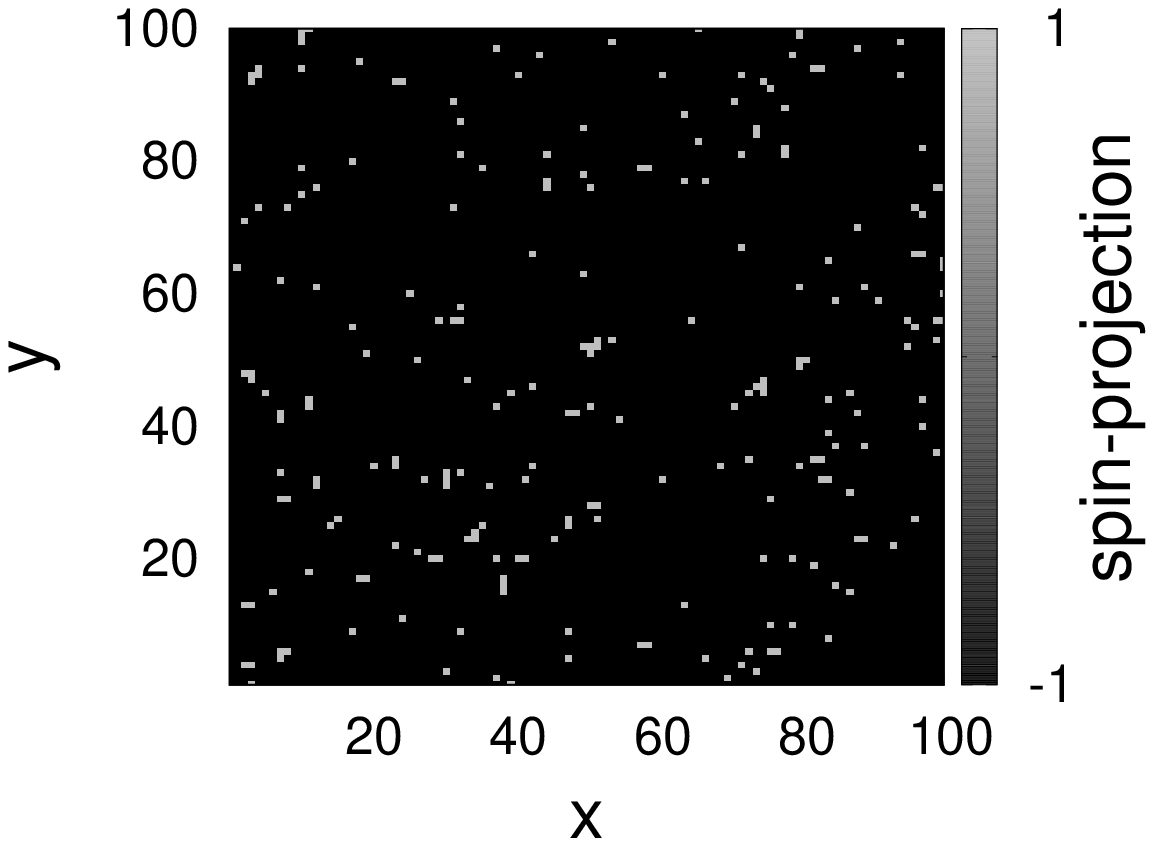}}
			
			\resizebox{5.0cm}{!}{\includegraphics[angle=0]{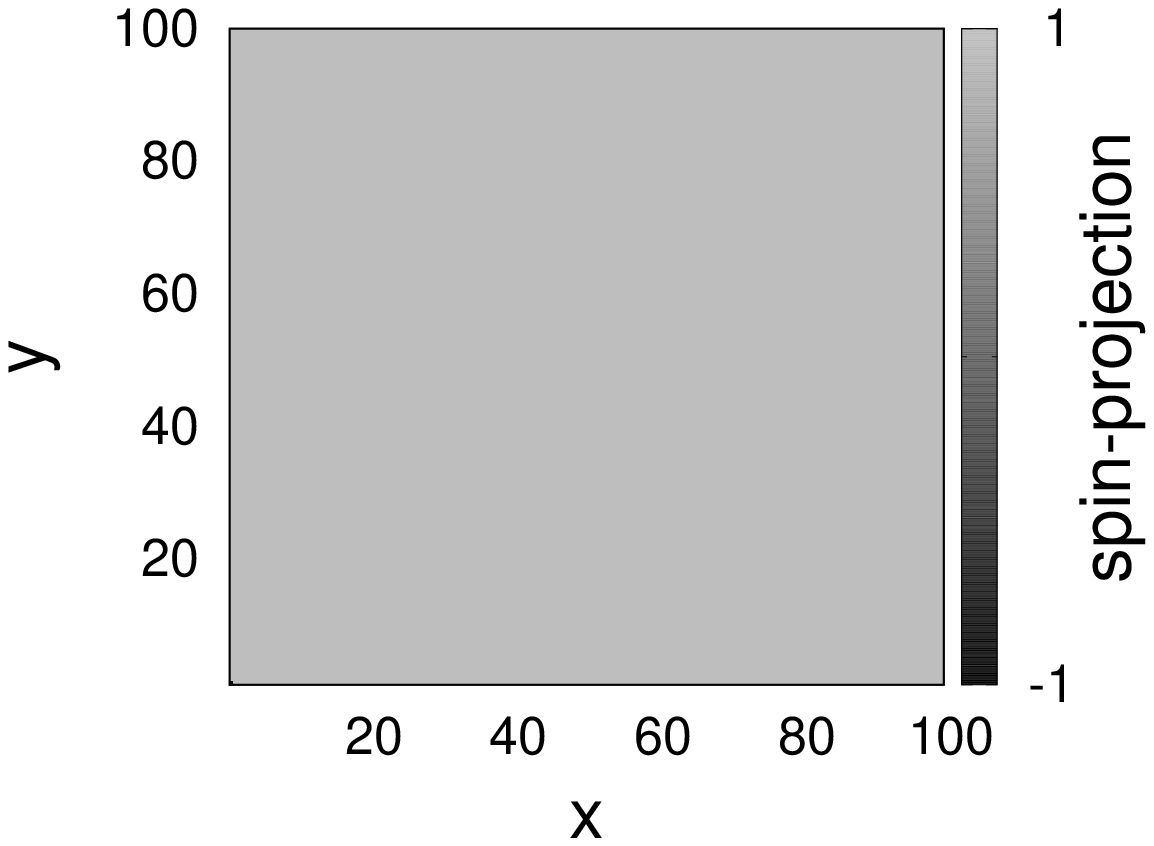}}\\
			
			\phantom{}
			\hspace{0.0cm} $M_{t}(t_{morph})=-0.208$ \hspace{1.5cm} $M_{m}(t_{morph})=-0.954$ \hspace{1.75cm} $M_{b}(t_{morph})=+1.00$\\
			
			\\
		\end{tabular}
		\caption{ \textbf{For AAB configuration}: Lattice morphologies of \textbf{top layer (at Left)}; \textbf{mid layer (at Middle)} and \textbf{bottom layer (at Right)} at $t=t_{morph}=10^{5}$ $MCS$ for a few selected coupling strengths and external field. The destruction of compensation in the cases (a) and (c) is due to the significant reduction of magnetic ordering in the top and bottom layers i.e. \textit{surface layers}. In the case (d), the increase in the value of total magnetisation from $-0.333$ (in field-free cases) to $-0.096$ (with $\sigma=1.00$) at the lowest temperature is due to the partial loss of magnetic order and it is responsible for the change in the nature of evolution of total magnetisation of the system.}
		\label{fig_morphology_AAB}
	\end{center}
\end{figure*}

Now we are in a position to understand why, in some cases, there is a change in the magnitude of total magnetisation at the lowest temeperature. For example, in Figure \ref{fig_mag_response}(b), even when compensation is present with $J_{AA}/J_{BB}=0.04$ and $J_{AB}/J_{BB}=-1.00$, for $\sigma=0.20$ onwards, we see changes in the nature of evolution (with respect to temperature) as well the absolute value. The nature of evolution is governed by the competition of energy scales between spin-external field and in-plane cooperative energies of the surface layers. But from $\sigma=0.76$ onwards, the strength of the external field is such that the spin-field energy per site is comparable to the combined cooperative energy (in-plane ferromagnetic and inter-plane anti-ferromagnetic) per site in the surface layers. This leads to a pronounced randomisation, similar in nature to the Figure \ref{fig_morphology_ABA}(c). The only difference is, the surface layers retain much of the magnetic ordering (for being partially randomised) to cancel out the magnetization of the midlayer at a shifted compensation temperature ($<T_{comp}(\sigma=0.00)$). In these two cases, the two competing energy scales are again comparable.

Figure \ref{fig_morphology_ABA}(d) supports the above argument in favour of different nature of evolution of total magnetisation [Figure \ref{fig_mag_response}(b)] which comes out to be the competition between the spin-field and cooperative energies. Here, with $J_{AA}/J_{BB}=0.04$ ; $J_{AB}/J_{BB}=-1.00$ and $\sigma=1.00$, the per site cooperative energy in surface A-layers is comparable to the per site coupling energy with the external field.

At this point, in light of the explanations provided in the Section \ref{subsec_response} and above for the ABA configuration, we can understand the behaviour of spin-density plots of AAB configuration in Figure \ref{fig_morphology_AAB}, for a few interesting selected cases.
\subsection{Phase Diagram and Scaling}

\begin{figure*}[!htb]
	\begin{center}
		\begin{tabular}{c}
			\textbf{(a)}
			\resizebox{7cm}{!}{\includegraphics[angle=0]{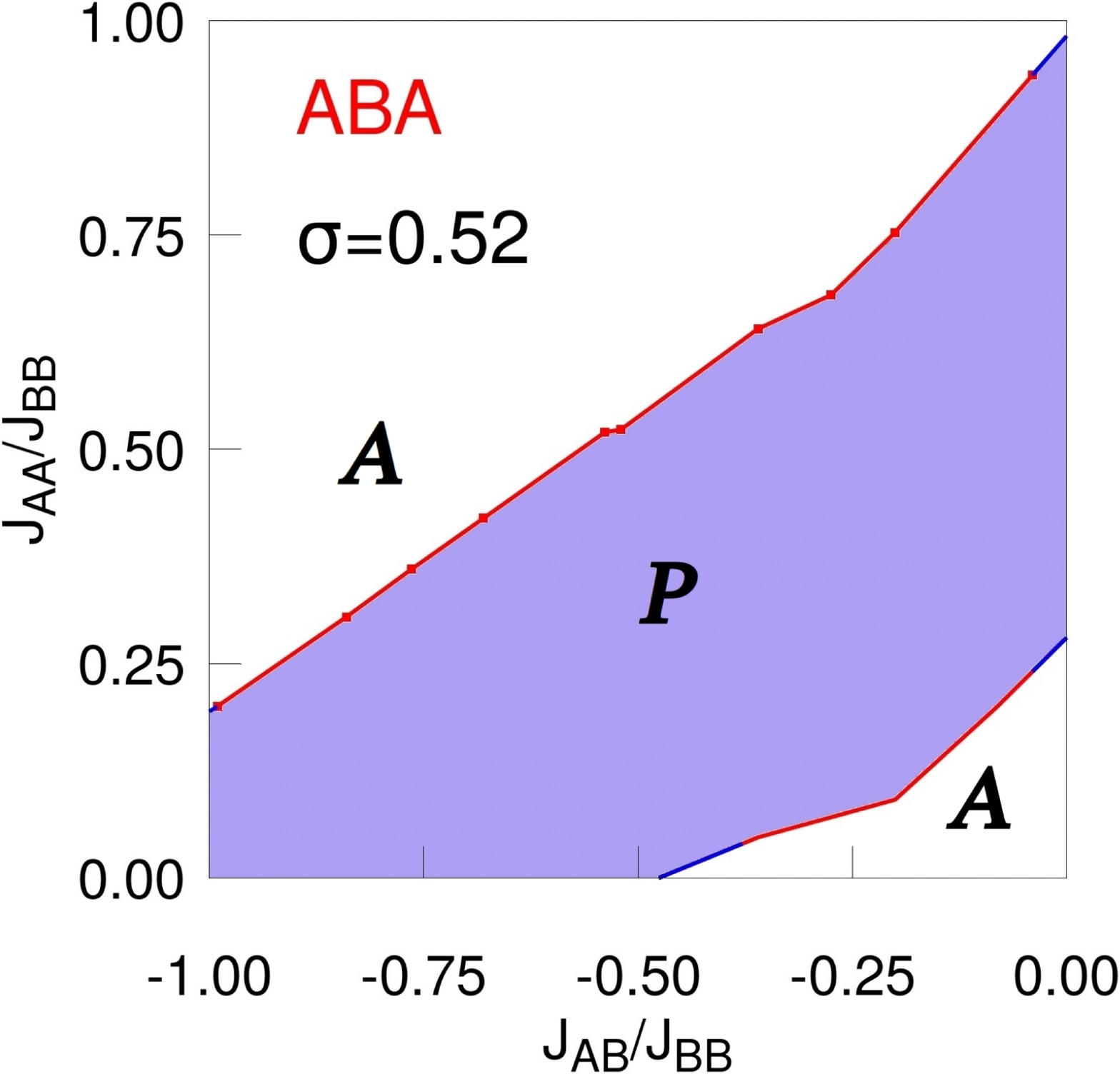}}
			\hfill
			\textbf{(b)}
			\resizebox{7cm}{!}{\includegraphics[angle=0]{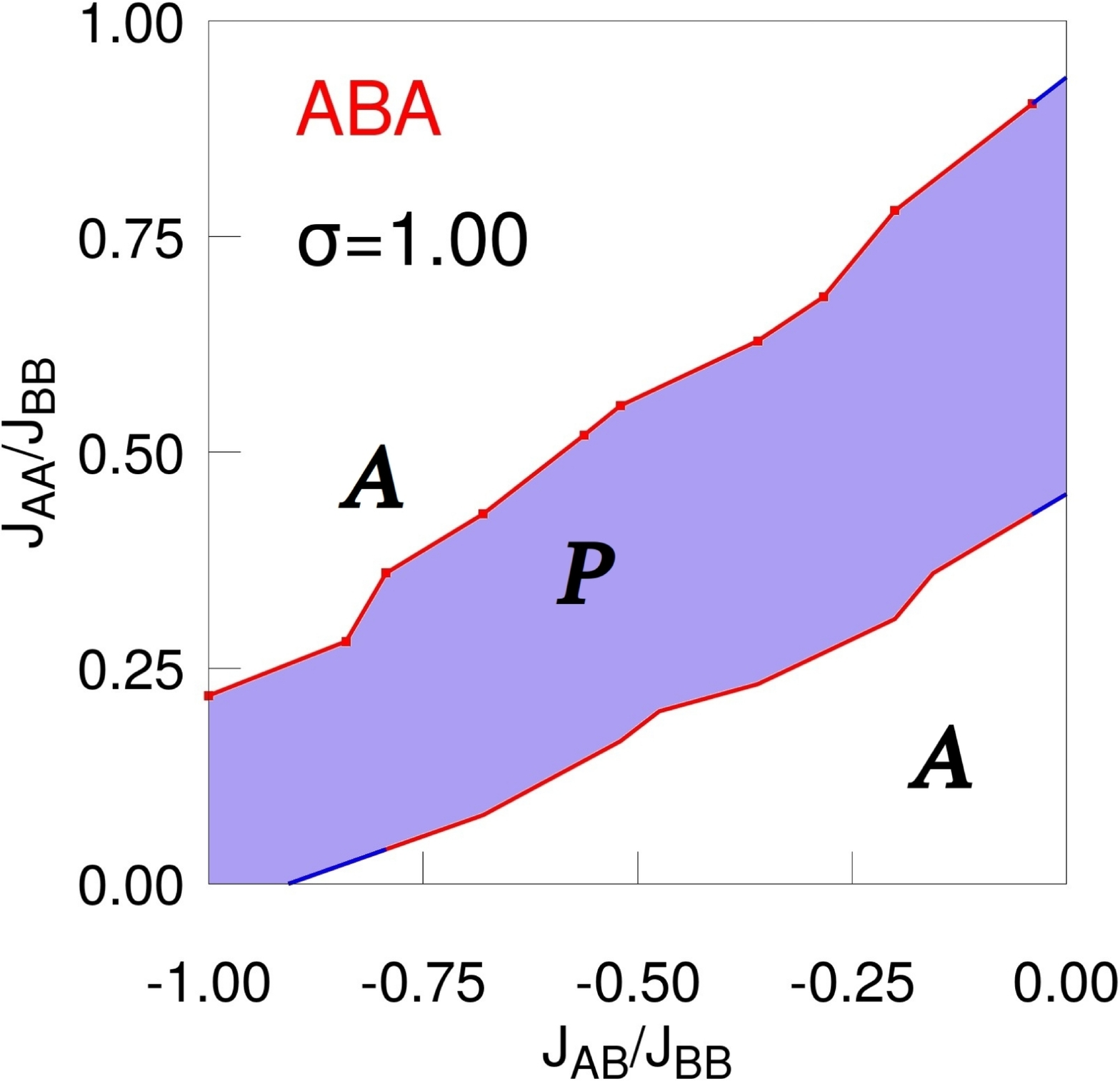}}\\
			
			\textbf{(c)}
			\resizebox{7cm}{!}{\includegraphics[angle=0]{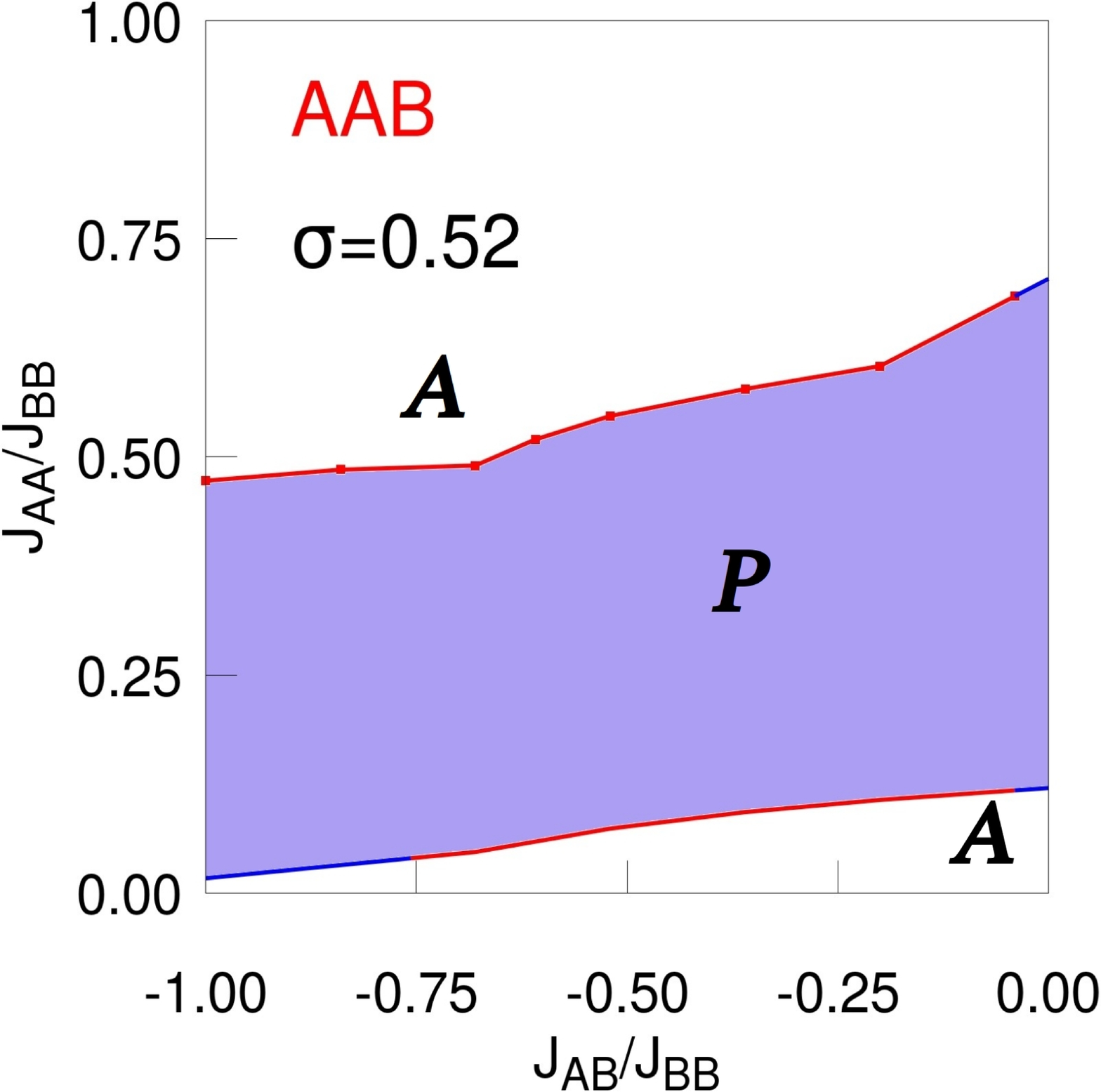}}
			\hfill
			\textbf{(d)}
			\resizebox{7cm}{!}{\includegraphics[angle=0]{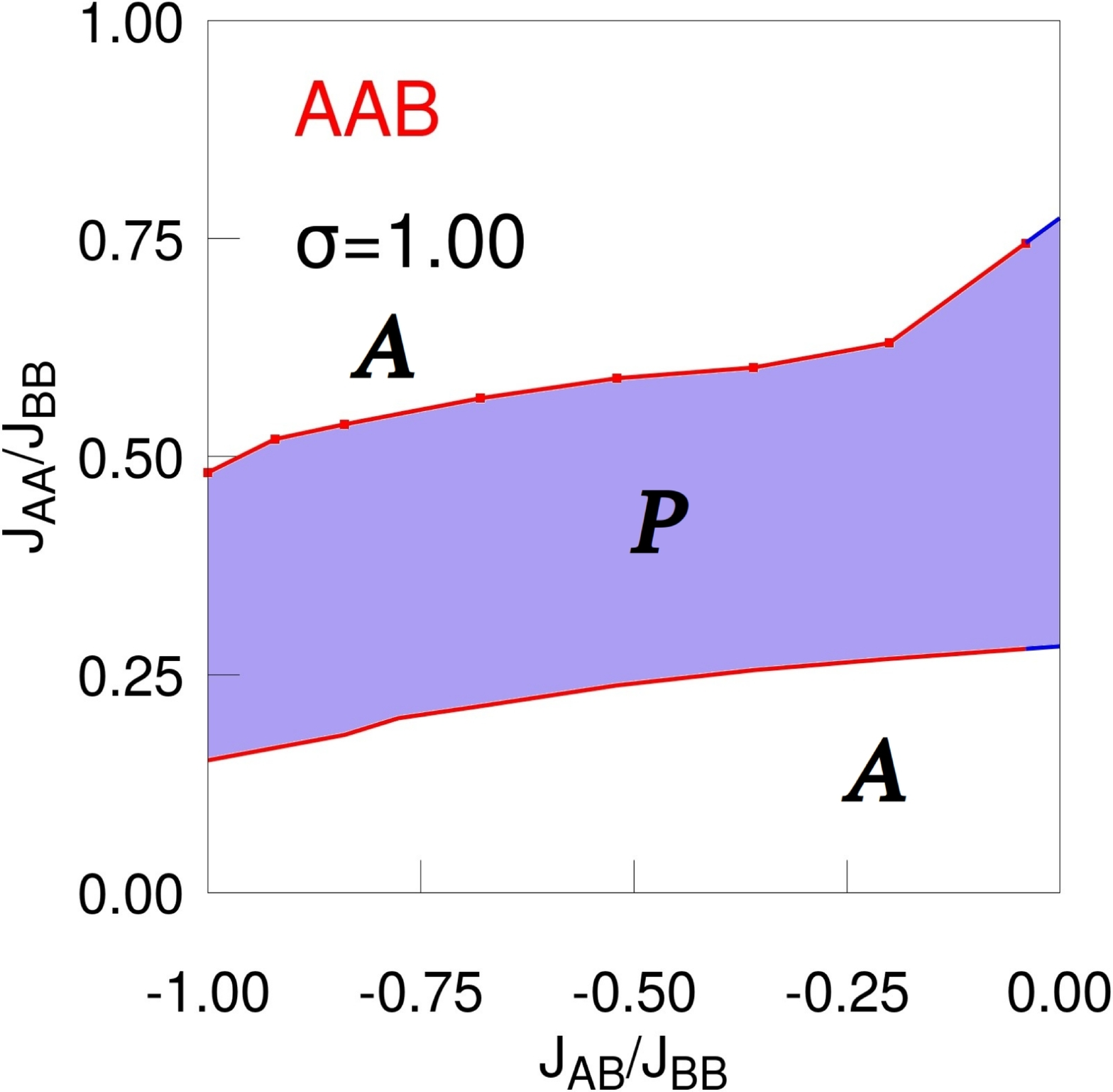}}
			
		\end{tabular}
		\caption{ (Colour Online) Phase diagram for the: ABA trilayered ferrimagnetic system when: (a) $\sigma=0.52$; (b) $\sigma=1.00$ and AAB trilayered ferrimagnetic system when: (c) $\sigma=0.52$; (d) $\sigma=1.00$, in presence of the uniform random external magnetic field. A: Compensation is absent; P: Compensation is present. With increase in the standard deviation of the external field, the magnitude of the area of the no-compensation island have grown. The blue segment of the phase separation curves are obtained via linear extrapolation. All these plots are obtained for a system of $3\times100\times100$ sites. Where the errorbars are not visible, they are smaller than the point markers.}
		\label{fig_phasecurve_2d}
	\end{center}
\end{figure*}

\indent For a fixed standard deviation or strength of the external uniform random field with spatio-temporal variation, in both the configurations: ABA and AAB spin-1/2 Ising trilayered ferrimagnet on square lattice, we have the following common observations:\\
(a) In presence of the field and in those cases \textit{with compensation}, compensation temperature merges with the critical temperature for higher values of $J_{AA}/J_{BB}$ when $|J_{AB}/J_{BB}|$ is fixed or vice-versa, just like in the zero field case. This implies the phase diagrams, in presence of the field, can be drwan by following similar procedures as in the references \cite{Diaz,Chandra}. In Figure \ref{fig_phasecurve_2d}, the areas coloured in pale-blue are where compensation is \textit{present} (marked by P) and the area(s) in white are the regions where compensation is absent (marked by A). The qualitative features of the phase separation curves are same like that in the field-free cases. These curves divide the entire area of the parameter space into: one contains ferrimagnetic phases with compensation and the others contain ferrimagnetic phases without compensation. \\
(b) But in cases where the external field is present [From $\sigma$=0.2, onwards], we see the field makes compensation disappear for a certain range of values of the coupling strengths [Refer to Figure \ref{fig_phasecurve_2d}]. This observation is a novel one. In the phase diagrams, the range of values of such coupling strengths for which compensation is absent are confined in areas within the region where compensation is present. The appearance of such closed areas resemble like islands or enclaves. \textit{The magnitude of areas of such an island also grows as the randomness of the external field increases}. 

The islands of \textit{no compensation} starts to appear from the low coupling strength part of the phase diagrams (when $\sigma=0.20$). Then the area of the No-Compensation island (NCI) gradually increases (reaches higher values of coupling ratios) with increase in the standard deviation of the field.
\begin{figure*}[!htb]
	\begin{center}
		\begin{tabular}{c}
			\textbf{(a)}
			\resizebox{8.5cm}{!}{\includegraphics[angle=-90]{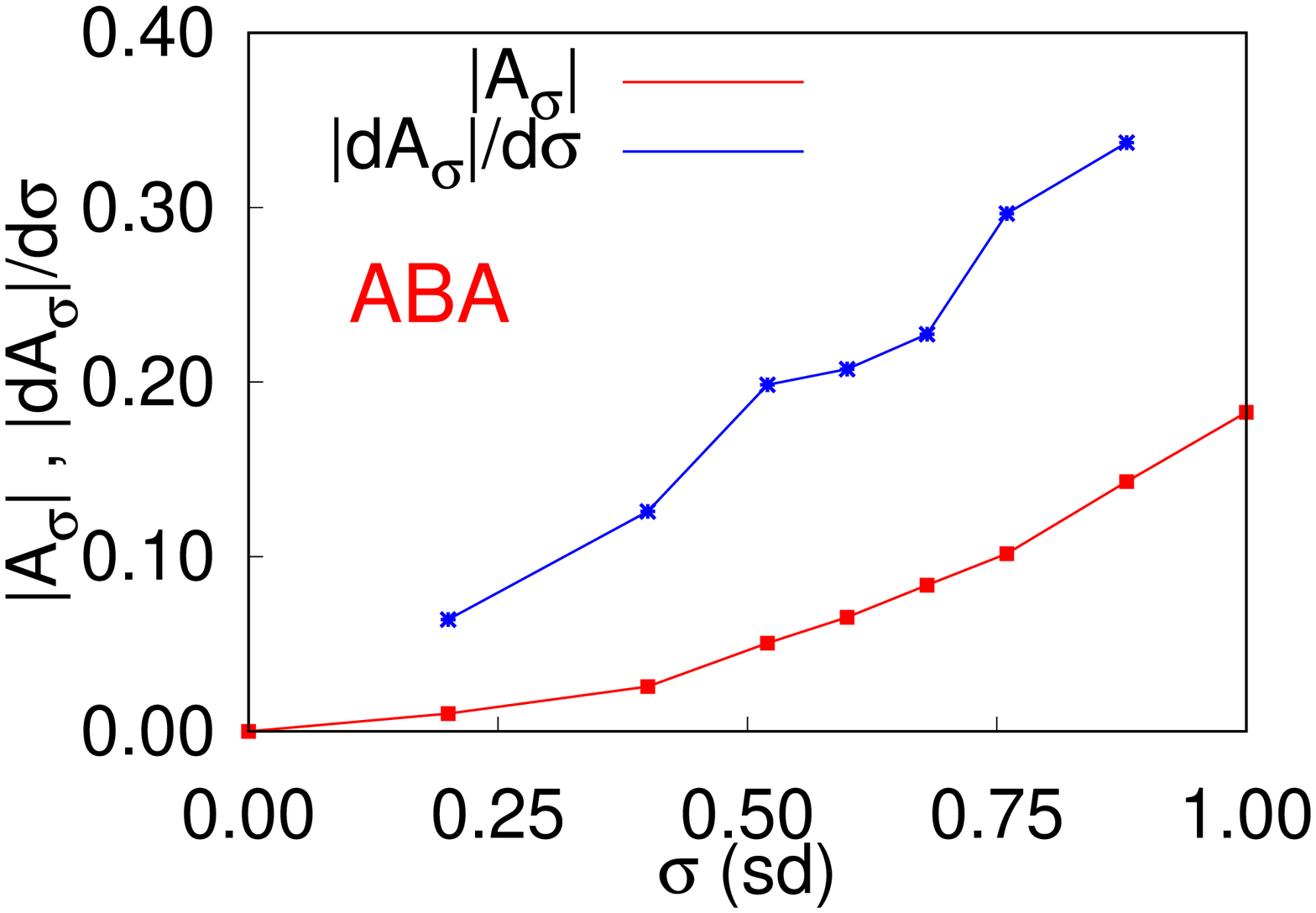}}
			\hfill
			\textbf{(b)}
			\resizebox{8.5cm}{!}{\includegraphics[angle=-90]{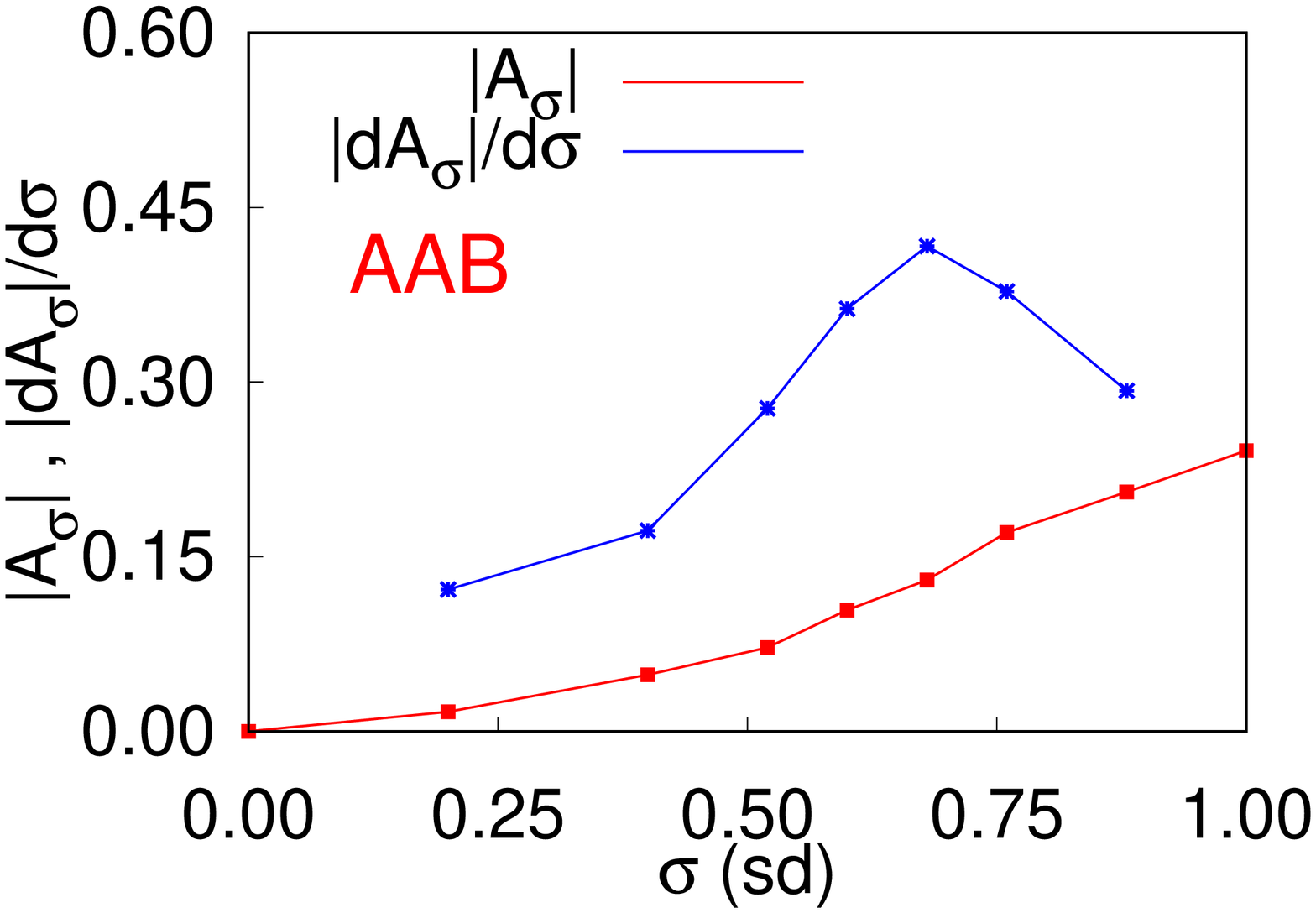}}
			
		\end{tabular}
		\caption{ (Colour Online) Plots of: Magnitude of the area of the no-compensation islands versus standard deviation of the field (in RED) and the rate of increase in the magnitude of the area of the no-compensation islands versus standard deviation of the field (in BLUE) for (a) ABA and (b) AAB configurations. All these plots are obtained for a system of $3\times100\times100$ sites.}
		\label{fig_area_slope_field}
	\end{center}
\end{figure*}
In Figure \ref{fig_area_slope_field}, we see the plots of absolute area and rate of increase of absolute area versus the sd of the applied field. With reference to Figure \ref{fig_mag_fr_afr}, let's discuss \textbf{how the magnitude of the area of NCIs are determined:} 
\begin{itemize}
	\item[(a)] Variation of the magnetization at the lowest simulational temeperature with coupling strengths is noted where external field destroys compensation at any of its strengths.
	\item[(b)] The variation in (a) is linearly approximated to find out the value of the coupling strength where the magnetization at the lowest temperature is zero. Only above this approximated magnitude of relative coupling strength, we should expect compensation. 
	\item[(c)] Intermediate points and the leading and trailing parts of the separation curves are again linearly approximated. This provides us with a closed curve.
	\item[(d)] Monte Carlo integration is now employed \cite{Krauth} to find out the fractional area under the curve obtained in (c).
	\item[(e)] Central difference formula is used to find out the rate of increase of the area of NCIs. 
\end{itemize}

A few comments with respect to the nature of the curves of absolute area versus the sd are in order. First, the curve (in RED), comes out to be mostly a \textit{superlinear} one for the ABA configuration and mixture of a \textit{superlinear and sublinear} ones for the AAB configuration. The plots of the slopes (in BLUE) also confirms it. Second, where the spin-field energies per site are greater than the in-plane cooperative energies per site, the compensation disappers. For higher strengths, the spin-field interaction energies can greatly influence the higher ends of the cooperative energies. For the islands, the area unavailable for compensation for both the configurations are mostly affected at the lower in-plane coupling strength region. The regions with lowest in-plane coupling ratios are affected by all the strengths of the external random field, starting from $\sigma=0.20$. as the cooperative energies in this region are smaller than spin-field terms for most of the field-strengths.

Now an interesting exercise can be carried out on how the magnitude of the area of NCIs scale with the sd of the field. This would be an $1D$ scaling and let's assume the following scaling function:
\begin{equation}
\label{eq_scaling}
	f(A(\sigma), \sigma)= \sigma^{-b}A(\sigma) 
\end{equation}
The scaling exponent comes out to be, for ABA: $b_{ABA}=1.958\pm0.122$ and for AAB: $b_{AAB}=1.783\pm0.118$. The estimate of error quoted in scaling exponent, $b$, is obtained by the standard deviation among all the sets of data. The agreement is a little poor at the two extreme ends ($\sigma$ is either $0.20$ or $1.00$ ) for both the configurations. At the low ends, the effect of the field is not much pronounced while at the highest end, the absolute values of the area of NCIs tend to saturate.
\section{Summary}
\label{sec_summary}

\indent From the zero-field cases \cite{Chandra}(b), it is already established from lattice morphologies for both the configurations of Figure \ref{fig_lattice_structure} that magnetic ordering develops considerably at the compensation temperatures in the A-layers (for cases with compensation and weaker $J_{AA}$) and sublayers are almost completely ordered at the lowest temperature. With the increase in the field-strength in temeperatures much lower than $T_{crit}$ (nearly athermal), for both the ABA and AAB configurations, magnetic ordering gradually diminishes in the A-layers in its steady state (Refer to Figures \ref{fig_mag_response},\ref{fig_fluc_mageng},\ref{fig_morphology_ABA} and \ref{fig_morphology_AAB}; when the magnitudes of per site spin-field energies is comparable to in-plane cooperative energies per site). This leads to having a much lower magnitude of sublattice magnetisations in the A-layers than in the field-free case. In moderate-to-high field strengths, such low sublattice magnetisations in the A-layers are unable to cancel out the magnetization in the mid B-layer at even in the nearly athermal case. This is the reason why compensation disappears when the field strength is comparably higher than in-plane coupling strengths of the surface A-layers. This can also be judged from the Figure \ref{fig_fluc_mageng}, where the plateau around compensation point flattens and loses its smeared peak (which happens at the location of compensation) with increasing strength of the random field. Even at the lowest point of tempearture in the simulation, both the fluctuations increase from zero with increase in the standard deviation of the external field. This indicates gradual increase in the randomness in the orientations of spin projections in the surface A-layers. The origin of \textit{dynamical field-driven vanishing of compensation} can be attributed to the \textit{diminishing of magnetic ordering in the A-layers} when the spin-field interaction energy is larger than in-plane coupling, for the trlayered spin-1/2 Ising ferrimagnetic systems of Figure \ref{fig_lattice_structure}. Figure \ref{fig_phasecurve_2d} shows how the enclave of No-Compensation comes up within the phase diagram and evolves with the increase of the strength of the external uniform random field. A proposed scaling relation is found to hold good [Equation \ref{eq_scaling}] with exponent $b_{ABA}=1.958\pm 0.122$ and $b_{ABA}=1.783\pm 0.118$. So the temporally varying and site dependent external random field, $h_{i}(t)$'s in Equation \ref{eq_Hamiltonian} can reveal novel features in the phase diagrams [Figure \ref{fig_phasecurve_2d}] and represent systems where the Hamiltonian is itself a dynamic one. The coupling constants in the Ising model is traditionally understood to be translationally invariant. Under such a constraint, if competing ferromagnetic and anti-ferromagnetic interactions are taken into account, the Ising model exhibits a remarkable complexity. For an example, the equilibrium studies of the system studied in the current article can be mentioned \cite{Diaz,Chandra}. In reality, compositional disorder, impurities, vacancies, lattice dislocations etc. lead to modifications in the Hamiltonian, which, with Ising mechanics, may be characterized by changes which are no longer translationally invariant, but random quantities, characterized by their probability distributions. It would be interesting to study how the system behaves when it is exposed to other types of random magnetic fields with probabilty distribution functions. This is planned for the future. 
\section*{Acknowledgements}
The author acknowledges financial assistance from University Grants Commission, India in the form of Research Fellowship and extends his thanks to Tamaghna Maitra for technical assistance. Useful comments and suggestions provided by the anonymous referees are also acknowledged.
\vspace{1cm}
\begin{center} {\Large \textbf {References}} \end{center}

\end{document}